\documentclass[twocolumn]{openjournal}

\usepackage{lipsum}

\usepackage{xcolor}
\usepackage{textgreek}
\usepackage[utf8]{inputenc}
\usepackage[english]{babel}

\usepackage{hyperref}
\hypersetup{
    unicode, 
    colorlinks=true,
    linkcolor=linkcolor,
    citecolor=linkcolor,
    filecolor=linkcolor,
    urlcolor=linkcolor,
}
\usepackage{color,colortbl}
\definecolor{linkcolor}{rgb}{0.0,0.3,0.5}
\usepackage{tensind}
\tensordelimiter{?}
\DeclareGraphicsExtensions{.bmp,.png,.jpg,.pdf}
\usepackage{verbatim}
\usepackage[normalem]{ulem}
\usepackage{orcidlink}
\usepackage{soul}

\usepackage{makecell}
\usepackage{subfigure}
\usepackage{float}
\usepackage{graphicx}	
\usepackage{amsmath}	
\usepackage{amssymb}	
\usepackage{newtxtext,newtxmath}
\usepackage{multirow}

\usepackage{placeins}

\begin{document}
\title{Profile Analysis of the Multiwavelength 2.1-year Oscillations of PG 1553+113}

\author{P. Pe\~nil$^{1,\ast}$\orcidlink{0000-0003-3741-9764},
        N. Torres$-$Alb\`a$^{2,3,\ast}$\orcidlink{0000-0003-3638-8943}, 
        A. Rico$^{1,\ast}$\orcidlink{0000-0001-5233-7180},
        L. Marcotulli$^{1,4}$\orcidlink{0000-0002-8472-3649},
        A. Dom\'inguez,$^{5}$\orcidlink{0000-0002-3433-4610},
        M. Ajello$^{1}$\orcidlink{0000-0002-6584-1703},        
        S. Buson$^{4,6}$\orcidlink{0000-0002-3308-324X},        
        S. Adhikari$^{1}$\orcidlink{0009-0006-1029-1026}
        }
\thanks{$^\ast$ppenil@clemson.edu, nuria@virginia.edu, aricoro@clemson.edu}

\affiliation{$^1$Department of Physics and Astronomy, Clemson University, Kinard Lab of Physics, Clemson, SC 29634-0978, USA}
\affiliation{$^2$Department of Astronomy, University of Virginia, P.O. Box 400325, Charlottesville, VA 22904, USA}
\affiliation{$^3$GECO Fellow}
\affiliation{$^4$Deutsches Elektronen-Synchrotron DESY, Platanenallee 6, 15738 Zeuthen, Germany}
\affiliation{$^5$IPARCOS and Department of EMFTEL, Universidad Complutense de Madrid, E-28040 Madrid, Spain}
\affiliation{$^6$Julius-Maximilians-Universit\"at W\"urzburg, Fakultät f\"ur Physik und Astronomie, Emil-Fischer-Str. 31, D-97074 W\"urzburg, Germany}

\begin{abstract}
We investigate the morphology of the oscillation profiles of the blazar PG~1553+113 in relation to its well-known $\sim$2.1 yr periodicity. We identify individual cycles in the $\gamma$-ray, X-ray, UV, and optical light curves and characterize their temporal profiles using analytical models for single- and multi-peaked events.

We find that the oscillations are generally described by a broad activity envelope with shorter-timescale substructure, showing that the $\sim$2.1 yr signal is not a strictly sinusoidal or self-similar modulation. The internal morphology varies across cycles and energy bands. This is particularly evident in X-rays, where all analyzed cycles show a strong formal preference for multi-component profiles, unlike the $\gamma$-ray band, where several cycles admit statistically comparable empirical descriptions.

The contemporaneous MWL oscillations show broadly aligned activity episodes, but the timing and relative amplitudes of secondary components are not systematically repeated. This suggests that a common long-term modulation affects the broadband emission, while additional local or energy-dependent processes shape individual cycles. Such a picture is compatible with a geometric, jet-related contribution to the broad recurrent envelope, with intrinsic variability superimposed on it.

We also identify new cycles with a dominant peak accompanied by weaker twin-peak-like features, similar to structures previously discussed in a supermassive black hole binary scenario for PG~1553+113. Although our results do not provide definitive evidence for this interpretation, the recurrence of comparable morphologies in newly analyzed cycles keeps this scenario viable.
\end{abstract}


\maketitle

%

\section{Introduction}
PG~1553+113 is a BL Lac object that has become one of the benchmark sources for long-term periodic variability studies in blazars. Its relevance is mainly due to the $\sim$2.1 yr modulation first reported in \textit{Fermi}-LAT $\gamma$-ray observations by \citet{ackermann_pg1553}, which led to its interpretation as a possible supermassive black hole binary system \citep[SMBHB; e.g.,][]{ackermann_pg1553, tavani_pdm_pg_1553, stefan_pg1553_2024, sagar_pg1553}. Subsequent analyses using \textit{Fermi}-LAT data spanning a larger number of years have found consistent evidence for periodic pattern at a similar period, consolidating PG~1553+113 as one of the most significant $\gamma$-ray periodicity candidates among blazars \citep[e.g.,][]{penil_2020, alba_ssa, stefan_pg1553_2024, magic_pg1553_2024, penil_24candidates_2025, penil_17fermi_candidates_2026}.

Evidence for the same periodic pattern has also been investigated at other wavelengths. \citet{ackermann_pg1553} also reported similar behavior in the optical and radio bands. Subsequent multiwavelength (MWL) analyses, including broader optical and UV coverage, reached similar conclusions and reinforced the view that the periodicity of PG 1553+113 is not limited to the $\gamma$-ray band \citep[][]{penil_mwl_pg1553, magic_pg1553_2024, stefan_pg1553_2024}. In the X-ray band, however, the situation remains less clear. Some studies have reported a less significant signal at a period compatible with the $\gamma$-ray period \citep[][]{gao_pg1553_2023}, whereas others have found indications of a different timescale near $\sim$1.5 yr \citep[][]{penil_mwl_pg1553, aniello_pg1553_xray_2024}. More recently, \citet{penil_xray_pg1553_2026} reported hints of a 2.1-year periodic pattern in X-rays, although the combination of strong variability and substantial gaps in the light curve prevented a robust confirmation.

The MWL correlations between energy bands have also been studied. Cross-correlation analyses between the $\gamma$-ray and X-ray/UV/optical light curves have reported significant ($\geq$3$\sigma$) inter-band correlations, while the inferred time delays are generally consistent with zero within the uncertainties. This behavior suggests that the variability across these bands is closely connected, either through co-spatial emission regions or through regions responding on timescales shorter than the temporal resolution of the data \citep[e.g.,][]{penil_mwl_pg1553, magic_pg1553_2024, penil_xray_pg1553_2026}.

In addition to the year-scale modulation, recent studies have identified a longer-term trend in the MWL emission of PG 1553+113 over approximately the last decade of observations \citep[][]{penil_mwl_pg1553, alba_ssa, penil_xray_pg1553_2026}. By extending the optical coverage to nearly a century using archival data of about 100 years, \citet{sagar_pg1553} interpreted this long-term behavior as part of a longer modulation of about 22 yr. That longer modulation was discussed within a binary-system framework and linked to the so-called lump model in circumbinary accretion flows \citep[][]{farris_2014, WS2022+}. Although this interpretation remains model dependent, it further highlights PG 1553+113 as an important source for exploring possible binary-driven variability in blazars.

Beyond the question of whether a characteristic period is genuine, a key question is what the detailed morphology of the oscillations can reveal about the underlying physical mechanism. A dedicated analysis of the temporal morphology of the oscillations is therefore a natural next step toward improving our understanding of the physical processes responsible for the MWL emission in this candidate SMBHB system. For PG 1553+113, previous discussions have already raised the possibility that substructure in the oscillations, including double-peaked features, may offer clues about the physical driver of the modulation \citep[][]{tavani_pdm_pg_1553}. This characterization may also have methodological implications for periodicity searches. Common approaches such as the Lomb--Scargle periodogram are most sensitive to approximately sinusoidal modulations \citep[][]{lomb_1976, scargle_1982}, whereas complex or asymmetric oscillation profiles can distribute the signal power between the fundamental frequency and its harmonics, potentially reducing the significance of the primary periodicity peak. Identifying recurrent non-sinusoidal morphologies may therefore provide the basis for future searches using physically or empirically motivated profile templates better matched to the observed variability. Our aim in this paper is to characterize the oscillations in each band and to determine whether their profiles are mutually consistent. 

The structure of this paper is organized as follows. In $\S$~\ref{sec:profiles_oscillations}, we introduce the oscillation profiles used to model the periodic MWL emission of PG~1553+113. $\S$~\ref{sec:methodology} describes the MWL datasets analyzed in this work and presents the mathematical expressions of the selected profile models. $\S$~\ref{sec:results} presents the results of the profile analysis. In $\S$~\ref{sec:tuples}, we analyze contemporaneous oscillations identified across the different energy bands. $\S$~\ref{sec:physical} discusses possible physical interpretations, highlighting the connection with the SMBHB scenario proposed for PG~1553+113. Finally, $\S$~\ref{sec:summary} summarizes the main results of this work.

\section{Temporal morphology of the oscillations}\label{sec:profiles_oscillations}
The morphology of the oscillations of a periodic pattern depends on the nature of the underlying physical process. As such, an accurate modeling of the repeated shape of the oscillation can provide powerful constraints on the physical mechanisms of emission. However, previous studies have shown that flares display a wide range of temporal behaviors, from nearly symmetric events to asymmetric outbursts \citep[e.g.,][]{moraitis_lorentzian_2011}, and in some cases the observed long-duration activity appears to be the superposition of multiple shorter flaring episodes rather than a single pulse \citep[e.g.,][]{hovatta_multi_exponential_2009}. 

Given the heterogeneity of previous results, this work does not attempt to directly model the potentially complex structure of the flares. Rather, as a first step toward this modeling, we attempt to characterize the substructure of the flares of a particularly bright source, PG 1553+113, which is also the strongest candidate for periodic emissions among blazars. Before any attempts at modeling, it is essential to characterize the substructure of each flare, whether they repeat or change in the (potentially) periodic oscillations, and whether they share properties across different wavelengths. Therefore, the models described below provide empirical descriptions of the observed temporal morphology.

\subsection{Single-peak profiles}

To characterize isolated peaks, we adopt a set of analytical profiles covering the main morphological possibilities reported in AGN and blazar variability studies (Figure~\ref{fig:example_profiles_single_peak}). In particular, \textit{Fermi}-LAT observations of bright $\gamma$-ray flares reveal both nearly symmetric and asymmetric events, while optical studies likewise characterize flare profiles through their rise and decay behavior \citep[e.g.,][]{nalewajko_flares_shapes_2013, roy_multiple_flares_2019, bhatta_flares_2023}.

For symmetric peaks, we consider Gaussian and Lorentzian profiles. The Gaussian provides a smooth, symmetric template, with its centroid identifying the time of maximum and its width characterizing the duration of the event. Gaussian-like profiles have also been used as injected flare templates for testing transient-event recovery in AGN variability studies \citep{penil_flares_2025}. The Lorentzian offers an alternative with heavier tails, allowing a more extended flux excess away from the central maximum \citep{moraitis_lorentzian_2011}. Together, these profiles allow us to explore differences in the concentration of the emission around the peak and the extent of its wings.

To account for temporal asymmetry, we consider exponential rise--decay profiles with independent rise and decay timescales \citep[e.g.,][]{abdo_exponential_flares_2010, blinov_exponential_flares_2018, li_multiple_flares_2018}. Comparable timescales produce approximately symmetric events, whereas unequal timescales allow either a rapid rise followed by a slower decline \citep[known as fast-rise exponential-decay, or FRED, morphology when the asymmetry is pronounced; ][]{fenimore_fred_1996, yan_fred_2018} or a gradual rise followed by a faster decline. We also consider a Gaussian-rise plus exponential-decay profile, which combines a smooth approach to the maximum with an exponential decline. This form has been used as a generic flare model in AGN light-curve studies \citep{graham_gaussian_expo_2020, veronesi_gaussian_expo_2025}.

\subsection{Multiple-peak profiles}
Beyond single-peaked profiles, some oscillation events require models with multiple local maxima. Such structured profiles may arise when a long-duration activity episode is composed of several flares occurring within the same active phase of the jet, rather than from a single elementary pulse \citep[e.g.,][]{li_multiple_flares_2018}. We therefore consider multi-component empirical models to describe oscillations showing distinct peaks or a dominant component accompanied by secondary enhancements. This decomposition is motivated by AGN variability studies in which complex activity episodes are described as the superposition of individual flares with similar or different rise and decay timescales \citep[][]{valtaoja_multi_exponential_1999, hovatta_multi_exponential_2009, vlasyuk_multi_exponential_2024}.

For PG~1553+113, the possibility of such substructure is particularly relevant. \citet{tavani_pdm_pg_1553} reported a MWL oscillation profile consisting of a dominant flare accompanied by secondary ``twin peaks'' on both sides of the main maximum. They argued that this morphology is qualitatively consistent with an SMBHB scenario and discussed two possible origins: additional instabilities induced in the primary jet along the inner orbital arc of the companion, or emission from a weaker secondary jet associated with the smaller black hole and affected by the primary SMBH jet \citep[see Figure~5 in][]{tavani_pdm_pg_1553}. Multi-peaked oscillations have also been reproduced in hydrodynamic simulations of SMBHB systems over a range of orbital eccentricities and mass ratios, supporting the idea that binary-driven variability can naturally generate complex periodic profiles \citep[][]{WS2022+}. 

However, a multi-peaked morphology is not unique to the SMBHB interpretation. Similar structures can also arise from other processes within the jet. For example, long-term flare profiles may be shaped by the time required for radiation, particles, or a disturbance such as a shock to cross the emitting region \citep[][]{roy_multiple_flares_2019, finke_shocks_2024}. In this scenario, the broad component traces the global response of the disturbed region, while smaller side peaks may be produced by internal substructure, multiple closely spaced shocks, or localized enhancements within the same active zone. The observed profile can therefore be interpreted as a long-duration flare envelope with superposed subflares.

To capture this variety empirically, we consider sums of two or three Gaussian, Lorentzian, or exponential rise--decay components (Figure~\ref{fig:example_profiles_multi_peak}). Gaussian and Lorentzian combinations describe subpeaks with symmetric individual shapes but different wing structures \citep[e.g.,][]{sarkar_double_lorentz_2021}, whereas exponential rise--decay combinations allow the individual components to have different degrees of temporal asymmetry \citep{valtaoja_multi_exponential_1999, hovatta_multi_exponential_2009, vlasyuk_multi_exponential_2024}. The double- and triple-component configurations are thus extensions of the same underlying profile families. For oscillations showing a dominant asymmetric component accompanied by two secondary peaks, we also consider mixed profiles. These combine either a skewed Gaussian or a FRED component with two Gaussian components representing the secondary peaks. 

\section{Methodology}\label{sec:methodology}
Using the profile models described above, we performed an individual profile analysis of the oscillations identified in the MWL emission of PG~1553+113. In total, we analyzed eight $\gamma$-ray oscillations, five X-ray oscillations, five UV oscillations, and eight optical oscillations. To make this procedure clear and reproducible, the following subsections describe the datasets included in the analysis, the mathematical form in which the profile models used to fit the identified oscillations of each MWL light curves, and the criteria used to identify the profile model that best represents each oscillation profile.

\subsection{MWL datasets}
In this study, we considered light curves of four energy bands: $\gamma$ ray, X-ray, ultraviolet (UV) and optical. For the $\gamma$-ray analysis, we relied on the publicly available \textit{Fermi}-LAT Light Curve Repository \citep[][]{fermi_repository}\footnote{\url{https://fermi.gsfc.nasa.gov/ssc/data/access/lat/LightCurveRepository/about.html}}. This database provides more than 17 years of \textit{Fermi}-LAT monitoring, covering the period from August 2008 through April 2026. In the present work, we adopted the 30-day binned light curve to trace and describe the long-term variability of each oscillation. We also examined the 7-day \textit{Fermi}-LAT light curve to assess whether the 30-day binning could hide relevant substructure. The 30-day light curve contains no upper limits, while the 7-day light curve contains only four; their prevalence was therefore not the main factor motivating our choice of binning. Although the weekly data provide higher temporal resolution, their typical relative flux uncertainty is approximately twice that of the monthly data (26.9\% versus 12.9\%), making individual weekly fluctuations less reliable for identifying subcomponents. When averaged over 30-day intervals, the weekly light curve closely reproduces the monthly evolution in all the analyzed cycles. Moreover, the weekly fits do not provide a more stable identification of individual subcomponents. We therefore use the 30-day light curve as the best compromise between temporal resolution and statistical precision for characterizing the broad oscillation profiles.

We use X-ray and UV observations from \textit{Swift}-XRT and UVOT, respectively. The procedure used to derive those light curves was described by \citet{saple_2026} and was carried out with the semi-automatic \textit{Swift} Analysis Pipeline for Lightcurve Extraction (SAPLE)\footnote{SAPLE GitHub link: \url{https://github.com/leamarcotulli/saple}}. Together, the \textit{Swift}-XRT and UVOT observations cover the period from 2005 to 2026.

Finally, the optical light curve was assembled by combining measurements from several public surveys. In particular, we used data from the CSS \citep[Catalina Sky Survey;][]{drake2009}\footnote{\url{http://nesssi.cacr.caltech.edu/DataRelease/}}, ASAS-SN \citep[All-Sky Automated Survey for Supernovae;][]{shappee2014,kochanek2017}\footnote{\url{http://www.astronomy.ohio-state.edu/asassn/index.shtml}}, and ZTF \citep[Zwicky Transient Facility;][]{zwicky_observatory}\footnote{\url{https://www.ztf.caltech.edu/}} databases. By merging the V-band observations from these databases, we obtained an optical data set spanning the interval 2005--2026. The UV and optical profiles are fitted in magnitude space and are used as empirical descriptions of their temporal morphology. 

\begin{figure}
        \centering
        \includegraphics[scale=0.14]{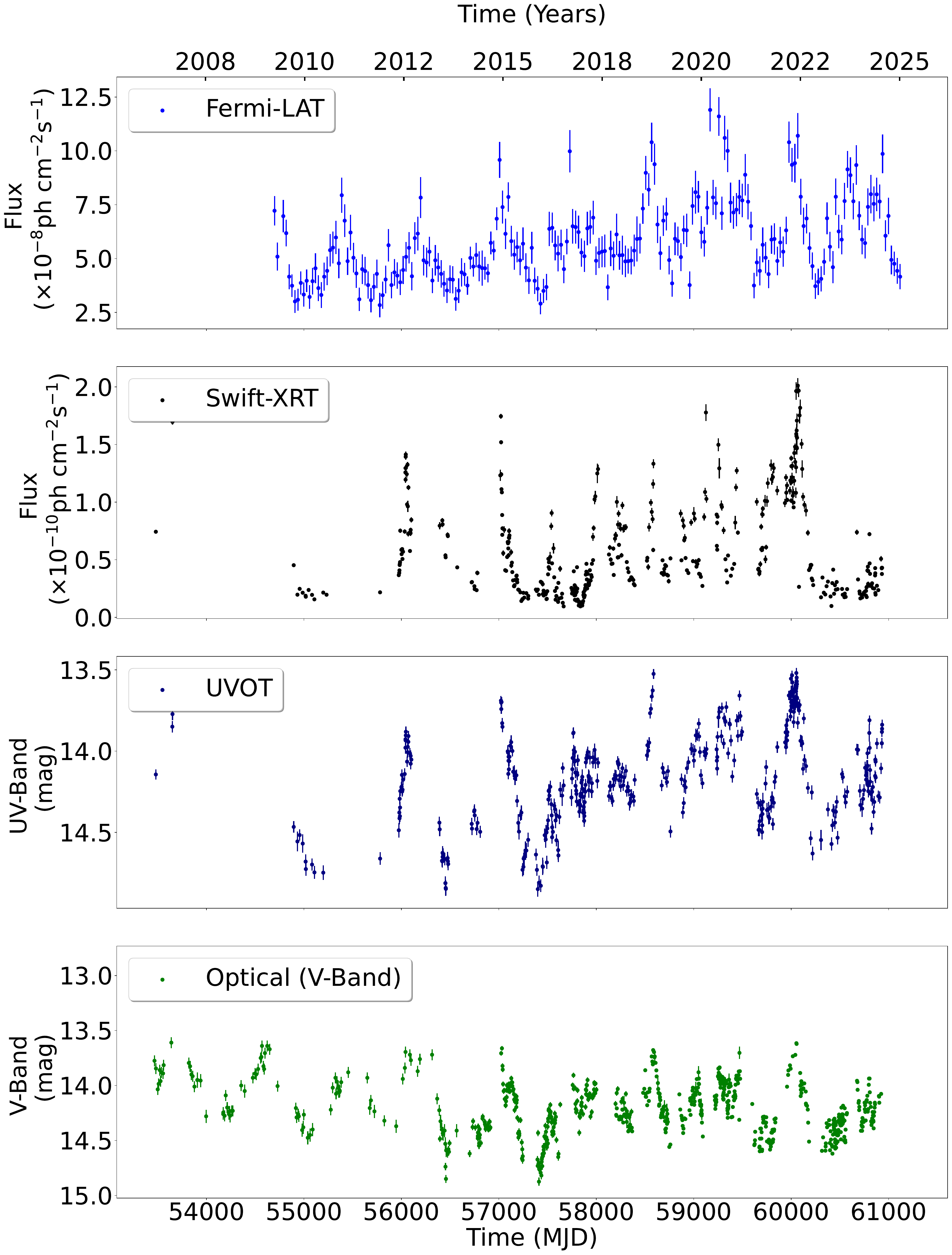}
        \caption{Light curves used in this study. From top to the bottom: \textit{Fermi}-LAT ($\gamma$ rays), Swift-XRT (X-rays), UVOT (UV), and optical (V-band).} \label{fig:mwl_lcs}
\end{figure}

The MWL light curves considered in this analysis are shown in Figure~\ref{fig:mwl_lcs}.

\subsection{Profile model expressions}
We describe the individual oscillations using the set of analytical profile families presented in  $\S$\ref{sec:profiles_oscillations}. Multiple-peaked configurations are constructed by combining these elementary profiles. Below, we define the individual components. Table~\ref{tab:profile_models} summarizes the configurations considered and their identifiers.

\paragraph{Gaussian and Lorentzian profiles.}
We use Gaussian and Lorentzian functions as symmetric representations of individual peaks:
\begin{equation}
G_i(t)=A_i\exp\left[
-\frac{(t-t_{0,i})^2}{2\sigma_i^2}
\right],
\label{eq:gaussian_component}
\end{equation}
and
\begin{equation}
L_i(t)=A_i
\frac{\Gamma_i^2}{(t-t_{0,i})^2+\Gamma_i^2}.
\label{eq:lorentzian_component}
\end{equation}
Here, $A_i$ is the peak amplitude above the baseline and $t_{0,i}$ is the time of maximum of component $i$. The parameters $\sigma_i$ and $\Gamma_i$ are the Gaussian standard deviation and the Lorentzian half-width at half-maximum, respectively. The Lorentzian provides an alternative description with heavier tails than the Gaussian.

\paragraph{Exponential rise--decay profiles.}
To describe both symmetric and asymmetric events, we use the exponential rise--decay profile introduced by \citet{abdo_exponential_flares_2010}:
\begin{equation}
e_i(t)=
\frac{2A_i}{
\exp\left(\dfrac{t_{0,i}-t}{T_{{\rm r},i}}\right)
+
\exp\left(\dfrac{t-t_{0,i}}{T_{{\rm d},i}}\right)
}.
\label{eq:exp_component}
\end{equation}
The rise and decay timescales, $T_{{\rm r},i}$ and $T_{{\rm d},i}$, are independent. Equal timescales produce a symmetric profile, whereas unequal timescales allow either a faster rise followed by a slower decay or the reverse. The parameter $t_{0,i}$ is a reference time at which $e_i(t_{0,i})=A_i$; for unequal timescales, it does not generally coincide with the maximum.

FRED behaviour corresponds to the regime $T_{{\rm r},i}\ll T_{{\rm d},i}$ \citep{fenimore_fred_1996} (Figure~\ref{fig:example_profiles_single_peak}). We consider the following FRED parameterization:
\begin{equation}
FRED_i(t)=
\begin{cases}
A_i\exp\left(\dfrac{t-t_{0,i}}{T_{{\rm r},i}}\right),
& t<t_{0,i}, \\[6pt]
A_i\exp\left(-\dfrac{t-t_{0,i}}{T_{{\rm d},i}}\right),
& t\geq t_{0,i}.
\end{cases}
\label{eq:fred_piecewise}
\end{equation}
In this expression, $A_i$ is the peak amplitude and $t_{0,i}$ is the time of maximum. This parameterization joins the exponential rise and decay branches directly at the peak and is mathematically distinct from the expression in Equation~\ref{eq:exp_component}.

\paragraph{Gaussian-rise and exponential-decay profile.}
To allow a Gaussian approach to the maximum followed by an exponential decline, we use
\citep[e.g.,][]{graham_gaussian_expo_2020}
\begin{equation}
h_i(t)=
\begin{cases}
A_i\exp\left[
-\dfrac{(t-t_{0,i})^2}{2\sigma_{{\rm r},i}^2}
\right],
& t\leq t_{0,i}, \\[6pt]
A_i\exp\left(-\dfrac{t-t_{0,i}}{\tau_{{\rm d},i}}\right),
& t>t_{0,i}.
\end{cases}
\label{eq:gred_component}
\end{equation}
Here, $A_i$ is the peak amplitude, $t_{0,i}$ is the time of maximum, $\sigma_{{\rm r},i}$ controls the width of the Gaussian rise, and $\tau_{{\rm d},i}$ is the exponential decay timescale.

\paragraph{Skewed-Gaussian profile.}
We additionally use a skewed Gaussian to represent an asymmetric central component in composite profiles:
\begin{equation}
s_i(t)=A_i\exp\left[
-\frac{(t-t_{0,i})^2}{2\sigma_i^2}
\right]
\left[
1+\operatorname{erf}\left(
\frac{\alpha_i(t-t_{0,i})}{\sqrt{2}\sigma_i}
\right)
\right].
\label{eq:skewed_gaussian_component}
\end{equation}
The parameter $A_i$ sets the normalization, $t_{0,i}$ is a location parameter, $\sigma_i$ controls the width, and $\alpha_i$ determines the degree and direction of the asymmetry. For $\alpha_i=0$, this expression reduces to the Gaussian component defined above. For nonzero $\alpha_i$, $t_{0,i}$ is not generally the time of maximum.

\paragraph{Multiple-component configurations.}
For oscillations with multiple subpeaks, we construct the complete profile as
\begin{equation}
F(t)=F_{\rm c}+\sum_{i=1}^{N_{\rm p}}q_i(t),
\label{eq:multiple_components}
\end{equation}
where $N_{\rm p}$ is the number of components and $q_i(t)$ is one of the elementary functions defined above. We test two- and three-component configurations of the Gaussian, Lorentzian, and smooth exponential rise--decay profiles \citep[e.g.,][]{sarkar_double_lorentz_2021, vlasyuk_multi_exponential_2024}. These configurations belong to the same underlying profile families as their single-component counterparts, but differ in their number of fitted components and parameters.

We also consider two mixed configurations in which a central component is combined with two Gaussian components representing adjacent subpeaks. The central component is either a skewed Gaussian or a FRED profile.

The oscillation profiles used in this work are summarized in Table~\ref{tab:profile_models}, which lists their identifiers, configurations, and mathematical expressions.

\begin{table*}
\centering
\caption{
Summary of the profile configurations. The $S$ and $M$ identifiers denote single-component and multiple-component configurations, respectively. Figure \ref{fig:example_profiles_single_peak} and Figure \ref{fig:example_profiles_multi_peak} shows examples of single-component and multiple-component oscillations, respectively. $F_{\rm c}$ represent the baseline of the profile.
}
\label{tab:profile_models}
\begin{tabular}{llcl}
\hline
\hline
Identifier & Profile configuration & Components & Expression \\
\hline
S1 & Gaussian
   & 1 & $F_{\rm c}+G_1(t)$ \\
S2 & Lorentzian
   & 1 & $F_{\rm c}+L_1(t)$ \\
S3 & Exponential rise--decay
   & 1 & $F_{\rm c}+e_1(t)$ \\
S4 & FRED
   & 1 & $F_{\rm c}+FRED_1(t)$ \\
S5 & Gaussian rise and exponential decay
   & 1 & $F_{\rm c}+h_1(t)$ \\
\hline
M1 & Double Gaussian
   & 2 & $F_{\rm c}+\sum_{i=1}^{2}G_i(t)$ \\
M2 & Double Lorentzian
   & 2 & $F_{\rm c}+\sum_{i=1}^{2}L_i(t)$ \\
M3 & Double exponential rise--decay
   & 2 & $F_{\rm c}+\sum_{i=1}^{2}e_i(t)$ \\
M4 & Triple Gaussian
   & 3 & $F_{\rm c}+\sum_{i=1}^{3}G_i(t)$ \\
M5 & Triple Lorentzian
   & 3 & $F_{\rm c}+\sum_{i=1}^{3}L_i(t)$ \\
M6 & Triple exponential rise--decay
   & 3 & $F_{\rm c}+\sum_{i=1}^{3}e_i(t)$ \\
M7 & Skewed Gaussian plus two Gaussians
   & 3 & $F_{\rm c}+s_0(t)+G_1(t)+G_2(t)$ \\
M8 & FRED plus two Gaussians
   & 3 & $F_{\rm c}+FRED_0(t)+G_1(t)+G_2(t)$ \\
\hline
\end{tabular}
\end{table*}

\subsection{Additional morphological parameters}
To complement the profile classification, we characterize each oscillation using a set of morphological parameters derived from the selected profile: the main-peak time and amplitude, oscillation width, rise and decay times, asymmetry, and, for multi-component profiles, structure fraction. The main-peak time corresponds to the epoch of maximum observed flux within the analyzed interval for the $\gamma$-ray and X-ray bands, and to maximum observed brightness (minimum magnitude) for the UV and optical bands. Its amplitude is measured relative to the estimated local baseline, in flux units for the $\gamma$-ray and X-ray bands and in magnitudes for the UV and optical bands. The width characterizes the duration of the oscillation, and the rise ($T_{\rm r}$) and decay ($T_{\rm d}$) times describe its temporal evolution. The asymmetry is derived from these times. We define the asymmetry as \(A=(T_{\rm d}-T_{\rm r})/(T_{\rm d}+T_{\rm r})\). Positive values indicate a slower decay, negative values indicate a slower rise, and values near zero correspond to approximately symmetric profiles. The asymmetry is not reported when either timescale does not converge. Finally, the structure fraction measures the relative contribution of the secondary fitted components, distinguishing oscillations dominated by one main peak from those with substantial internal substructure. 

\subsection{Profile model comparison}\label{sec:profile_comparision}
The comparison among the different profile models is based primarily on the Bayesian Information Criterion \citep[BIC;][]{bic_schwarz}. The BIC is particularly useful in this context because it evaluates the quality of the fit while explicitly penalizing the number of free parameters. For each oscillation, the model with the minimum BIC is adopted as the empirical profile.

To quantify the uncertainty in the model ranking, we define \(\Delta{\rm BIC}_i={\rm BIC}_i-{\rm BIC}_{\min}\). Following \citet{Kass_Raftery_1995}, values of \(\Delta{\rm BIC}<2\) are considered weakly distinguishable from the best-ranked model and are therefore treated as empirically competitive. This restrictive threshold avoids grouping disfavored models with the best-ranked profile and is applied consistently throughout the analysis.

As a complementary diagnostic, we also compute the coefficient of determination, R$^{2}$, in order to assess the goodness of fit of the selected profile with the minimum BIC. Its values lie between 0 and 1, with higher values corresponding to a better descriptive fit. Nevertheless, the interpretation of R$^{2}$ depends on the specific application, and there is no universal threshold that separates acceptable from unacceptable fits. As discussed by \citet{hair_r2_2011}, values near 0.25, 0.50, and 0.75 can be interpreted, in general terms, as indicating weak, moderate, and substantial fit, respectively.

\section{Results}\label{sec:results}
In this section, we report the results of the characterization of each analyzed oscillation based on the profile models described in $\S$\ref{sec:profiles_oscillations} and using the methodology presented in $\S$\ref{sec:methodology} for each energy band included in Figure \ref{fig:mwl_lcs}.

\subsection{$\gamma$ rays}
In the $\gamma$-ray light curve of Figure~\ref{fig:mwl_lcs}, we identified eight oscillations associated to the periodic pattern of $\approx$2.1 yr obtained in multiple works \citep[][]{ackermann_pg1553, magic_pg1553_2024, penil_mwl_pg1553,penil_24candidates_2025}. The results of the profile study are  presented in Table \ref{tab:oscillation_formology}. Each oscillation's profiling is shown in Figure \ref{fig:best_fit_gamma}.  

The results of the profile analysis indicate that the eight $\gamma$-ray oscillations do not share a single common morphology, but instead exhibit different profile shapes (Table~\ref{tab:oscillation_formology} and Figure~\ref{fig:best_fit_gamma}). In particular, three of them are well described by single-component models, $S2$, and $S4$ functions for cycles 2, 3, and 5. The remaining cycles require more complex multi-component structures, including $M3$, $M5$, $M7$, and $M8$. The amplitudes span from $\sim$3 to $\sim$9\(\times10^{-8}\) ph cm\(^{-2}\) s\(^{-1}\), while the widths range from $\sim$200 to $\sim$700 days. This indicates that some oscillations are relatively compact, whereas others evolve over substantially longer timescales.

The ranking based on $\Delta{\rm BIC}$ further shows that the preferred profile is not always unique (see Figure \ref{fig:best_fit_gamma}). In several cycles, more than one model lies within $\Delta{\rm BIC}$$<$2, indicating that different empirical descriptions provide statistically comparable fits \citep[][]{Kass_Raftery_1995}. This is particularly clear in cycles 1, 3, 5, and 8. In Cycle 1, the preferred profile is $M7$, corresponding to a skewed-Gaussian plus twin-Gaussian profile, but the single-component Lorentzian ($S2$) and FRED ($S4$) profiles also remain close to the best-ranked model, with $\Delta{\rm BIC}$$=$0.9 and 1.8, respectively. Cycle 3 is best ranked as a FRED-like profile, although the double-Lorentzian ($M2$) and FRED plus twin-Gaussian models ($M8$) are also comparable. Similarly, Cycle 5 is formally best described by a Lorentzian profile, but Gaussian ($S1$), Gaussian-rise exponential-decay ($S5$), and $M8$ profiles all have $\Delta{\rm BIC}$$<$2. Cycle 8 also shows a non-unique description, with the double exponential rise--decay model ($M3$), double-Lorentzian ($M2$), and double-Gaussian profiles ($M1$) all providing comparable representations. These cases indicate that the distinction between simple and structured profiles is sometimes gradual rather than sharply defined.

In contrast, cycles 6 and 7 show a clearer preference for multi-component descriptions. For Cycle 6, the best-ranked model is $M7$, corresponding to a skewed-Gaussian plus twin-Gaussian profile, with the next closest model already outside the $\Delta{\rm BIC}$$<$2 range. For Cycle 7, the $M8$ profile is clearly preferred over the single-component alternatives, which have larger $\Delta{\rm BIC}$ values. Cycle 4 represents an intermediate case: the triple-Lorentzian profile ($M5$) is preferred but the model $M4$ (triple-Gaussian profile) remains statistically close. Overall, the $\Delta{\rm BIC}$ ranking confirms that some oscillations require genuinely structured profiles, while others can be described by several competing single- or multi-component templates.

This diversity is also reflected in the symmetry properties of the oscillations. Five of the eight cycles are approximately symmetric, namely cycles 1, 2, 5, 6, and 7. In contrast, cycles 3, 4, and 8 show positive asymmetry, indicating that their decay times are longer than their rise times. These profiles are therefore characterized by a relatively fast rise followed by a more gradual decline. The multi-component cycles, cycles 1, 4, 6, 7, and 8, show non-negligible structure fractions $\sim$50\%, which indicates that, in these cases, secondary components make a contribution to the overall oscillation morphology. 

The profile diversity therefore appears to be mainly associated with the internal structure of each oscillation rather than with the existence of the oscillation itself. The repeated presence of broad profiles, with widths from $\sim$200 to $\sim$700 days, suggests that the $\gamma$-ray events are governed by a common long-timescale envelope. Within this envelope, some cycles remain relatively simple and are adequately represented by single-component Lorentzian or FRED-like profiles, whereas others require multiple components to reproduce secondary peaks. This behavior suggests that the $\approx$2.1 yr modulation may define the recurrence of the broad activity episodes, while the detailed profile shape is modulated by cycle-to-cycle changes in the short-timescale substructure.

\subsection{X-rays}
The X-ray profile analysis must be interpreted with caution because the light curve is sparsely and irregularly sampled. Even so, the five cycles identified that can be examined show several notable morphological properties. In all cases, the preferred profile is $M6$, indicating that the X-ray oscillations exhibit structured events with multiple contributions (see Table \ref{tab:oscillation_formology_xray} and Figure \ref{fig:best_fit_xray}). The amplitudes range from $\sim$0.9 to $\sim$2.4 flux units, while the widths span from $\sim$60 to $\sim$300 days.  

The $\Delta{\rm BIC}$ ranking shows that this preference for the triple exponential rise--decay profile is not marginal. In all five X-ray cycles, $M6$ is the only model within $\Delta{\rm BIC}$$<$2, while the closest competing profiles are already strongly disfavored. This is especially evident in cycles 2 and 4, where the next-best alternatives have $\Delta{\rm BIC}$ values of several thousand, and in cycles 1, 3, and 5, where the closest alternatives still lie far outside the comparable-model regime. Therefore, unlike the $\gamma$-ray case, where several cycles admit more than one statistically comparable empirical description, the X-ray profiles show a much stronger formal preference for a single multi-component morphology. This result supports the interpretation that the X-ray oscillations are dominated by structured flare complexes rather than by isolated single-peaked events, although the large $\Delta{\rm BIC}$ separations should still be interpreted in light of the uneven sampling and the different number of points available in each cycle.

The rise and decay times show that the X-ray oscillations are often asymmetric. Cycles 1 and 3 are nearly symmetric. By contrast, Cycles 2, 4, and 5 show negative asymmetry, implying that the rise time is longer than the decay time and therefore that these events follow a slow-rise, fast-decay pattern. This indicates that the X-ray oscillations more often appear to develop gradually and decline more rapidly.

The structure fraction further shows that all five X-ray oscillations contain a substantial contribution from secondary components. The inferred values range from 31.4\% to 56.1\%, showing that the oscillations are systematically structured rather than being dominated by a single main peak. 

The triple-exponential function is sufficiently flexible to reproduce structured events with more than one local component, while at the same time remaining smooth enough to fit data sets affected by large gaps. Nevertheless, the results suggest that the X-ray variability is composed of multiple-structured flares. In particular, the better-sampled flare near MJD 60000 resulting from a dedicated monitoring campaign \citep[][]{penil_xray_pg1553_2026} supports the view that at least some X-ray cycles are genuinely extended and morphologically complex. 

\subsection{UV}
The UV light curve is also affected by strong irregular sampling and numerous gaps, so the morphological results must be interpreted with caution (see Table \ref{tab:oscillation_formology_uv} and Figure \ref{fig:best_fit_uv}). Even so, the five oscillations that can be examined reveal several clear trends. The UV oscillations are preferentially described by multi-component models, with two cycles favoring the $M6$ profile and one the $M4$, one cycle a $M3$ profile, and only one cycle a single $S2$ profile. This already suggests that the UV events are generally not well represented by simple isolated peaks. The peak amplitudes range from ~0.53 to ~1.01 mag relative to the local baseline, while the widths vary from $\sim$200 to $\sim$700 days. 

The $\Delta{\rm BIC}$ ranking shows that the UV model classification is not equally robust for all cycles. In cycles 1 and 2, the best-ranked profiles are not unique: Cycle 1 is formally best described by a $M4$ profile, but the $M6$ remains statistically comparable with $\Delta{\rm BIC}$$=$1.6; similarly, Cycle 2 favors a $M6$ profile, although the $M4$ model is very close, with $\Delta{\rm BIC}$$=$0.9. These two cases therefore indicate that the data require a structured multi-component morphology, but do not uniquely distinguish between different multi-component parameterizations. By contrast, Cycle 4 and 5 show a much clearer preference for their best-ranked models, with all alternative profiles lying far outside the $\Delta{\rm BIC}$$<$2 range. Cycle 3 is also formally best described by a $S2$, although the $S5$ lies just above the comparable-model threshold, with $\Delta{\rm BIC}$$=$2.2. 

The rise and decay times further show that the UV oscillations exhibit a broad range of asymmetries. Cycles 3 and 5 are nearly symmetric. By contrast, Cycle 1 shows a clearly positive asymmetry, with a relatively rapid rise followed by a much longer decay. The opposite behavior is seen in Cycles 2 and especially 4, where the rise time is substantially longer than the decay time. The structure fractions have values range from 46.1\% to 55.9\%, which implies that secondary components account for roughly half of the total oscillation morphology, denoting that the UV oscillations are not dominated exclusively by the main component. 

The conclusions about the morphology of the UV oscillations should be interpreted with caution because of the large gaps in the light curve. As in the X-ray case, the frequent preference for exponential multi-component profiles may partly reflect the fact that such functions are flexible enough to reproduce structured events while remaining smooth across poorly sampled intervals. Cycle 5, which was also covered by a dedicated monitoring campaign \citep[][]{penil_xray_pg1553_2026}, is best fitted by a $M3$ that may indicate that the multi-component structure is not purely an artifact of sparse sampling, but could reflect genuine substructure within at least some UV oscillations.

\subsection{Optical}
The optical oscillations are more frequently described by multi-component models (see Table~\ref{tab:oscillation_formology_optical} and Figure~\ref{fig:best_fit_optical}). In this band, profiles such as $M1$, $M4$, $M5$, $M6$, and $M7$ are favored, while only Cycle~1 is best described by the simpler single-component $S4$ profile.

The peak amplitudes show substantial cycle-to-cycle variation, ranging from ~0.39 to ~1.23 mag relative to the local baseline, whereas the widths span from $\sim$200 to $\sim$600 days. The rise and decay times show that most optical oscillations are either nearly symmetric or skewed toward slower rises and faster decays. The structure fractions further support the presence of significant internal complexity. In all multi-component cases, the secondary components contribute significantly to the total morphology, from $\sim$41\% to $\sim$63\%.

The $\Delta{\rm BIC}$ ranking shows that the optical profile classification is only ambiguous in the first two cycles. In Cycle 1, the formally preferred model is the $S4$, but the profile $S5$ is statistically comparable, with $\Delta{\rm BIC}$$=$0.5. Similarly, Cycle 2 is best described by the $M7$, although the $M8$ provides an almost equivalent description, with $\Delta{\rm BIC}$$=$0.3. In both cases, the competing models correspond to closely related asymmetric or structured morphologies, so the exact profile label should not be interpreted as unique. By contrast, Cycles 3--8 show no alternative models within $\Delta{\rm BIC}$$<$2, indicating a clearer formal preference for the best-ranked profile in those cases. This is particularly evident for Cycles 4, 6, 7, and 8, where the closest competing models are strongly disfavored. Therefore, the optical light curve confirms the predominance of structured profiles, while also showing that only a small subset of cycles admits statistically comparable empirical descriptions.

\subsection{Global Analysis}
\begin{figure}
        \centering
        \includegraphics[scale=0.25]{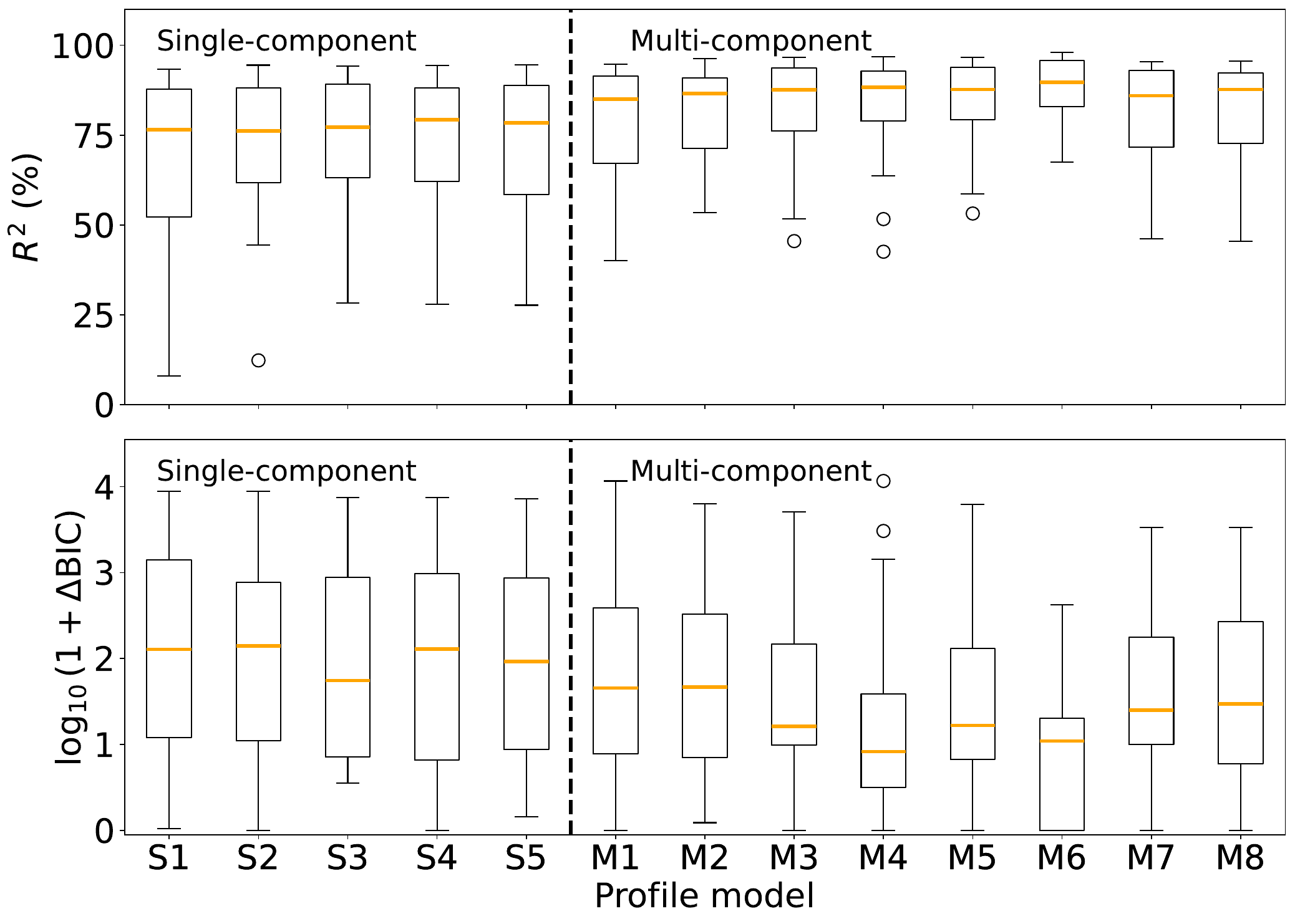}
        \includegraphics[scale=0.245]{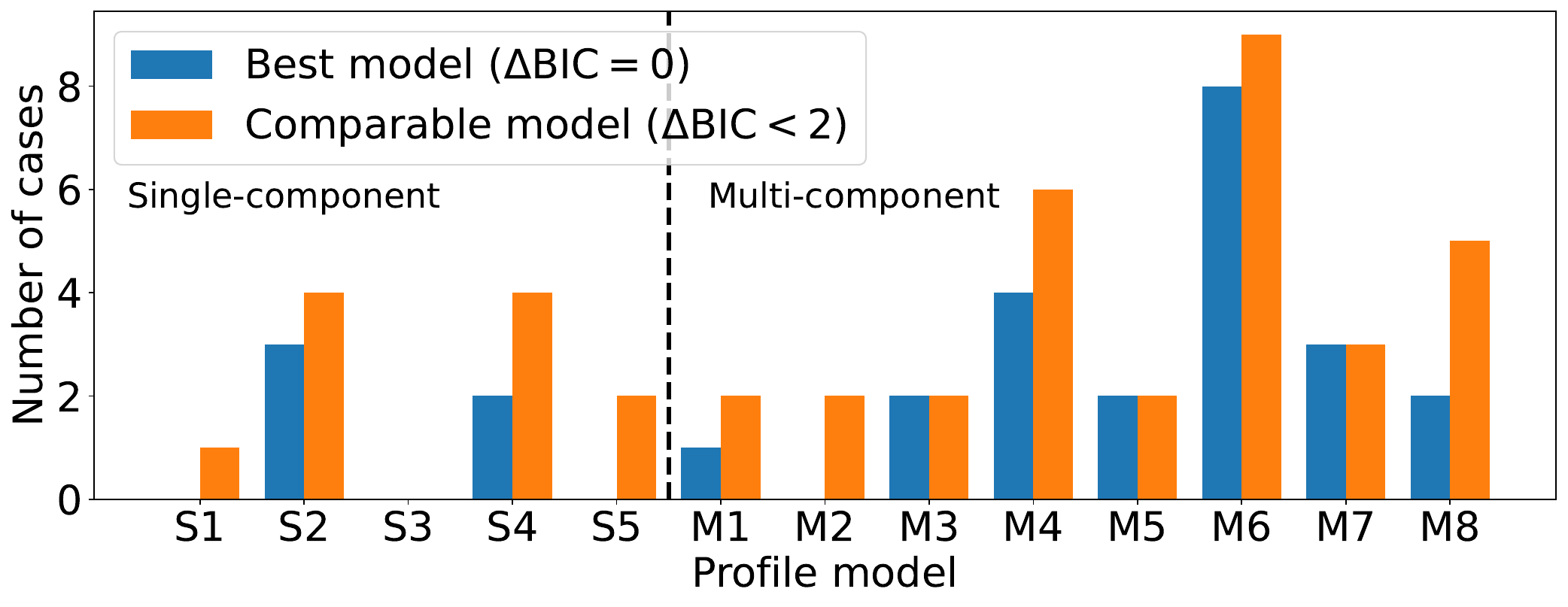}        
        \caption{Global comparison of the empirical profile models across all analyzed oscillations and wavelength bands. Top: distributions of the $R^{2}$ values and $\log_{10}(1+\Delta{\rm BIC})$ for each profile model, where ($\Delta{\rm BIC}$) is computed relative to the minimum BIC obtained for each individual oscillation and band. Bottom: number of cases in which each model is selected as the best-ranked profile $\Delta{\rm BIC}$$=$0 and number of cases in which it remains statistically comparable to the best-ranked model $\Delta{\rm BIC}$$<$2. The vertical dashed line separates single-component models from multi-component models. In the boxplots, isolated points indicate outliers, corresponding to individual oscillation/band cases in which a given model performs substantially differently from its typical behavior. Multi-component profiles generally yield higher $R^{2}$ values and lower $\Delta{\rm BIC}$, indicating that structured morphologies provide a better global description of the MWL oscillations.} \label{fig:global_model}
\end{figure}

The band-by-band analysis shows that the preferred oscillation profiles vary from cycle to cycle and from one wavelength range to another. However, individual best-fit classifications do not by themselves reveal whether some profile families perform systematically better across the full data set. To assess this, we carried out a global comparison of all fitted models by examining the distributions of $R^{2}$, $\Delta{\rm BIC}$, and the number of cases in which each model is either the best-ranked solution or statistically comparable to it. The results are shown in Figure~\ref{fig:global_model}.

The distributions of $R^{2}$ and $\Delta{\rm BIC}$ show that multi-component models generally provide better descriptions than single-component profiles, with systematically higher $R^{2}$ values and lower relative BIC values. The generally high $R^{2}$ values across most profile families indicate that the main long-timescale modulation is a robust feature of the data and can be captured by different empirical templates. This is expected because the fits are dominated by the broad oscillation envelope, which is sampled over several points in most cycles. Differences among models therefore arise mainly from how they reproduce secondary peaks, shoulders, or asymmetric wings. In this context, $R^{2}$ quantifies how well each model follows the overall oscillation, while $\Delta{\rm BIC}$ provides the complementary information needed to assess whether the additional flexibility of multi-component profiles is statistically favored. 

This trend is also reflected in the model-frequency analysis: $M6$ is the most frequently selected model and the model most often found within $\Delta{\rm BIC}<2$ of the best-ranked solution  \citep[][]{Kass_Raftery_1995}. Its performance is not reproduced by the other three-component profiles, suggesting that the preference for $M6$ is not solely driven by the number of fitted components. Unlike the triple-Gaussian and triple-Lorentzian models, $M6$ allows each exponential component to have independent rise and decay times, providing additional flexibility to reproduce asymmetric substructures. Nevertheless, because this additional complexity is penalized by the BIC, the systematic preference for $M6$ may indicate that asymmetric rise--decay components provide a genuinely better empirical description of the observed oscillation profiles. Other multi-component profiles, such as $M4$ and $M8$ (Triple-Gaussian and FRED plus twin-Gaussian profiles, respectively), also appear repeatedly as competitive descriptions. By contrast, among the single-component models, only the $S2$ and $S4$ profiles (Lorentzian and FRED, respectively) are selected in more than isolated cases. The preference for multi-component models should nevertheless be interpreted empirically. Dense sampling can resolve secondary peaks and favor more complex profiles, whereas large gaps may weaken the constraints on a smooth single-envelope model and allow multi-component functions to absorb unresolved variability or sampling-induced discontinuities.

\subsection{Morphological Parameters}
\begin{figure*}
        \centering
        \includegraphics[scale=0.3]{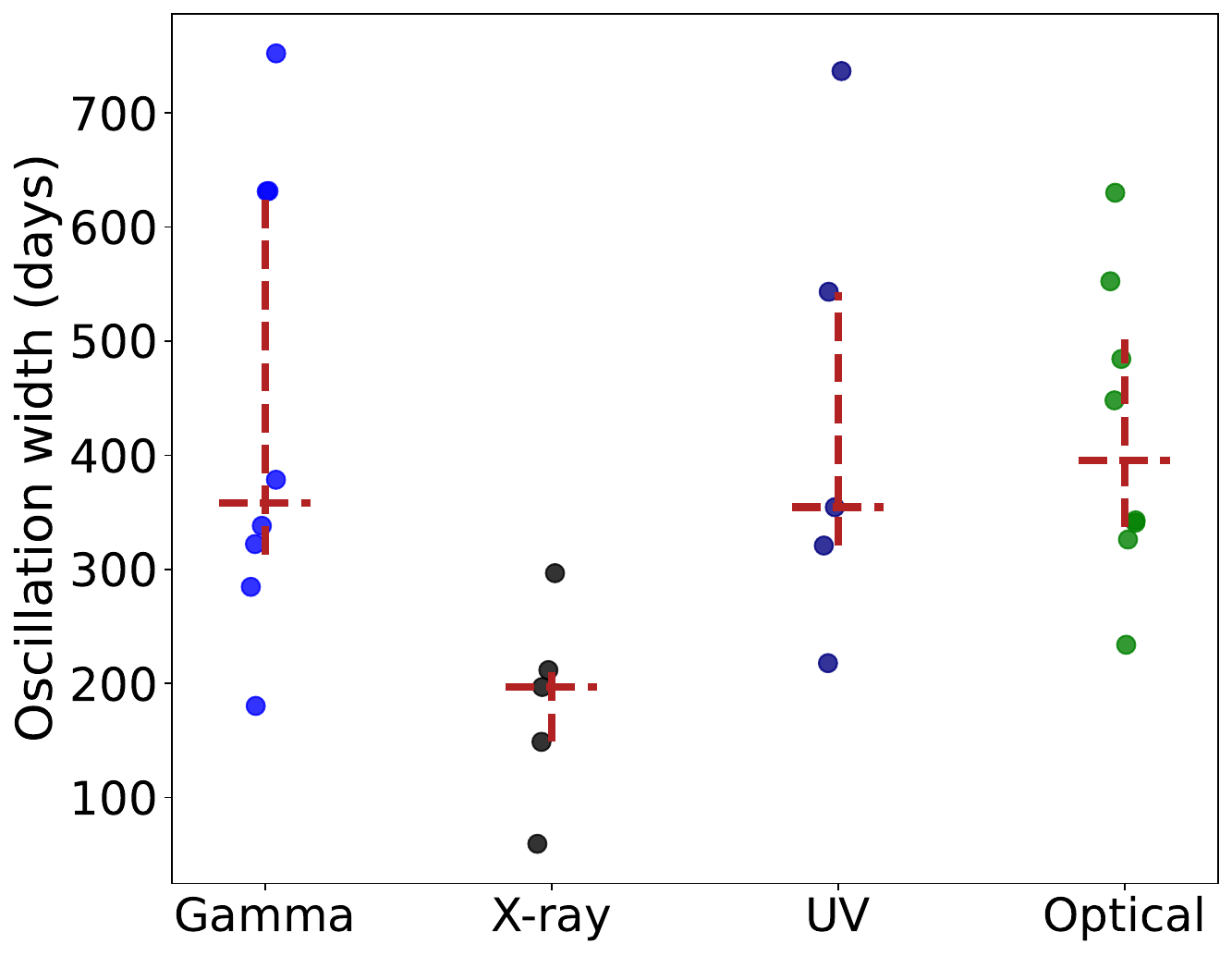}
        \includegraphics[scale=0.3]{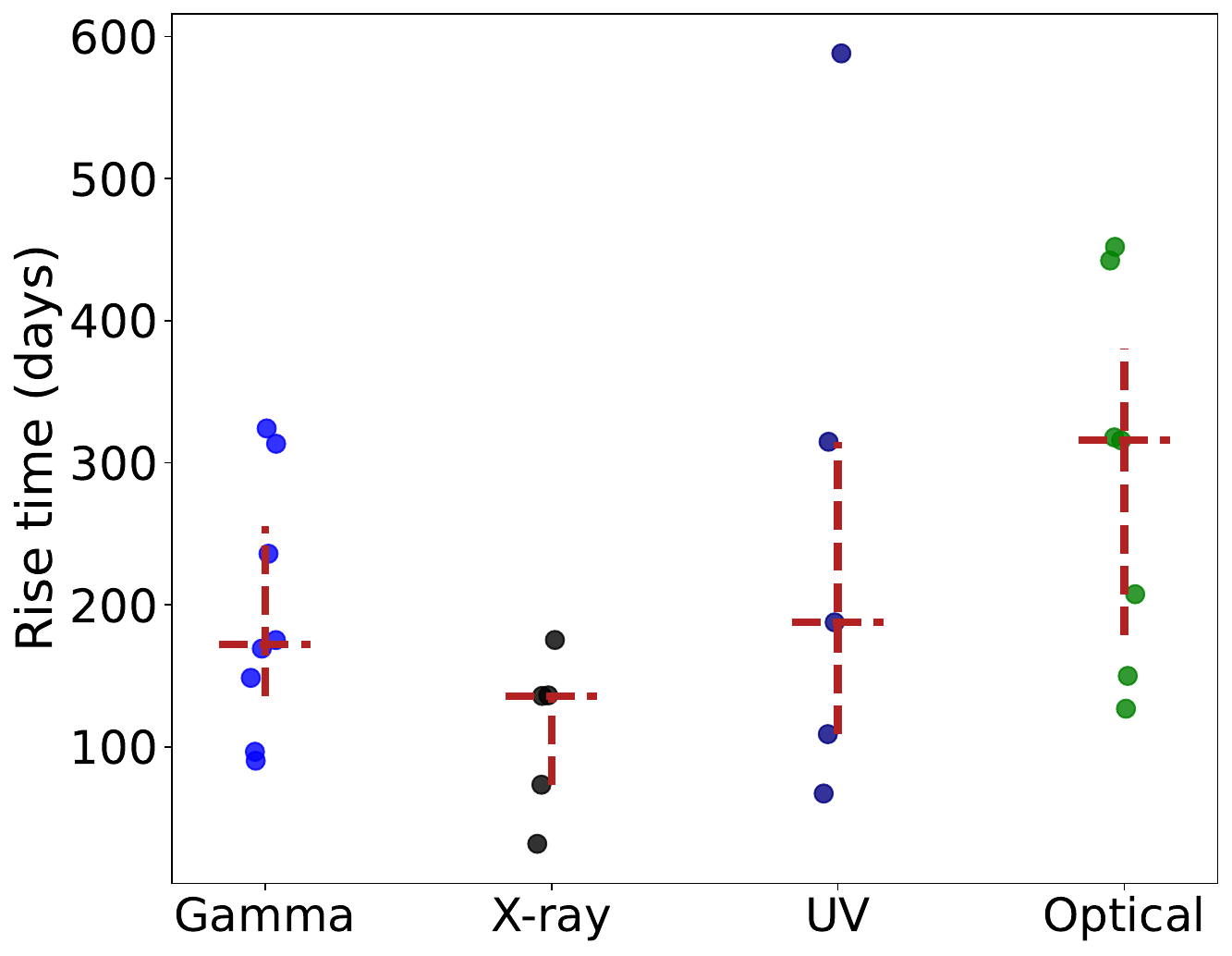} 
        \includegraphics[scale=0.3]{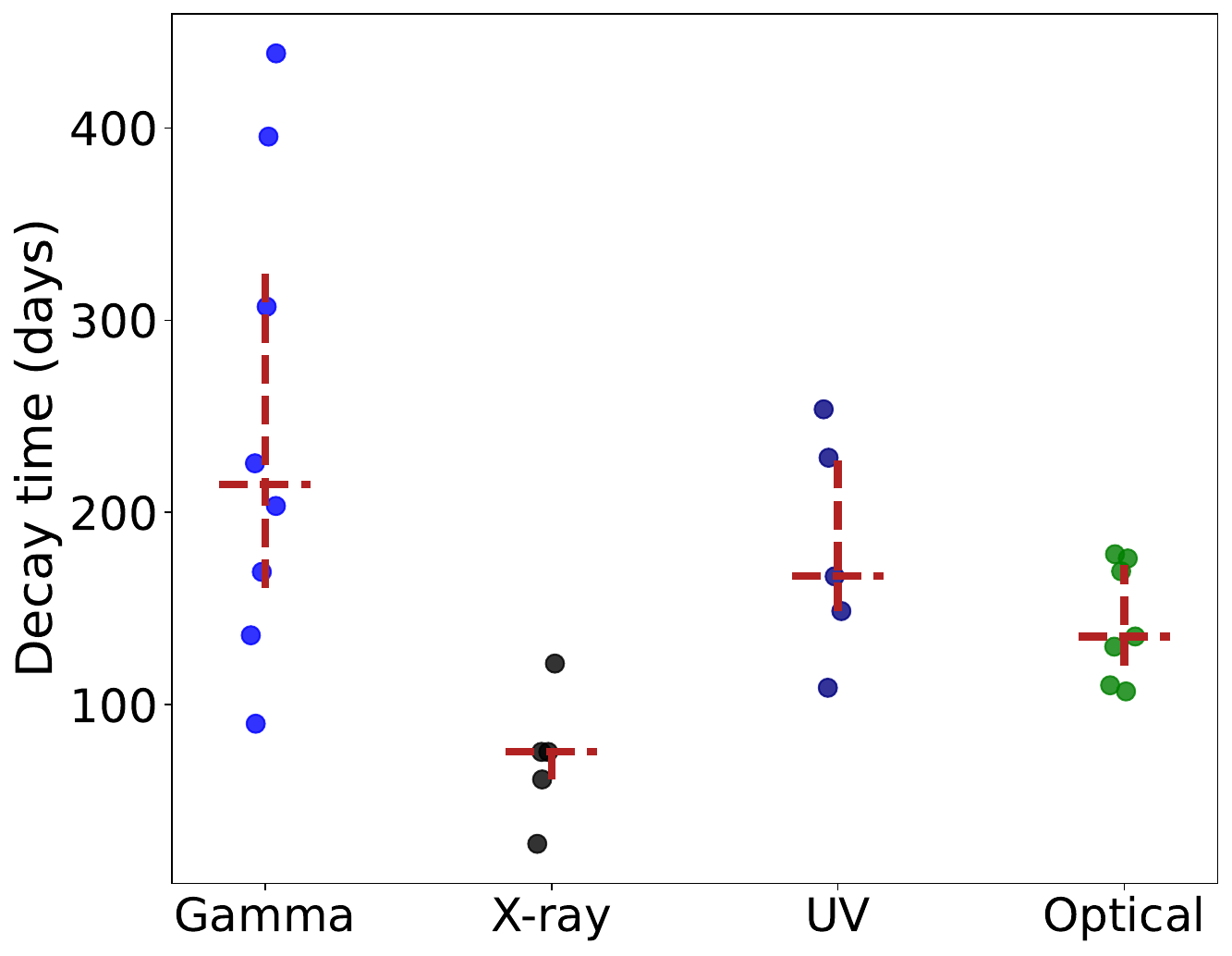}   
        \includegraphics[scale=0.3]{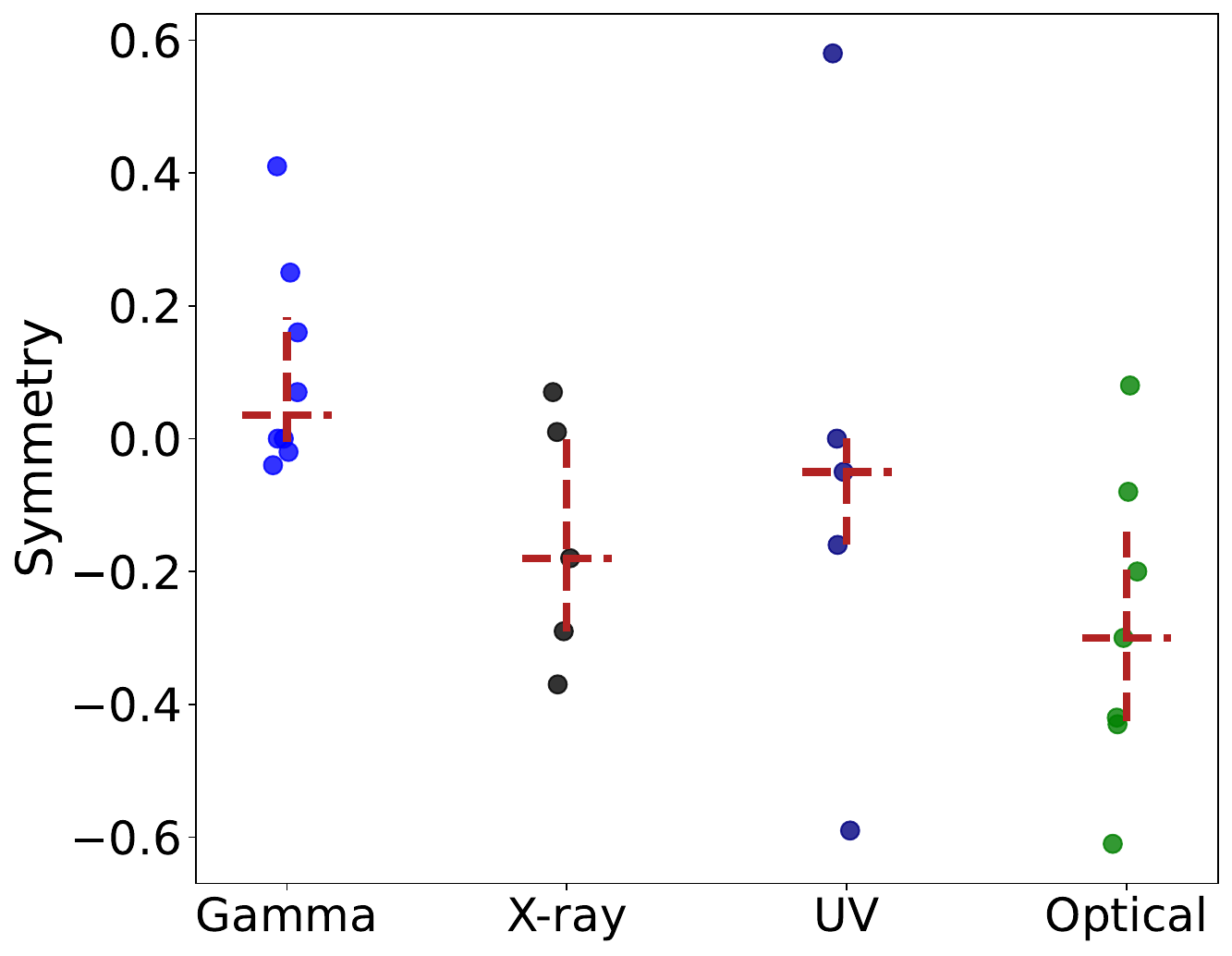}   
        \includegraphics[scale=0.3]{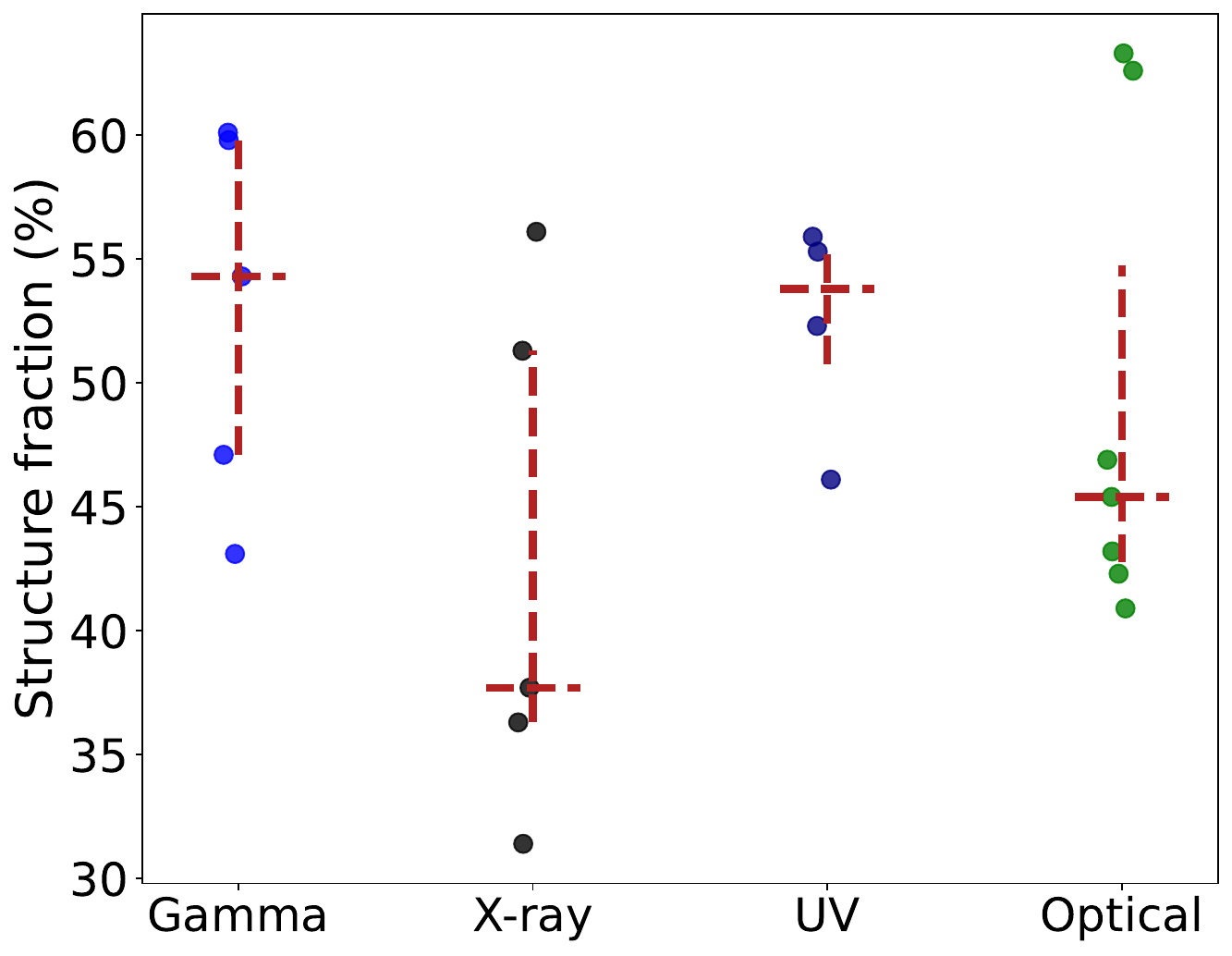}   
        \caption{Distribution of the morphological parameters derived from the best-fit oscillation profiles in the $\gamma$-ray, X-ray, UV, and optical bands of Table \ref{tab:oscillation_formology}, Table \ref{tab:oscillation_formology_xray}, Table \ref{tab:oscillation_formology_uv}, and Table \ref{tab:oscillation_formology_optical}, respectively. From top left to bottom: oscillation width, rise time, decay time, profile symmetry, and structure fraction. Individual points represent the fitted oscillations in each energy band. Parameters that did not converge are omitted from the corresponding distribution, while the other successfully determined parameters of the same cycle are retained. The horizontal red dashed bars indicate the median, while the vertical red dashed bars show the interquartile range. The comparison highlights differences in the characteristic temporal scales and symmetry of the oscillations across energy bands, while the structure fraction shows a broader overlap between the MWL distributions.} \label{fig:morphology_distribution} 
\end{figure*}

To investigate whether the oscillation profiles show systematic morphological differences with energy, we compare the distributions of the width, rise time, decay time, symmetry, and structure fraction derived from the best-fit profiles (Figure \ref{fig:morphology_distribution}). Individual oscillations are shown together with the median and interquartile range for each energy band.

The clearest difference is found in the temporal scales of the profiles. The X-ray oscillations are generally narrower than those observed at the other wavelengths and show shorter rise and, particularly, decay times. In contrast, the $\gamma$-ray and UV profiles span a broader range of widths, while the optical oscillations tend to show longer rise times. The decay-time distributions show a progressive shift from the shortest values in X-rays to longer timescales in the UV and $\gamma$-ray bands, although the distributions overlap and the number of oscillations is limited.

Differences are also visible in the profile symmetry. The $\gamma$-ray oscillations are predominantly distributed around zero symmetry, indicating approximately symmetric main peaks, whereas the X-ray and optical profiles tend toward negative values. This behavior indicates longer rise than decay times in these bands. The UV profiles show a wider range of symmetry values and do not exhibit a similarly clear concentration. This tendency toward asymmetric profiles is consistent with the frequent preference for the $M6$ model, whose exponential components can independently accommodate different rise and decay times.

The structure fraction does not show a simple monotonic dependence on wavelength. The $\gamma$-ray and UV oscillations generally show relatively large contributions from secondary components, whereas the X-ray distribution extends to lower structure fractions. The optical band spans a broad range, including some of the largest values in the sample. Thus, the presence of internal substructure appears common across the MWL profiles, but its relative contribution varies between individual oscillations and energy bands.

Overall, the distributions provide tentative evidence that the oscillation morphology is not identical across wavelength. In particular, the shorter X-ray timescales and the tendency toward asymmetric X-ray and optical profiles may indicate energy-dependent differences in the temporal evolution of the emission. However, these trends should be interpreted cautiously because of the small number of oscillations and the different sampling and temporal coverage of the light curves.

\section{Analysis of contemporaneous oscillations}\label{sec:tuples}
Although the sampling and observational coverage differ among the energy bands, several of the analyzed oscillations are contemporaneous, as reported by \citet[][]{penil_xray_pg1553_2026}. This temporal overlap provides additional context for interpreting the results presented in $\S$\ref{sec:results}.

To examine this correspondence, we group contemporaneous oscillations into tuples based on the peak times reported in Table~\ref{tab:oscillation_formology}, Table~\ref{tab:oscillation_formology_xray}, Table~\ref{tab:oscillation_formology_uv}, and Table~\ref{tab:oscillation_formology_optical}. These groupings are also consistent with the temporal associations discussed by \citet[][]{penil_xray_pg1553_2026}. We then retain only those tuples with sufficient temporal coverage to allow a detailed comparison. For each selected tuple, we compare the profiles across the different energy bands using three morphological parameters: the fitted component peak times, defined by the centers of the individual model components; the temporal separation between these components; and their relative amplitude ratios. For multi-component profiles, the dominant component is defined as that with the largest fitted amplitude. Its peak time may differ from the maximum of the total profile because of the overlap between components.

\begin{figure*}
        \centering
        \includegraphics[scale=0.18]{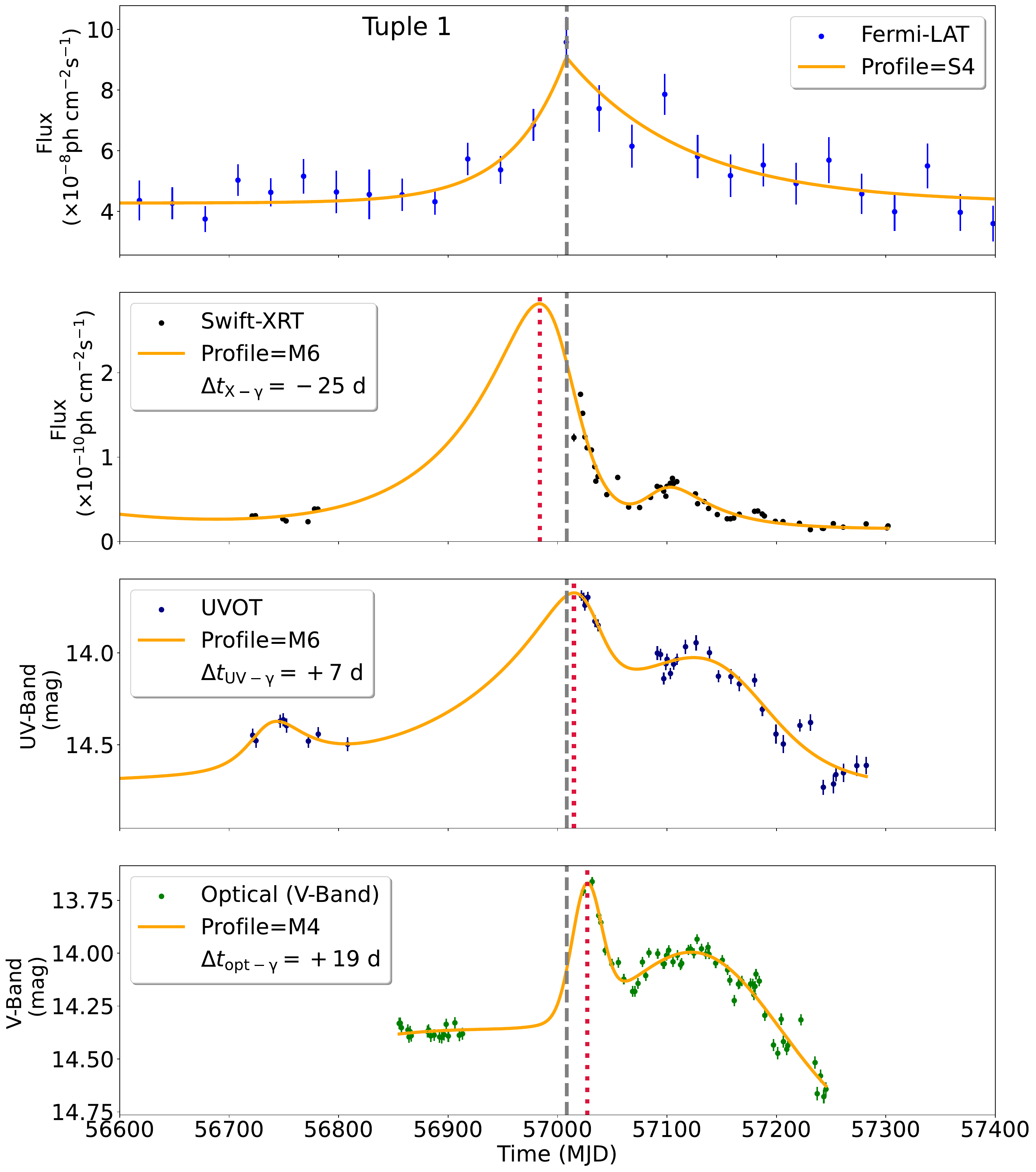}
        \includegraphics[scale=0.18]{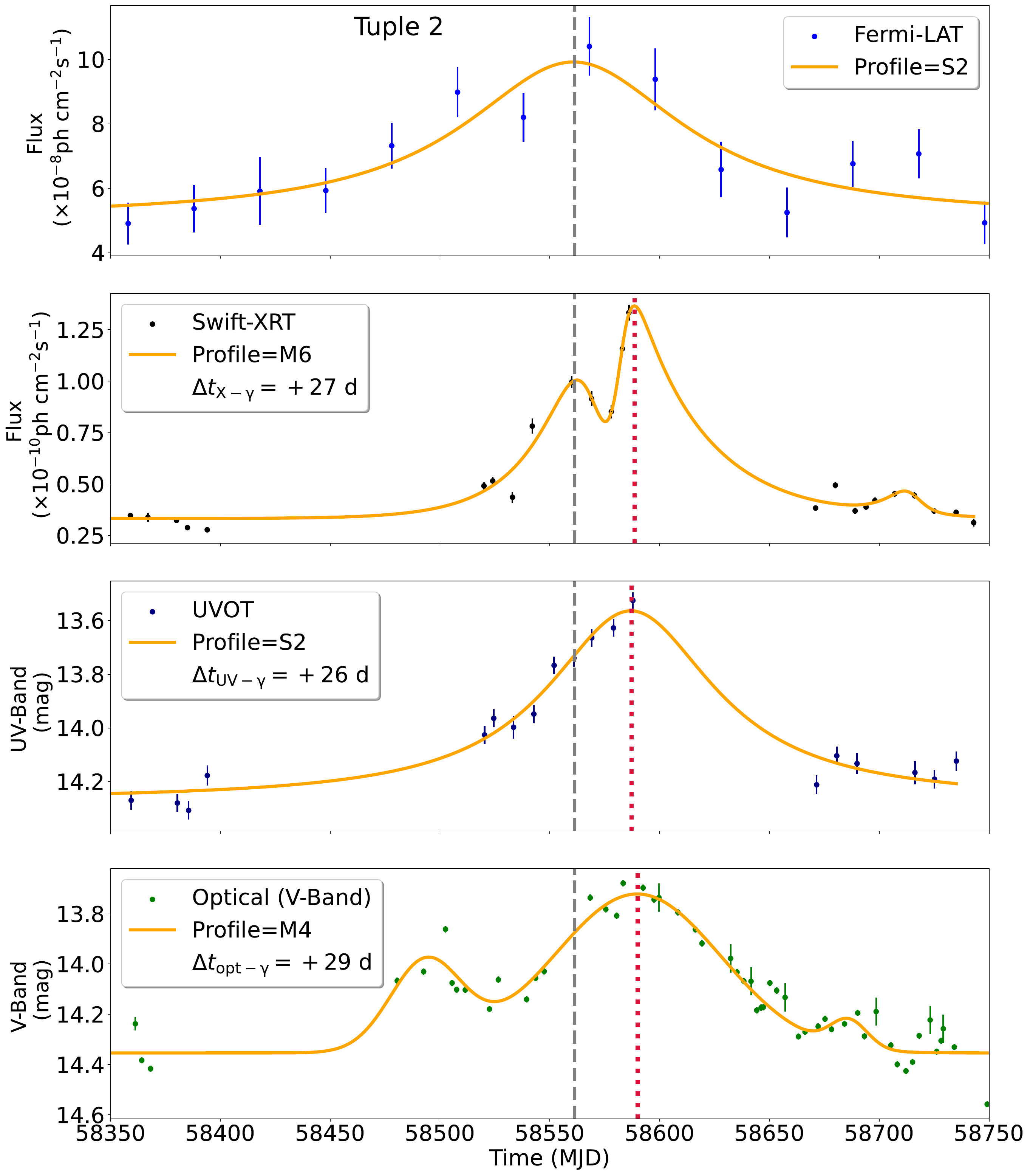}
        \includegraphics[scale=0.18]{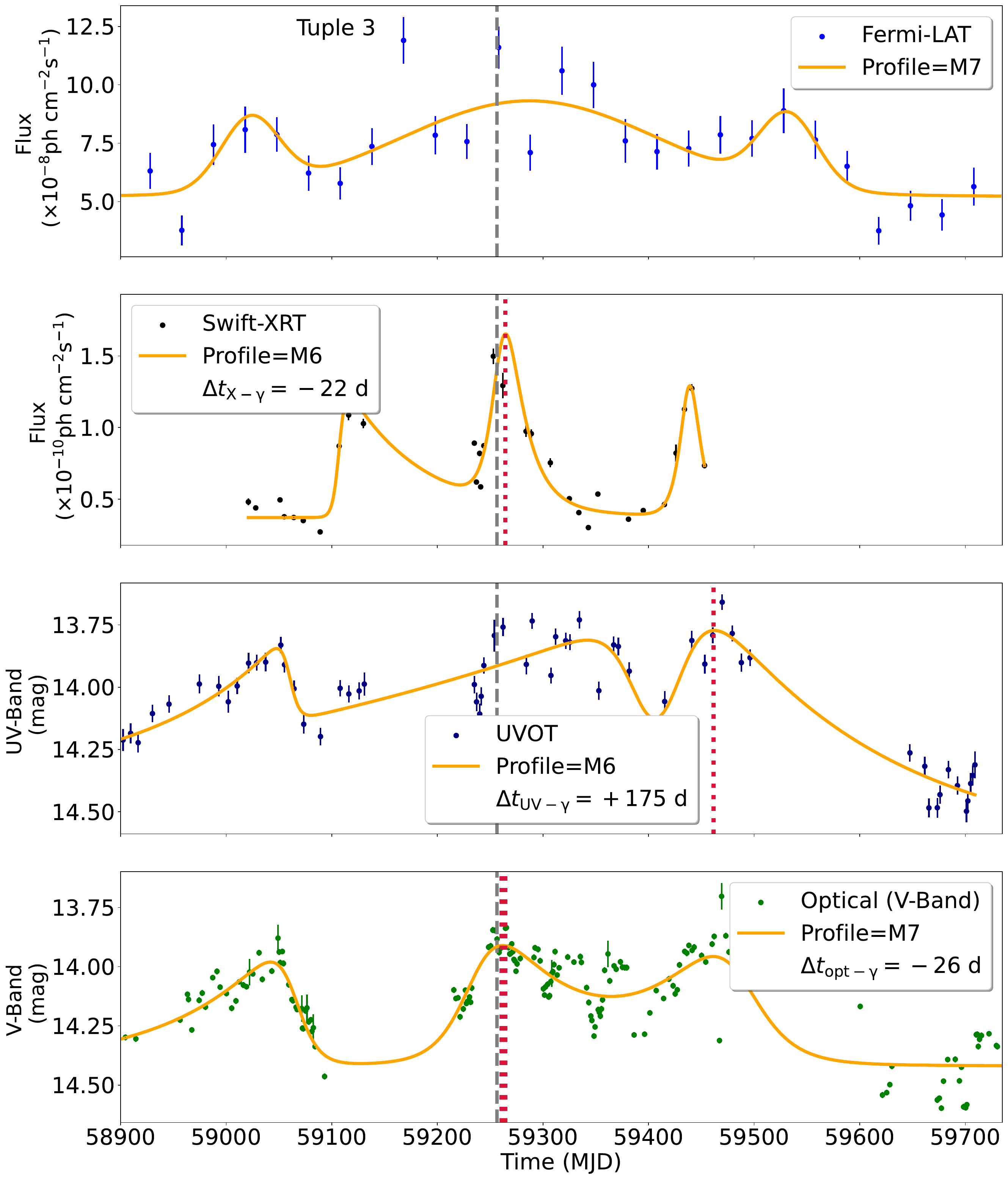}
        \includegraphics[scale=0.18]{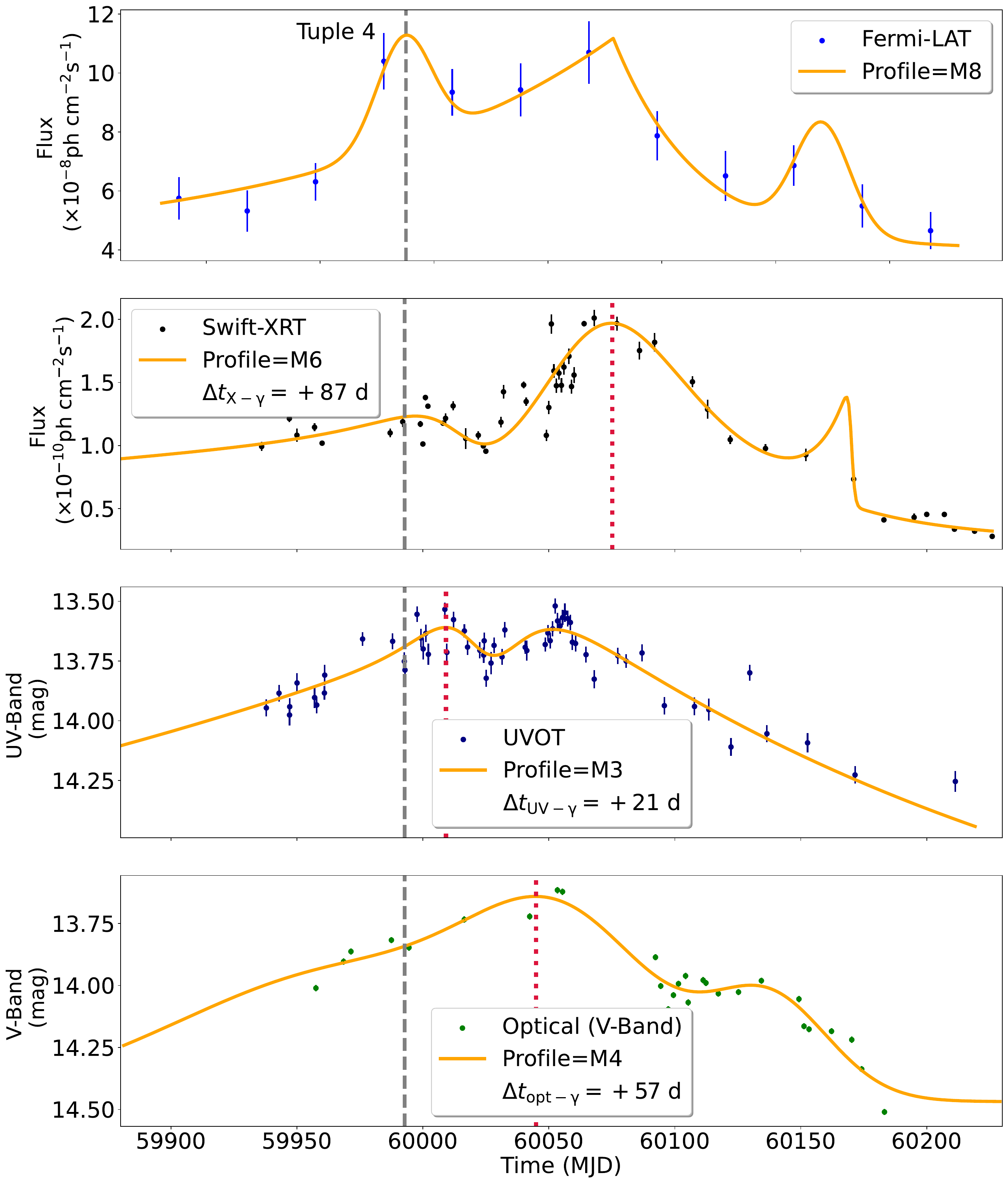}
        \caption{Tuples of contemporaneous multiwavelength oscillations. The gray dashed vertical line marks the highest-amplitude fitted component of $\gamma$-ray profile, while the red dotted vertical lines mark the corresponding highest-amplitude fitted components at the other wavelengths. The corresponding model-based delays relative to the $\gamma$-ray peak, defined as $\Delta t_{\rm band-\gamma}=t_{\rm peak,band}-t_{\rm peak,\gamma}$, are reported in the legends of the X-ray, UV, and optical panels. Thus, negative values indicate that the band peaks before the $\gamma$ rays, whereas positive values indicate a delayed peak. Top left: $Tuple_{1}$. The best-fit profile models are labeled as $S4$, denoting a FRED component; $M6$, denoting a triple-exponential model; and $M4$, for triple-Gaussian profile. Top right: $Tuple_{2}$. The best-fit profile models are labeled as $S2$, denoting a Lorentzian component; $M6$ and $M4$. Bottom left: $Tuple_{3}$. The best-fit profile models are labeled as $M7$, denoting skewed-Gaussian plus twin-Gaussian profile; and $M6$, denoting a triple-exponential model. Bottom right: $Tuple_{4}$, corresponding to the dedicated monitoring campaign presented by \citet{penil_xray_pg1553_2026}. The best-fit profile models are labeled as $M8$, denoting a FRED component plus twin-Gaussian profile; $M6$ profile; $M3$, double exponential model; and $M4$ denoting a triple-Gaussian model.}  .\label{fig:tuple_example}
\end{figure*}

The tuples are defined as $Tuple(Cycle_{i\_\gamma-ray}, Cycle_{j\_X-ray}, Cycle_{k\_UV}, Cycle_{l\_optical})$ following the order of the oscillations in each MWL band listed in Table~\ref{tab:oscillation_formology}, Table~\ref{tab:oscillation_formology_xray}, Table~\ref{tab:oscillation_formology_uv}, and Table~\ref{tab:oscillation_formology_optical}, respectively. We identify four tuples with sufficient temporal coverage for the comparative analysis (Figure~\ref{fig:tuple_example}): $Tuple_{1}(Cycle_{3}, Cycle_{2}, Cycle_{2}, Cycle_{4})$, $Tuple_{2}(Cycle_{5}, Cycle_{3}, Cycle_{3}, Cycle_{6})$, $Tuple_{3}(Cycle_{6}, Cycle_{4}, Cycle_{4}, Cycle_{7})$, and $Tuple_{4}(Cycle_{7}, Cycle_{5}, Cycle_{5}, Cycle_{8})$. The results of this dedicated study of contemporaneous oscillations are presented in Table~\ref{tab:tuple_5}.

\subsection{Results}

The comparison of contemporaneous oscillations shows that the degree of MWL correspondence changes from one tuple to another (Table~\ref{tab:tuple_5}). In general, the peak times indicate that several structures are approximately aligned across the different energy bands, although the level of agreement depends on the temporal coverage and on the complexity of the fitted profiles.

\paragraph{$Tuple_{1}$}
$Tuple_{1}$ shows a relative correspondence of the dominant MWL peaks (Figure \ref{fig:tuple_example}). The model-based peak offsets relative to the $\gamma$-ray maximum are $-25$, $+7$, and $+19$~days for the X-ray, UV, and optical bands, respectively. A later secondary peak is also detected in the X-ray, UV, and optical profiles, whereas the first UV peak has no clear counterpart in the other bands. In general, the amplitude ratios indicate that the central peak is the dominant component in the bands where multiple peaks are detected, while the later secondary peak is weaker, reaching 10.9\%, 64.9\%, and 48.7\% of the main-peak amplitude in the X-ray, UV, and optical bands, respectively. This result suggests that $Tuple_{1}$ contains a common MWL envelope, but that not all internal substructures are shared across the full data set. In particular, the X-ray, UV, and optical bands show evidence of a coincident double-peaked structure, whereas the $\gamma$-ray profile is better described by a single dominant peak. This difference may indicate either that the secondary component is weaker or smoothed out in the 30-day binned $\gamma$-ray light curve, or that the lower-energy bands trace additional substructure not clearly detected at $\gamma$-ray energies.

\paragraph{$Tuple_{2}$}
$Tuple_{2}$ is less complete because the $\gamma$-ray and UV profiles show only one identified peak (Figure \ref{fig:tuple_example}). The X-ray, UV, and optical maxima are delayed by $+27$, $+26$, and $+29$~days relative to the $\gamma$-ray peak, respectively. The optical profile shows a three-peak structure that is nearly symmetric in peak separation, with the secondary peaks located 95~days from the main peak on either side. Their amplitudes, however, differ substantially, reaching 60.3\% and 21.8\% of the main-peak amplitude. In contrast, the X-ray profile is more asymmetric, with peak separations of 26 and 124~days, although the relative amplitudes of the secondary components, 64.9\% and 12.4\%, are broadly similar to those observed in the optical band. These results suggest that the same broad activity episode is detected across multiple bands, while its internal substructure is not recovered uniformly. In particular, the optical and X-ray profiles reveal additional short-timescale structure, whereas the $\gamma$-ray and UV emission is dominated by a single peak.

\paragraph{$Tuple_{3}$}
$Tuple_{3}$ provides evidence for a coincident multi-component MWL oscillation, with three identifiable peaks in all four bands (Figure \ref{fig:tuple_example}). The main X-ray and optical model peaks occur 22 and 26~days before the $\gamma$-ray maximum, respectively. In contrast, the UV model peak is delayed by $+175$~days because the highest-amplitude fitted peak corresponds to the later component of the oscillation. This large UV offset should therefore not be interpreted directly as an interband delay, since the selected main peak can correspond to a different subcomponent and may also be affected by sampling and intrinsic variability.

The first and third peaks are also present in all bands, suggesting that the oscillation is structured rather than dominated by a single isolated maximum. However, the detailed timing is not identical across the MWL data set: the $\gamma$-ray, X-ray, and optical profiles show a more similar and approximately symmetric three-peak pattern, while the UV profile displays a less symmetric structure, with peak separations that differ more strongly from the other bands. The relatively high amplitude ratios, especially in the $\gamma$-ray, UV, and optical bands where the secondary peaks reach $\gtrsim80\%$ of the dominant peak, further support the presence of pronounced internal structure. Therefore, the oscillation shows a recurrent multi-component structure, although the relative dominance and timing of the individual peaks differ between bands.

\paragraph{$Tuple_{4}$}
$Tuple_{4}$ shows a temporal agreement between the $\gamma$-ray and X-ray profiles (Figure \ref{fig:tuple_example}). Both bands display three approximately aligned peaks with similar separations. However, the highest-amplitude X-ray model peak occurs $+87$~days after the $\gamma$-ray maximum because the dominant fitted component is not the same in the two profiles. This offset therefore reflects the selection of different dominant components and does not contradict the close component-by-component correspondence of the three-peak structures.

The UV and optical maxima occur $+21$ and $+57$~days after the $\gamma$-ray peak, respectively. Both bands show contemporaneous activity with fewer detected components, and the optical emission appears smoother and more delayed in its second component. The amplitude ratios further show that the first two $\gamma$-ray peaks have comparable values ($\sim$100\%), whereas the third peak is substantially weaker ($\sim$45\%). A similar behavior is observed in the UV and optical bands, where the detected peaks are associated with the strongest part of the oscillation, while the later component is weaker. Therefore, $Tuple_{4}$ provides evidence for a common MWL activity episode, although the dominant fitted peak is not always the same morphological component in each band.

The approximate agreement between the main $\gamma$-ray and X-ray activity peaks suggests that both bands may trace a related variability episode, although the detailed substructure is not equally well defined. The UV and optical profiles also cover a similar broad activity interval but show smoother or less pronounced secondary structure. The temporal widths further suggest a broad envelope across all bands. While the X-ray profile appears more compact ($\sim$200 days), the overall duration of the activity episode is comparable in the $\gamma$-ray, UV, and optical bands ($\sim$350 days).

\subsection{Discussion}
Overall, the tuple analysis suggests that the contemporaneous MWL oscillations share a broadly comparable temporal envelope, while their internal morphology can differ substantially across energy bands. In several cases, the same activity episode is described by a smooth component profile in one band, such as a Lorentzian or FRED-like function, and by a combination of exponential or Gaussian components in another band. These differences involve not only the best-fit analytical profile, but also the number of components, their relative timing, peak separations, and amplitude ratios. Although sampling and data coverage can affect the level of detail recovered, the observed morphological diversity suggests that the internal structure is not simply a common pattern reproduced identically at all energies

The model-based offsets between the highest-amplitude peak in each band and the $\gamma$-ray maximum provide an additional measure of this morphological diversity. $Tuple_{2}$ shows similar positive offsets of $\sim$26--29~days in the lower-energy bands, whereas the other tuples display a wider range of values. However, these offsets should not always be interpreted as physical interband delays. In multi-component oscillations, the highest-amplitude fitted peak can correspond to different substructures in different bands, as illustrated by the UV profile of $Tuple_{3}$ and the X-ray profile of $Tuple_{4}$. The inferred offsets can also be affected by sampling, data gaps, and intrinsic short-timescale variability.

This point is particularly important when comparing the $\gamma$-ray and lower-energy bands. Due to the 30-day binning of the \textit{Fermi}-LAT light curve, it can smooth short-timescale variability and can merge nearby subpeaks into a broader structure. As a result, the $\gamma$-ray profiles can tend to emphasize the global envelope of the oscillation rather than a potential multi-peak structure. By contrast, the X-ray, UV, and optical light curves are not uniformly sampled. When the temporal coverage is dense, these bands can reveal sharper or multiple components within the same broad oscillation. However, when the coverage contains large gaps, multi-component models, especially combinations of exponential rise/decay profiles, may partly reflect the flexibility needed to connect incomplete segments of the light curve. 

In the tuples, the broad modulation is accompanied by shorter-timescale substructure, observed as secondary peaks. Some of these features occur at comparable epochs in different bands, suggesting that part of the short-timescale variability may be connected across the MWL emission. However, this correspondence is not systematic. In $Tuple_{4}$, the $\gamma$-ray and X-ray profiles show approximately aligned main activity peaks, although the detailed substructure is not equally pronounced in the two bands. The UV and optical profiles trace the same broad activity interval but display different numbers and timings of secondary components. $Tuple_{3}$ shows a three-peak structure across all bands, but with differences in peak timing, separation, and relative amplitudes. In contrast, $Tuple_{1}$ and $Tuple_{2}$ show only partial correspondence, with broadly contemporaneous dominant peaks but secondary components that differ between bands or are detected only at specific energies. Overall, this diversity indicates that the variability within the broad recurrent envelope is not itself strictly periodic or reproduced in the same way across wavelengths. Instead, the changing peak timing, relative amplitudes, and profile shapes suggest that multiple variability processes contribute to the detailed morphology of each oscillation.

\section{Physical Interpretation}\label{sec:physical}
The profile analysis of the contemporaneous MWL tuples suggests that the oscillations in PG~1553+113 are not described by a single, strictly self-similar profile repeated from cycle to cycle. Instead, they appear to consist of a broad, long-timescale modulation with shorter-timescale substructure, manifested as secondary peaks, superimposed on it. Although the $\approx$2.1 yr modulation provides the dominant recurrent timescale, the non-repeating internal morphology suggests that additional variability processes shape individual activity episodes. These processes may vary from cycle to cycle and across energy bands.

Previous studies have reported significant MWL correlations with lags globally consistent with zero \citep[][]{ackermann_pg1553, magic_pg1553_2024, penil_mwl_pg1553, penil_xray_pg1553_2026}. These correlations support a broadband connection in the long-term variability. Our morphological analysis provides a complementary view: the dominant activity peaks can be approximately contemporaneous, whereas individual substructures show temporal offsets whose magnitude and direction vary between oscillation episodes. Thus, a global lag consistent with zero does not necessarily imply strictly simultaneous variability within each individual cycle. The band-dependent profile differences and shorter-timescale substructure instead suggest the presence of additional local or energy-dependent variability processes that modulate the detailed emission within each episode.

This picture is consistent with the dual-variability scenario proposed by \citet{madero_dominguez_2026}, in which the long-term modulation has a geometric origin, while shorter-timescale variability is driven by intrinsic plasma processes. In this framework, the broad MWL envelope could arise when the emitting region becomes more closely aligned with the line of sight, increasing the Doppler boosting across several energy bands \citep[e.g.,][]{camenzind_jet,rieger_2004,penil_2025_linear,madero_dominguez_2026}. Such an interpretation is compatible with the contemporaneous enhancements observed from $\gamma$-ray to optical wavelengths in the analyzed tuples.

Consistent with this scenario, the diversity of the fitted profiles suggests that intrinsic jet variability shapes the substructure superimposed on the geometrically modulated envelope. Several cycles are better reproduced by multi-component models, with pronounced secondary structure detected in more than one band. Shock-in-jet scenarios and multi-zone turbulent models provide natural interpretations of this additional variability, since they allow a major activity episode to comprise multiple emitting zones or dissipation events, producing asymmetric and multi-peaked temporal profiles \citep[e.g.,][]{marscher_gear_1985,marcher_turbulent_2014}. This picture is further supported by analyses of bright $\gamma$-ray blazar flares, in which apparently coherent long-duration events can arise from the superposition of shorter flaring episodes \citep{saito_flares_2013,nalewajko_flares_shapes_2013}.

Nevertheless, the role of the sampling pattern cannot be neglected. Differences in sampling, data gaps, and uneven temporal coverage across energy bands can alter the apparent shape of individual oscillations and affect the preferred profile model. However, multi-component substructure is clearly recovered in several well-sampled episodes, and the differences in peak timing, separation, and relative amplitude between energy bands do not follow a consistent pattern from one episode to another. Therefore, although sampling may influence the detailed fitted morphology, the observed diversity of the internal structure is unlikely to be explained solely by observational coverage. The physical interpretation of individual components should remain cautious, but the overall picture is that the recurrent $\sim2.1$ yr modulation defines a broad activity envelope, while the detailed evolution within each episode is shaped by additional variability processes and is not repeated identically across cycles or energy bands.


\subsection{The Binary hypotheses}
The rise--decay parameters show that the $\gamma$-ray oscillations cannot be divided simply into two distinct classes, namely purely symmetric single flares and asymmetric complex events. Instead, the symmetry parameter appears to characterize the global envelope of each oscillation. This is illustrated by Cycles~1, 2, 5, 6, and 7 in $\gamma$ rays, for which the overall profiles are approximately symmetric, even though some of them require multi-component fits. In particular, Cycles~1, 6, and 7 are best reproduced by multi-component models, but their symmetry parameters remain close to zero (Table~\ref{tab:oscillation_formology}). This indicates that secondary components may be present without strongly disturbing the global envelope. The regular 30-day $\gamma$-ray binning must also be considered. Closely spaced subpeaks can be blended into a smooth envelope, reducing the apparent structure of the profile, potentially not capturing subpeaks in some oscillations.

In this context, \citet{tavani_pdm_pg_1553} interpreted the long-term modulation of PG~1553+113 within a SMBHB scenario. In their model, the interaction between the two black holes can produce weaker secondary features, or ``twin peaks,'' located approximately symmetrically on both sides of the principal maxima. These twin peaks were interpreted as evidence that the broad oscillation envelope contains physically distinct subcomponents rather than constituting a single smooth event. At the same time, \citet{tavani_pdm_pg_1553} did not present the $\gamma$-ray twin peaks as strong and obvious features in every cycle. Their limited prominence in the \textit{Fermi}-LAT light curve is plausibly related to their lower amplitude relative to the main outbursts, the larger statistical noise of the $\gamma$-ray data, and the fact that the proposed secondary pattern is repetitive but not strictly periodic. Our results are broadly consistent with this picture. In particular, the intermittent appearance of twin-peak-like structure in the $\gamma$-ray data agrees with the fact that such substructure is not expected to be equally evident in all cycles. Within the interpretation of \citet{tavani_pdm_pg_1553}, the intermittent prominence of these secondary peaks may reflect cycle-to-cycle variations in the plasma instabilities triggered by the binary dynamics, so that the twin-peak components need not reach comparable amplitudes in every oscillation. 

The time interval analyzed by \citet{tavani_pdm_pg_1553} covers the Fermi-LAT data from 2008 to 2018, corresponding approximately to our cycles~1--4. Within our analysis, cycles~1 and 4 show evidence of a dominant main peak accompanied by weaker flanking components in Figure~\ref{fig:best_fit_gamma}. Although these $\gamma$-ray cycles are not included in the MWL tuples discussed in \S\ref{sec:results}, they correspond approximately to optical cycles~2 and 5, where a multi-peak structure is also observed (Figure~\ref{fig:best_fit_optical}). In addition, optical cycle~6 shows a similar morphology, even though its counterpart is not clearly resolved in the $\gamma$ rays (see $Tuple_{2}$ in Figure~\ref{fig:tuple_example}).

The later cycles, extending beyond the time range considered by \citet{tavani_pdm_pg_1553}, show comparable morphological diversity. In particular, cycles~6 and 7, corresponding to the MWL $Tuple_{3}$ and $Tuple_{4}$ in Figure~\ref{fig:tuple_example}, display a dominant peak accompanied by weaker secondary components. This morphology is especially evident in $Tuple_{3}$ and is recovered in more than one energy band, showing that it is not confined to the epochs studied by \citet{tavani_pdm_pg_1553}. Although its statistical recurrence and physical origin cannot yet be established, its appearance in later cycles is phenomenologically consistent with the scenario proposed in that work, in which secondary structures may arise from binary-related perturbations, and provides additional motivation for examining the SMBHB interpretation of PG~1553+113.

\section{Summary} \label{sec:summary}
In this paper, we analyzed the MWL oscillations of the blazar PG~1553+113, focusing on the morphology of the individual oscillation profiles associated with the previously reported $\sim 2.1$-yr modulation. We studied the $\gamma$-ray, X-ray, UV, and optical bands, and modeled the identified oscillations using a set of profile functions designed to describe both single-peaked and multi-peaked structures, based on forms commonly adopted in the literature for flares and oscillatory events.

Our results show that the oscillations are generally characterized by a broad, approximately symmetric envelope, but that this global structure is often accompanied by internal substructure in the form of secondary peaks. Therefore, the MWL oscillations of PG~1553+113 are not well described as purely smooth isolated events, but rather as structured modulations combining a global recurrent pattern with additional internal complexity. A particularly notable result is the close component-by-component correspondence between the $\gamma$-ray and X-ray profiles in one of the contemporaneous oscillations, where both bands recover a similar multi-peak structure. This level of morphological agreement is not observed in the corresponding UV and optical profiles. Although differences in sampling may contribute to this result, the markedly different lower-energy profiles leave open the possibility of a genuine connection between the $\gamma$-ray and X-ray variability.

In particular, we identify new oscillations showing a dominant main peak accompanied by secondary twin-peak-like features. Similar structures were previously discussed in the context of a SMBHB scenario for PG~1553+113. The presence of comparable morphologies in newly analyzed oscillations shows that this pattern is not restricted to the epochs examined in earlier studies. Although our results do not provide definitive proof of the binary interpretation, they show that this scenario remains plausible for explaining at least part of the observed MWL oscillation morphology.

\section*{Acknowledgements}
\begin{acknowledgements}
P.P. and M.A. acknowledge funding under NASA contract 80NSSC20K1562. L.M. acknowledges that this work was supported by the Initiative and Networking Fund of the Helmholtz Association under the Helmholtz Investigator Groups Programme, call 2025 (VH-NG-21-01). LM acknowledges support from DESY (Zeuthen, Germany), a member of the Helmholtz Association HGF. AD acknowledges support from MCIN/AEI under grant PID2025-171816NB-I00. This work was supported by the European Research Council, ERC Starting grant \textit{MessMapp}, S.B. Principal Investigator, under contract no. 949555, and by the German Science Foundation DFG, research grant “Relativistic Jets in Active Galaxies” (FOR 5195, grant No. 443220636). 
Generative AI tools were used solely to improve the English language and clarity of the manuscript. 
\end{acknowledgements}

\bibliographystyle{mnras}
\bibliography{oja_template}

@ARTICLE{ackermann_pg1553,
       author = {{Ackermann}, M. and {Ajello}, M. and {Albert}, A. and {Atwood}, W.~B. and {Baldini}, L. and {Ballet}, J. and {Barbiellini}, G. and {Bastieri}, D. and {Becerra Gonzalez}, J. and {Bellazzini}, R. and {Bissaldi}, E. and {Blandford}, R.~D. and {Bloom}, E.~D. and {Bonino}, R. and {Bottacini}, E. and {Bregeon}, J. and {Bruel}, P. and {Buehler}, R. and {Buson}, S. and {Caliandro}, G.~A. and {Cameron}, R.~A. and {Caputo}, R. and {Caragiulo}, M. and {Caraveo}, P.~A. and {Cavazzuti}, E. and {Cecchi}, C. and {Chekhtman}, A. and {Chiang}, J. and {Chiaro}, G. and {Ciprini}, S. and {Cohen-Tanugi}, J. and {Conrad}, J. and {Cutini}, S. and {D'Ammando}, F. and {de Angelis}, A. and {de Palma}, F. and {Desiante}, R. and {Di Venere}, L. and {Dom{\'\i}nguez}, A. and {Drell}, P.~S. and {Favuzzi}, C. and {Fegan}, S.~J. and {Ferrara}, E.~C. and {Focke}, W.~B. and {Fuhrmann}, L. and {Fukazawa}, Y. and {Fusco}, P. and {Gargano}, F. and {Gasparrini}, D. and {Giglietto}, N. and {Giommi}, P. and {Giordano}, F. and {Giroletti}, M. and {Godfrey}, G. and {Green}, D. and {Grenier}, I.~A. and {Grove}, J.~E. and {Guiriec}, S. and {Harding}, A.~K. and {Hays}, E. and {Hewitt}, J.~W. and {Hill}, A.~B. and {Horan}, D. and {Jogler}, T. and {J{\'o}hannesson}, G. and {Johnson}, A.~S. and {Kamae}, T. and {Kuss}, M. and {Larsson}, S. and {Latronico}, L. and {Li}, J. and {Li}, L. and {Longo}, F. and {Loparco}, F. and {Lott}, B. and {Lovellette}, M.~N. and {Lubrano}, P. and {Magill}, J. and {Maldera}, S. and {Manfreda}, A. and {Max-Moerbeck}, W. and {Mayer}, M. and {Mazziotta}, M.~N. and {McEnery}, J.~E. and {Michelson}, P.~F. and {Mizuno}, T. and {Monzani}, M.~E. and {Morselli}, A. and {Moskalenko}, I.~V. and {Murgia}, S. and {Nuss}, E. and {Ohno}, M. and {Ohsugi}, T. and {Ojha}, R. and {Omodei}, N. and {Orlando}, E. and {Ormes}, J.~F. and {Paneque}, D. and {Pearson}, T.~J. and {Perkins}, J.~S. and {Perri}, M. and {Pesce-Rollins}, M. and {Petrosian}, V. and {Piron}, F. and {Pivato}, G. and {Porter}, T.~A. and {Rain{\`o}}, S. and {Rando}, R. and {Razzano}, M. and {Readhead}, A. and {Reimer}, A. and {Reimer}, O. and {Schulz}, A. and {Sgr{\`o}}, C. and {Siskind}, E.~J. and {Spada}, F. and {Spandre}, G. and {Spinelli}, P. and {Suson}, D.~J. and {Takahashi}, H. and {Thayer}, J.~B. and {Thompson}, D.~J. and {Tibaldo}, L. and {Torres}, D.~F. and {Tosti}, G. and {Troja}, E. and {Uchiyama}, Y. and {Vianello}, G. and {Wood}, K.~S. and {Wood}, M. and {Zimmer}, S. and {Berdyugin}, A. and {Corbet}, R.~H.~D. and {Hovatta}, T. and {Lindfors}, E. and {Nilsson}, K. and {Reinthal}, R. and {Sillanp{\"a}{\"a}}, A. and {Stamerra}, A. and {Takalo}, L.~O. and {Valtonen}, M.~J.},
        title = "{Multiwavelength Evidence for Quasi-periodic Modulation in the Gamma-Ray Blazar PG 1553+113}",
      journal = {\apjl},
         year = 2015,
        month = nov,
       volume = {813},
       number = {2},
          eid = {L41},
        pages = {L41},
          doi = {10.1088/2041-8205/813/2/L41},
archivePrefix = {arXiv},
       eprint = {1509.02063},
 primaryClass = {astro-ph.HE},
       adsurl = {https://ui.adsabs.harvard.edu/abs/2015ApJ...813L..41A}
}

@ARTICLE{magic_pg1553_2024,
       author = {{MAGIC Collaboration} and {Abe}, H. and {Abe}, S. and {Abhir}, J. and {Acciari}, V.~A. and {Agudo}, I. and {Aniello}, T. and {Ansoldi}, S. and {Antonelli}, L.~A. and {Arbet Engels}, A. and {Arcaro}, C. and {Artero}, M. and {Asano}, K. and {Baack}, D. and {Babi{\'c}}, A. and {Baquero}, A. and {Barres de Almeida}, U. and {Batkovi{\'c}}, I. and {Baxter}, J. and {Becerra Gonz{\'a}lez}, J. and {Bernardini}, E. and {Bernete}, J. and {Berti}, A. and {Besenrieder}, J. and {Bigongiari}, C. and {Biland}, A. and {Blanch}, O. and {Bonnoli}, G. and {Bo{\v{s}}njak}, {\v{Z}}. and {Burelli}, I. and {Busetto}, G. and {Campoy-Ordaz}, A. and {Carosi}, A. and {Carosi}, R. and {Carretero-Castrillo}, M. and {Castro-Tirado}, A.~J. and {Chai}, Y. and {Cifuentes}, A. and {Cikota}, S. and {Colombo}, E. and {Contreras}, J.~L. and {Cortina}, J. and {Covino}, S. and {D'Amico}, G. and {D'Elia}, V. and {da Vela}, P. and {Dazzi}, F. and {de Angelis}, A. and {de Lotto}, B. and {Del Popolo}, A. and {Delfino}, M. and {Delgado}, J. and {Delgado Mendez}, C. and {Depaoli}, D. and {di Pierro}, F. and {di Venere}, L. and {Dominis Prester}, D. and {Donini}, A. and {Dorner}, D. and {Doro}, M. and {Elsaesser}, D. and {Emery}, G. and {Escudero}, J. and {Fari{\~n}a}, L. and {Fattorini}, A. and {Foffano}, L. and {Font}, L. and {Fukami}, S. and {Fukazawa}, Y. and {Garc{\'\i}a L{\'o}pez}, R.~J. and {Gasparyan}, S. and {Gaug}, M. and {Giesbrecht Paiva}, J.~G. and {Giglietto}, N. and {Giordano}, F. and {Gliwny}, P. and {Grau}, R. and {Green}, J.~G. and {Hadasch}, D. and {Hahn}, A. and {Heckmann}, L. and {Herrera}, J. and {Hovatta}, T. and {Hrupec}, D. and {H{\"u}tten}, M. and {Imazawa}, R. and {Inada}, T. and {Iotov}, R. and {Ishio}, K. and {Jimenez Mart{\'\i}nez}, I. and {Jormanainen}, J. and {Kerszberg}, D. and {Kluge}, G.~W. and {Kobayashi}, Y. and {Kouch}, P.~M. and {Kubo}, H. and {Kushida}, J. and {L{\'a}inez Lez{\'a}un}, M. and {Lamastra}, A. and {Leone}, F. and {Lindfors}, E. and {Liodakis}, I. and {Lombardi}, S. and {Longo}, F. and {L{\'o}pez-Moya}, M. and {L{\'o}pez-Oramas}, A. and {Loporchio}, S. and {Lorini}, A. and {Machado de Oliveira Fraga}, B. and {Majumdar}, P. and {Makariev}, M. and {Maneva}, G. and {Mang}, N. and {Manganaro}, M. and {Mannheim}, K. and {Mariotti}, M. and {Mart{\'\i}nez}, M. and {Mart{\'\i}nez-Chicharro}, M. and {Mas-Aguilar}, A. and {Mazin}, D. and {Menchiari}, S. and {Mender}, S. and {Miceli}, D. and {Miener}, T. and {Miranda}, J.~M. and {Mirzoyan}, R. and {Molero Gonz{\'a}lez}, M. and {Molina}, E. and {Mondal}, H.~A. and {Moralejo}, A. and {Morcuende}, D. and {Nakamori}, T. and {Nanci}, C. and {Neustroev}, V. and {Nigro}, C. and {Nikoli{\'c}}, L. and {Nilsson}, K. and {Nishijima}, K. and {Njoh Ekoume}, T. and {Noda}, K. and {Nozaki}, S. and {Ohtani}, Y. and {Okumura}, A. and {Otero-Santos}, J. and {Paiano}, S. and {Palatiello}, M. and {Paneque}, D. and {Paoletti}, R. and {Paredes}, J.~M. and {Pavlovi{\'c}}, D. and {Persic}, M. and {Pihet}, M. and {Pirola}, G. and {Podobnik}, F. and {Prada Moroni}, P.~G. and {Prandini}, E. and {Principe}, G. and {Priyadarshi}, C. and {Rhode}, W. and {Rib{\'o}}, M. and {Rico}, J. and {Righi}, C. and {Sahakyan}, N. and {Saito}, T. and {Satalecka}, K. and {Saturni}, F.~G. and {Schleicher}, B. and {Schmidt}, K. and {Schmuckermaier}, F. and {Schubert}, J.~L. and {Schweizer}, T. and {Sciaccaluga}, A. and {Sitarek}, J. and {Spolon}, A. and {Stamerra}, A. and {Stri{\v{s}}kovi{\'c}}, J. and {Strom}, D. and {Suda}, Y. and {Suutarinen}, S. and {Tajima}, H. and {Takeishi}, R. and {Tavecchio}, F. and {Temnikov}, P. and {Terauchi}, K. and {Terzi{\'c}}, T. and {Teshima}, M. and {Tosti}, L. and {Truzzi}, S. and {Tutone}, A. and {Ubach}, S. and {van Scherpenberg}, J. and {Ventura}, S. and {Verguilov}, V. and {Viale}, I. and {Vigorito}, C.~F. and {Vitale}, V. and {Walter}, R. and {Wunderlich}, C. and {Yamamoto}, T. and {MWL Collaborators}},
        title = "{The variability patterns of the TeV blazar PG 1553 + 113 from a decade of MAGIC and multiband observations}",
      journal = {\mnras},
         year = 2024,
        month = apr,
       volume = {529},
       number = {4},
        pages = {3894-3911},
          doi = {10.1093/mnras/stae649},
archivePrefix = {arXiv},
       eprint = {2403.02159},
 primaryClass = {astro-ph.HE},
       adsurl = {https://ui.adsabs.harvard.edu/abs/2024MNRAS.529.3894M}
}

@ARTICLE{penil_mwl_pg1553,
       author = {{Pe{\~n}il}, P. and {Westernacher-Schneider}, J.~R. and {Ajello}, M. and {Dom{\'\i}nguez}, A. and {Buson}, S. and {Otero-Santos}, J. and {Marcotulli}, L. and {Torres-Alb{\`a}}, N. and {Zrake}, J.},
        title = "{Multiwavelength analysis of Fermi-LAT blazars with high-significance periodicity: detection of a long-term rising emission in PG 1553+113}",
      journal = {\mnras},
         year = 2024,
        month = feb,
       volume = {527},
       number = {4},
        pages = {10168-10184},
          doi = {10.1093/mnras/stad3246},
archivePrefix = {arXiv},
       eprint = {2310.12754},
 primaryClass = {astro-ph.HE},
       adsurl = {https://ui.adsabs.harvard.edu/abs/2024MNRAS.52710168P}
}

@ARTICLE{penil_flares_2025,
       author = {{Pe{\~n}il}, P. and {Torres-Alb{\`a}}, N. and {Rico}, A. and {Ajello}, M. and {Buson}, S. and {Adhikari}, S.},
        title = "{Distortions in periodicity analysis of blazars: the impact of flares}",
      journal = {\mnras},
         year = 2025,
        month = may,
       volume = {539},
       number = {2},
        pages = {993-1014},
          doi = {10.1093/mnras/staf482},
archivePrefix = {arXiv},
       eprint = {2504.05092},
 primaryClass = {astro-ph.HE},
       adsurl = {https://ui.adsabs.harvard.edu/abs/2025MNRAS.539..993P}
}

@ARTICLE{penil_24candidates_2025,
       author = {{Pe{\~n}il}, P. and {Ajello}, M. and {Buson}, S. and {Dom{\'\i}nguez}, A. and {Westernacher-Schneider}, J.~R. and {Rico}, A. and {Adhikari}, S. and {Zrake}, J.},
        title = "{Search for periodic variability in {\ensuremath{\gamma}}-ray blazars Using Fermi-LAT}",
      journal = {\mnras},
         year = 2025,
        month = aug,
       volume = {541},
       number = {4},
        pages = {2955-2977},
          doi = {10.1093/mnras/staf1108},
archivePrefix = {arXiv},
       eprint = {2211.01894},
 primaryClass = {astro-ph.HE},
       adsurl = {https://ui.adsabs.harvard.edu/abs/2025MNRAS.541.2955P}
}

@ARTICLE{penil_xray_pg1553_2026,
       author = {{Pe{\~n}il}, P. and {Torres-Alb{\`a}}, N. and {Marcotulli}, L. and {Dom{\'\i}nguez}, A. and {Ajello}, M. and {Rico}, A. and {Buson}, S. and {Adhikari}, S.},
        title = "{Testing X-ray Periodicity and Long-Term Trend in PG 1553+113 via Targeted Swift-XRT Monitoring}",
       journal = {\aap},
         year = 2026,
        month = jun,
       volume = {710},
          eid = {A268},
        pages = {A268},
          doi = {10.1051/0004-6361/202659801},
archivePrefix = {arXiv},
       eprint = {2604.05905},
 primaryClass = {astro-ph.HE},
       adsurl = {https://ui.adsabs.harvard.edu/abs/2026A&A...710A.268P}
}

@ARTICLE{penil_2025_linear,
       author = {{Penil}, P. and {Otero-Santos}, O. and {Circiello}, A. and {Banerjee}, A. and {Buson}, S. and {Rico}, A. and {Ajello}, M. and {Adhikari}, S.},
        title = "{Transient QPOs of Fermi-LAT blazars with Linearly Multiplicative Oscillations}",
      journal = {The Open Journal of Astrophysics},
         year = 2025,
        month = dec,
       volume = {8},
        pages = {54123},
          doi = {10.33232/001c.154123},
archivePrefix = {arXiv},
       eprint = {2507.13906},
 primaryClass = {astro-ph.HE},
       adsurl = {https://ui.adsabs.harvard.edu/abs/2025OJAp....854123P}
}

@ARTICLE{penil_17fermi_candidates_2026,
       author = {{Pe{\~n}il}, P. and {Rico}, A. and {Dom{\'\i}nguez}, A. and {Ajello}, M. and {Buson}, S. and {Adhikari}, S.},
        title = "{Gamma-Ray Periodicity in Jetted AGN: Revisiting Periodicity Candidates with >17 years of Fermi-LAT Data}",
      journal = {\mnras},
         year = 2026,
        month = aug,
       volume = {550},
       number = {3},
          eid = {stag1141},
        pages = {stag1141},
          doi = {10.1093/mnras/stag1141},
archivePrefix = {arXiv},
       eprint = {2606.15676},
 primaryClass = {astro-ph.HE},
       adsurl = {https://ui.adsabs.harvard.edu/abs/2026MNRAS.550g1141P}
}

@ARTICLE{fermi_repository,
       author = {{Abdollahi}, S. and {Ajello}, M. and {Baldini}, L. and {Ballet}, J. and {Bastieri}, D. and {Becerra Gonzalez}, J. and {Bellazzini}, R. and {Berretta}, A. and {Bissaldi}, E. and {Bonino}, R. and {Brill}, A. and {Bruel}, P. and {Burns}, E. and {Buson}, S. and {Cameron}, R.~A. and {Caputo}, R. and {Caraveo}, P.~A. and {Cibrario}, N. and {Ciprini}, S. and {Cristarella Orestano}, P. and {Crnogorcevic}, M. and {Cutini}, S. and {D'Ammando}, F. and {De Gaetano}, S. and {Digel}, S.~W. and {Di Lalla}, N. and {Di Venere}, L. and {Dom{\'\i}nguez}, A. and {Ramazani}, V. Fallah and {Fegan}, S.~J. and {Ferrara}, E.~C. and {Fiori}, A. and {Fleischhack}, H. and {Franckowiak}, A. and {Fukazawa}, Y. and {Fusco}, P. and {Gammaldi}, V. and {Gargano}, F. and {Garrappa}, S. and {Gasbarra}, C. and {Gasparrini}, D. and {Giglietto}, N. and {Giordano}, F. and {Giroletti}, M. and {Green}, D. and {Grenier}, I.~A. and {Guiriec}, S. and {Gustafsson}, M. and {Hays}, E. and {Horan}, D. and {Hou}, X. and {J{\'o}hannesson}, G. and {Kerr}, M. and {Kocevski}, D. and {Kuss}, M. and {Latronico}, L. and {Li}, J. and {Liodakis}, I. and {Longo}, F. and {Loparco}, F. and {Lorusso}, L. and {Lott}, B. and {Lovellette}, M.~N. and {Lubrano}, P. and {Maldera}, S. and {Manfreda}, A. and {Mart{\'\i}-Devesa}, G. and {Mazziotta}, M.~N. and {Mereu}, I. and {Meyer}, M. and {Michelson}, P.~F. and {Mizuno}, T. and {Monzani}, M.~E. and {Morselli}, A. and {Moskalenko}, I.~V. and {Negro}, M. and {Omodei}, N. and {Orlando}, E. and {Ormes}, J.~F. and {Paneque}, D. and {Panzarini}, G. and {Perkins}, J.~S. and {Persic}, M. and {Pesce-Rollins}, M. and {Pillera}, R. and {Porter}, T.~A. and {Principe}, G. and {Racusin}, J.~L. and {Rain{\`o}}, S. and {Rando}, R. and {Rani}, B. and {Razzano}, M. and {Razzaque}, S. and {Reimer}, A. and {Reimer}, O. and {S{\'a}nchez-Conde}, M. and {Parkinson}, P.~M. Saz and {Scargle}, Jeff and {Scotton}, L. and {Serini}, D. and {Sgr{\`o}}, C. and {Siskind}, E.~J. and {Spandre}, G. and {Spinelli}, P. and {Suson}, D.~J. and {Tajima}, H. and {Thompson}, D.~J. and {Torres}, D.~F. and {Valverde}, J. and {Venters}, T. and {Wadiasingh}, Z. and {Wagner}, S. and {Wood}, K.},
        title = "{The Fermi-LAT Lightcurve Repository}",
      journal = {\apjs},
         year = 2023,
        month = apr,
       volume = {265},
       number = {2},
          eid = {31},
        pages = {31},
          doi = {10.3847/1538-4365/acbb6a},
archivePrefix = {arXiv},
       eprint = {2301.01607},
 primaryClass = {astro-ph.HE},
       adsurl = {https://ui.adsabs.harvard.edu/abs/2023ApJS..265...31A}
}

@ARTICLE{abdo_exponential_flares_2010,
       author = {{Abdo}, A.~A. and {Ackermann}, M. and {Ajello}, M. and {Antolini}, E. and {Baldini}, L. and {Ballet}, J. and {Barbiellini}, G. and {Bastieri}, D. and {Bechtol}, K. and {Bellazzini}, R. and {Berenji}, B. and {Blandford}, R.~D. and {Bloom}, E.~D. and {Bonamente}, E. and {Borgland}, A.~W. and {Bouvier}, A. and {Bregeon}, J. and {Brez}, A. and {Brigida}, M. and {Bruel}, P. and {Buehler}, R. and {Burnett}, T.~H. and {Buson}, S. and {Caliandro}, G.~A. and {Cameron}, R.~A. and {Caraveo}, P.~A. and {Carrigan}, S. and {Casandjian}, J.~M. and {Cavazzuti}, E. and {Cecchi}, C. and {{\c{C}}elik}, {\"O}. and {Chekhtman}, A. and {Cheung}, C.~C. and {Chiang}, J. and {Ciprini}, S. and {Claus}, R. and {Cohen-Tanugi}, J. and {Cominsky}, L.~R. and {Conrad}, J. and {Costamante}, L. and {Cutini}, S. and {Dermer}, C.~D. and {de Angelis}, A. and {de Palma}, F. and {Silva}, E. do Couto e. and {Drell}, P.~S. and {Dubois}, R. and {Dumora}, D. and {Farnier}, C. and {Favuzzi}, C. and {Fegan}, S.~J. and {Focke}, W.~B. and {Fortin}, P. and {Frailis}, M. and {Fukazawa}, Y. and {Funk}, S. and {Fusco}, P. and {Gargano}, F. and {Gasparrini}, D. and {Gehrels}, N. and {Germani}, S. and {Giebels}, B. and {Giglietto}, N. and {Giommi}, P. and {Giordano}, F. and {Glanzman}, T. and {Godfrey}, G. and {Grenier}, I.~A. and {Grondin}, M.-H. and {Grove}, J.~E. and {Guiriec}, S. and {Hadasch}, D. and {Hayashida}, M. and {Hays}, E. and {Healey}, S.~E. and {Horan}, D. and {Hughes}, R.~E. and {Itoh}, R. and {J{\'o}hannesson}, G. and {Johnson}, A.~S. and {Johnson}, W.~N. and {Kamae}, T. and {Katagiri}, H. and {Kataoka}, J. and {Kawai}, N. and {Kn{\"o}dlseder}, J. and {Kuss}, M. and {Lande}, J. and {Larsson}, S. and {Latronico}, L. and {Lemoine-Goumard}, M. and {Longo}, F. and {Loparco}, F. and {Lott}, B. and {Lovellette}, M.~N. and {Lubrano}, P. and {Madejski}, G.~M. and {Makeev}, A. and {Massaro}, E. and {Mazziotta}, M.~N. and {McEnery}, J.~E. and {Michelson}, P.~F. and {Mitthumsiri}, W. and {Mizuno}, T. and {Moiseev}, A.~A. and {Monte}, C. and {Monzani}, M.~E. and {Morselli}, A. and {Moskalenko}, I.~V. and {Mueller}, M. and {Murgia}, S. and {Nolan}, P.~L. and {Norris}, J.~P. and {Nuss}, E. and {Ohno}, M. and {Ohsugi}, T. and {Omodei}, N. and {Orlando}, E. and {Ormes}, J.~F. and {Ozaki}, M. and {Panetta}, J.~H. and {Parent}, D. and {Pelassa}, V. and {Pepe}, M. and {Pesce-Rollins}, M. and {Piron}, F. and {Porter}, T.~A. and {Rain{\`o}}, S. and {Rando}, R. and {Razzano}, M. and {Reimer}, A. and {Reimer}, O. and {Ritz}, S. and {Rodriguez}, A.~Y. and {Romani}, R.~W. and {Roth}, M. and {Ryde}, F. and {Sadrozinski}, H.~F.-W. and {Sander}, A. and {Scargle}, J.~D. and {Sgr{\`o}}, C. and {Shaw}, M.~S. and {Smith}, P.~D. and {Spandre}, G. and {Spinelli}, P. and {Starck}, J.-L. and {Strickman}, M.~S. and {Suson}, D.~J. and {Takahashi}, H. and {Takahashi}, T. and {Tanaka}, T. and {Thayer}, J.~B. and {Thayer}, J.~G. and {Thompson}, D.~J. and {Tibaldo}, L. and {Torres}, D.~F. and {Tosti}, G. and {Tramacere}, A. and {Uchiyama}, Y. and {Usher}, T.~L. and {Vasileiou}, V. and {Vilchez}, N. and {Vitale}, V. and {Waite}, A.~P. and {Wallace}, E. and {Wang}, P. and {Winer}, B.~L. and {Wood}, K.~S. and {Yang}, Z. and {Ylinen}, T. and {Ziegler}, M.},
        title = "{Gamma-ray Light Curves and Variability of Bright Fermi-detected Blazars}",
      journal = {\apj},
         year = 2010,
        month = oct,
       volume = {722},
       number = {1},
        pages = {520-542},
          doi = {10.1088/0004-637X/722/1/520},
archivePrefix = {arXiv},
       eprint = {1004.0348},
 primaryClass = {astro-ph.HE},
       adsurl = {https://ui.adsabs.harvard.edu/abs/2010ApJ...722..520A}
}

@ARTICLE{penil_2020,
       author = {{Pe{\~n}il}, P. and {Dom{\'\i}nguez}, A. and {Buson}, S. and {Ajello}, M. and {Otero-Santos}, J. and {Barrio}, J.~A. and {Nemmen}, R. and {Cutini}, S. and {Rani}, B. and {Franckowiak}, A. and {Cavazzuti}, E.},
        title = "{Systematic Search for {\ensuremath{\gamma}}-Ray Periodicity in Active Galactic Nuclei Detected by the Fermi Large Area Telescope}",
      journal = {\apj},
         year = 2020,
        month = jun,
       volume = {896},
       number = {2},
          eid = {134},
        pages = {134},
          doi = {10.3847/1538-4357/ab910d},
archivePrefix = {arXiv},
       eprint = {2002.00805},
 primaryClass = {astro-ph.HE},
       adsurl = {https://ui.adsabs.harvard.edu/abs/2020ApJ...896..134P}
}

@ARTICLE{alba_ssa,
       author = {{Rico}, Alba and {Dom{\'\i}nguez}, A. and {Pe{\~n}il}, P. and {Ajello}, M. and {Buson}, S. and {Adhikari}, S. and {Movahedifar}, M.},
        title = "{Singular Spectrum Analysis of Fermi-LAT Blazar Light Curves: A Systematic Search for Periodicity and Trends in the Time Domain}",
      journal = {arXiv e-prints},
         year = 2024,
        month = dec,
          eid = {arXiv:2412.05812},
        pages = {arXiv:2412.05812},
archivePrefix = {arXiv},
       eprint = {2412.05812},
 primaryClass = {astro-ph.HE},
       adsurl = {https://ui.adsabs.harvard.edu/abs/2024arXiv241205812R}
}

@ARTICLE{sagar_pg1553,
       author = {{Adhikari}, S. and {Pe{\~n}il}, P. and {Westernacher-Schneider}, J.~R. and {Dom{\'\i}nguez}, A. and {Ajello}, M. and {Buson}, S. and {Rico}, A. and {Zrake}, J.},
        title = "{Constraining the PG 1553+113 Binary Hypothesis: Interpreting Hints of a New, 22 yr Period}",
      journal = {\apj},
         year = 2024,
        month = apr,
       volume = {965},
       number = {2},
          eid = {124},
        pages = {124},
          doi = {10.3847/1538-4357/ad310a},
archivePrefix = {arXiv},
       eprint = {2307.11696},
 primaryClass = {astro-ph.HE},
       adsurl = {https://ui.adsabs.harvard.edu/abs/2024ApJ...965..124A}
}

@ARTICLE{aniello_pg1553_xray_2024,
       author = {{Aniello}, T. and {Antonelli}, L.~A. and {Tombesi}, F. and {Lamastra}, A. and {Middei}, R. and {Perri}, M. and {Saturni}, F.~G. and {Stamerra}, A. and {Verrecchia}, F.},
        title = "{Unveiling the periodic variability patterns of the X-ray emission from the blazar PG 1553+113}",
      journal = {\aap},
         year = 2024,
        month = jun,
       volume = {686},
          eid = {A300},
        pages = {A300},
          doi = {10.1051/0004-6361/202449515},
archivePrefix = {arXiv},
       eprint = {2404.07089},
 primaryClass = {astro-ph.HE},
       adsurl = {https://ui.adsabs.harvard.edu/abs/2024A&A...686A.300A}
}

@ARTICLE{tavani_pdm_pg_1553,
       author = {{Tavani}, M. and {Cavaliere}, A. and {Munar-Adrover}, Pere and {Argan}, A.},
        title = "{The Blazar PG 1553+113 as a Binary System of Supermassive Black Holes}",
      journal = {\apj},
         year = 2018,
        month = feb,
       volume = {854},
       number = {1},
          eid = {11},
        pages = {11},
          doi = {10.3847/1538-4357/aaa3f4},
archivePrefix = {arXiv},
       eprint = {1801.03335},
 primaryClass = {astro-ph.HE},
       adsurl = {https://ui.adsabs.harvard.edu/abs/2018ApJ...854...11T}
}

@ARTICLE{gao_pg1553_2023,
       author = {{Gao}, Quan-Gui and {Lu}, Fang-Wu and {Qin}, Long-hua and {Gong}, Yun-Lu and {Yu}, Gong-ming and {Li}, Huai-zhen and {Yi}, Ting-feng},
        title = "{A Geometric Model to Interpret the {\ensuremath{\gamma}}-Ray Quasiperiodic Oscillation of PG 1553+113}",
      journal = {\apj},
         year = 2023,
        month = mar,
       volume = {945},
       number = {2},
          eid = {146},
        pages = {146},
          doi = {10.3847/1538-4357/acbe3e},
       adsurl = {https://ui.adsabs.harvard.edu/abs/2023ApJ...945..146G}
}

@ARTICLE{camenzind_jet,
       author = {{Camenzind}, M. and {Krockenberger}, M.},
        title = "{The lighthouse effect of relativistic jets in blazars. A geometric originof intraday variability.}",
      journal = {\aap},
         year = 1992,
        month = feb,
       volume = {255},
        pages = {59-62},
       adsurl = {https://ui.adsabs.harvard.edu/abs/1992A&A...255...59C}
}

@ARTICLE{WS2022+,
       author = {{Westernacher-Schneider}, John Ryan and {Zrake}, Jonathan and {MacFadyen}, Andrew and {Haiman}, Zolt{\'a}n},
        title = "{Multiband light curves from eccentric accreting supermassive black hole binaries}",
      journal = {\prd},
         year = 2022,
        month = nov,
       volume = {106},
       number = {10},
          eid = {103010},
        pages = {103010},
          doi = {10.1103/PhysRevD.106.103010},
archivePrefix = {arXiv},
       eprint = {2111.06882},
 primaryClass = {astro-ph.HE},
       adsurl = {https://ui.adsabs.harvard.edu/abs/2022PhRvD.106j3010W}
}

@ARTICLE{farris_2014,
       author = {{Farris}, Brian D. and {Duffell}, Paul and {MacFadyen}, Andrew I. and {Haiman}, Zoltan},
        title = "{Binary Black Hole Accretion from a Circumbinary Disk: Gas Dynamics inside the Central Cavity}",
      journal = {\apj},
         year = 2014,
        month = mar,
       volume = {783},
       number = {2},
          eid = {134},
        pages = {134},
          doi = {10.1088/0004-637X/783/2/134},
archivePrefix = {arXiv},
       eprint = {1310.0492},
 primaryClass = {astro-ph.HE},
       adsurl = {https://ui.adsabs.harvard.edu/abs/2014ApJ...783..134F}
}

@ARTICLE{stefan_pg1553_2024,
       author = {{Abdollahi}, S. and {Baldini}, L. and {Barbiellini}, G. and {Bellazzini}, R. and {Berenji}, B. and {Bissaldi}, E. and {Blandford}, R.~D. and {Bonino}, R. and {Bruel}, P. and {Buson}, S. and {Cameron}, R.~A. and {Caraveo}, P.~A. and {Casaburo}, F. and {Cavazzuti}, E. and {Cheung}, C.~C. and {Chiaro}, G. and {Ciprini}, S. and {Cozzolongo}, G. and {Cristarella Orestano}, P. and {Cutini}, S. and {D'Ammando}, F. and {Di Lalla}, N. and {Dirirsa}, F. and {Di Venere}, L. and {Dom{\'\i}nguez}, A. and {Fegan}, S.~J. and {Ferrara}, E.~C. and {Fiori}, A. and {Fukazawa}, Y. and {Funk}, S. and {Fusco}, P. and {Gargano}, F. and {Garrappa}, S. and {Gasparrini}, D. and {Germani}, S. and {Giglietto}, N. and {Giordano}, F. and {Giroletti}, M. and {Green}, D. and {Grenier}, I.~A. and {Guiriec}, S. and {Hays}, E. and {Horan}, D. and {Kuss}, M. and {Larsson}, S. and {Laurenti}, M. and {Li}, J. and {Liodakis}, I. and {Longo}, F. and {Loparco}, F. and {Lott}, B. and {Lovellette}, M.~N. and {Lubrano}, P. and {Maldera}, S. and {Malyshev}, D. and {Manfreda}, A. and {Marcotulli}, L. and {Mart{\'\i}-Devesa}, G. and {Mazziotta}, M.~N. and {Mereu}, I. and {Michelson}, P.~F. and {Mitthumsiri}, W. and {Mizuno}, T. and {Monzani}, M.~E. and {Morselli}, A. and {Moskalenko}, I.~V. and {Negro}, M. and {Omodei}, N. and {Orienti}, M. and {Orlando}, E. and {Ormes}, J.~F. and {Paneque}, D. and {Perri}, M. and {Persic}, M. and {Pesce-Rollins}, M. and {Porter}, T.~A. and {Principe}, G. and {Rain{\`o}}, S. and {Rando}, R. and {Rani}, B. and {Razzano}, M. and {Reimer}, A. and {Reimer}, O. and {Saz Parkinson}, P.~M. and {Scotton}, L. and {Serini}, D. and {Sesana}, A. and {Sgr{\`o}}, C. and {Siskind}, E.~J. and {Spandre}, G. and {Spinelli}, P. and {Suson}, D.~J. and {Tajima}, H. and {Takahashi}, M.~N. and {Tak}, D. and {Thayer}, J.~B. and {Thompson}, D.~J. and {Torres}, D.~F. and {Valverde}, J. and {Verrecchia}, F. and {Zaharijas}, G.},
        title = "{Periodic Gamma-Ray Modulation of the Blazar PG 1553+113 Confirmed by Fermi-LAT and Multiwavelength Observations}",
      journal = {\apj},
         year = 2024,
        month = dec,
       volume = {976},
       number = {2},
          eid = {203},
        pages = {203},
          doi = {10.3847/1538-4357/ad64c5},
archivePrefix = {arXiv},
       eprint = {2501.08015},
 primaryClass = {astro-ph.HE},
       adsurl = {https://ui.adsabs.harvard.edu/abs/2024ApJ...976..203A}
}

@ARTICLE{lomb_1976,
       author = {{Lomb}, N.~R.},
        title = "{Least-Squares Frequency Analysis of Unequally Spaced Data}",
      journal = {\apss},
         year = 1976,
        month = feb,
       volume = {39},
       number = {2},
        pages = {447-462},
          doi = {10.1007/BF00648343},
       adsurl = {https://ui.adsabs.harvard.edu/abs/1976Ap&SS..39..447L}
}

@ARTICLE{scargle_1982,
       author = {{Scargle}, J.~D.},
        title = "{Studies in astronomical time series analysis. II. Statistical aspects of spectral analysis of unevenly spaced data.}",
      journal = {\apj},
         year = 1982,
        month = dec,
       volume = {263},
        pages = {835-853},
          doi = {10.1086/160554},
       adsurl = {https://ui.adsabs.harvard.edu/abs/1982ApJ...263..835S}
}

@article{hair_r2_2011,
author = {Hair, J.F. and Ringle, C.M. and Sarstedt, M.},
title = {PLS-SEM: Indeed a Silver Bullet},
journal = {Journal of Marketing Theory and Practice},
volume = {19},
number = {2},
pages = {139-152},
year = {2011},
publisher = {Routledge},
doi = {10.2753/MTP1069-6679190202},
URL = {https://doi.org/10.2753/MTP1069-6679190202},
}

@ARTICLE{drake2009,
       author = {{Drake}, A.~J. and {Djorgovski}, S.~G. and {Mahabal}, A. and {Beshore}, E. and {Larson}, S. and {Graham}, M.~J. and {Williams}, R. and {Christensen}, E. and {Catelan}, M. and {Boattini}, A. and {Gibbs}, A. and {Hill}, R. and {Kowalski}, R.},
        title = "{First Results from the Catalina Real-Time Transient Survey}",
      journal = {\apj},
         year = 2009,
        month = may,
       volume = {696},
       number = {1},
        pages = {870-884},
          doi = {10.1088/0004-637X/696/1/870},
archivePrefix = {arXiv},
       eprint = {0809.1394},
 primaryClass = {astro-ph},
       adsurl = {https://ui.adsabs.harvard.edu/abs/2009ApJ...696..870D}
}

@ARTICLE{shappee2014,
       author = {{Shappee}, B.~J. and {Prieto}, J.~L. and {Grupe}, D. and {Kochanek}, C.~S. and {Stanek}, K.~Z. and {De Rosa}, G. and {Mathur}, S. and {Zu}, Y. and {Peterson}, B.~M. and {Pogge}, R.~W. and {Komossa}, S. and {Im}, M. and {Jencson}, J. and {Holoien}, T.~W. -S. and {Basu}, U. and {Beacom}, J.~F. and {Szczygie{\l}}, D.~M. and {Brimacombe}, J. and {Adams}, S. and {Campillay}, A. and {Choi}, C. and {Contreras}, C. and {Dietrich}, M. and {Dubberley}, M. and {Elphick}, M. and {Foale}, S. and {Giustini}, M. and {Gonzalez}, C. and {Hawkins}, E. and {Howell}, D.~A. and {Hsiao}, E.~Y. and {Koss}, M. and {Leighly}, K.~M. and {Morrell}, N. and {Mudd}, D. and {Mullins}, D. and {Nugent}, J.~M. and {Parrent}, J. and {Phillips}, M.~M. and {Pojmanski}, G. and {Rosing}, W. and {Ross}, R. and {Sand}, D. and {Terndrup}, D.~M. and {Valenti}, S. and {Walker}, Z. and {Yoon}, Y.},
        title = "{The Man behind the Curtain: X-Rays Drive the UV through NIR Variability in the 2013 Active Galactic Nucleus Outburst in NGC 2617}",
      journal = {\apj},
         year = 2014,
        month = jun,
       volume = {788},
       number = {1},
          eid = {48},
        pages = {48},
          doi = {10.1088/0004-637X/788/1/48},
archivePrefix = {arXiv},
       eprint = {1310.2241},
 primaryClass = {astro-ph.HE},
       adsurl = {https://ui.adsabs.harvard.edu/abs/2014ApJ...788...48S}
}

@ARTICLE{kochanek2017,
       author = {{Kochanek}, C.~S. and {Shappee}, B.~J. and {Stanek}, K.~Z. and {Holoien}, T.~W. -S. and {Thompson}, Todd A. and {Prieto}, J.~L. and {Dong}, Subo and {Shields}, J.~V. and {Will}, D. and {Britt}, C. and {Perzanowski}, D. and {Pojma{\'n}ski}, G.},
        title = "{The All-Sky Automated Survey for Supernovae (ASAS-SN) Light Curve Server v1.0}",
      journal = {\pasp},
         year = 2017,
        month = oct,
       volume = {129},
       number = {980},
        pages = {104502},
          doi = {10.1088/1538-3873/aa80d9},
archivePrefix = {arXiv},
       eprint = {1706.07060},
 primaryClass = {astro-ph.SR},
       adsurl = {https://ui.adsabs.harvard.edu/abs/2017PASP..129j4502K}
}

@ARTICLE{zwicky_observatory,
       author = {{Masci}, Frank J. and {Laher}, Russ R. and {Rusholme}, Ben and {Shupe}, David L. and {Groom}, Steven and {Surace}, Jason and {Jackson}, Edward and {Monkewitz}, Serge and {Beck}, Ron and {Flynn}, David and {Terek}, Scott and {Landry}, Walter and {Hacopians}, Eugean and {Desai}, Vandana and {Howell}, Justin and {Brooke}, Tim and {Imel}, David and {Wachter}, Stefanie and {Ye}, Quan-Zhi and {Lin}, Hsing-Wen and {Cenko}, S. Bradley and {Cunningham}, Virginia and {Rebbapragada}, Umaa and {Bue}, Brian and {Miller}, Adam A. and {Mahabal}, Ashish and {Bellm}, Eric C. and {Patterson}, Maria T. and {Juri{\'c}}, Mario and {Golkhou}, V. Zach and {Ofek}, Eran O. and {Walters}, Richard and {Graham}, Matthew and {Kasliwal}, Mansi M. and {Dekany}, Richard G. and {Kupfer}, Thomas and {Burdge}, Kevin and {Cannella}, Christopher B. and {Barlow}, Tom and {Van Sistine}, Angela and {Giomi}, Matteo and {Fremling}, Christoffer and {Blagorodnova}, Nadejda and {Levitan}, David and {Riddle}, Reed and {Smith}, Roger M. and {Helou}, George and {Prince}, Thomas A. and {Kulkarni}, Shrinivas R.},
        title = "{The Zwicky Transient Facility: Data Processing, Products, and Archive}",
      journal = {\pasp},
         year = 2019,
        month = jan,
       volume = {131},
       number = {995},
        pages = {018003},
          doi = {10.1088/1538-3873/aae8ac},
archivePrefix = {arXiv},
       eprint = {1902.01872},
 primaryClass = {astro-ph.IM},
       adsurl = {https://ui.adsabs.harvard.edu/abs/2019PASP..131a8003M}
}

@ARTICLE{madero_dominguez_2026,
        author = {{Madero}, Elena and {Dom{\'\i}nguez}, Alberto},
        title = "{Coexistence of chromatic flares and an achromatic quasi-periodic oscillation in the gamma-ray blazar PG 1553+113}",
      journal = {\aap},
         year = 2026,
        month = mar,
       volume = {707},
          eid = {L18},
        pages = {L18},
          doi = {10.1051/0004-6361/202659239},
archivePrefix = {arXiv},
       eprint = {2603.04527},
 primaryClass = {astro-ph.HE},
       adsurl = {https://ui.adsabs.harvard.edu/abs/2026A&A...707L..18M}
}

@ARTICLE{rieger_2004,
       author = {{Rieger}, Frank M.},
        title = "{On the Geometrical Origin of Periodicity in Blazar-type Sources}",
      journal = {\apjl},
         year = 2004,
        month = nov,
       volume = {615},
       number = {1},
        pages = {L5-L8},
          doi = {10.1086/426018},
archivePrefix = {arXiv},
       eprint = {astro-ph/0410188},
 primaryClass = {astro-ph},
       adsurl = {https://ui.adsabs.harvard.edu/abs/2004ApJ...615L...5R}
}

@ARTICLE{roy_multiple_flares_2019,
       author = {{Roy}, Namrata and {Chatterjee}, Ritaban and {Joshi}, Manasvita and {Ghosh}, Aritra},
        title = "{Probing the jets of blazars using the temporal symmetry of their multiwavelength outbursts}",
      journal = {\mnras},
         year = 2019,
        month = jan,
       volume = {482},
       number = {1},
        pages = {743-757},
          doi = {10.1093/mnras/sty2748},
archivePrefix = {arXiv},
       eprint = {1810.06585},
 primaryClass = {astro-ph.GA},
       adsurl = {https://ui.adsabs.harvard.edu/abs/2019MNRAS.482..743R}
}

@ARTICLE{finke_shocks_2024,
AUTHOR={Finke, Justin D. },           
TITLE={Light travel time effects in blazar flares},          
JOURNAL={Frontiers in Astronomy and Space Sciences},          
VOLUME={Volume 11 - 2024},  
YEAR={2024},  
URL={https://www.frontiersin.org/journals/astronomy-and-space-sciences/articles/10.3389/fspas.2024.1384234},  
DOI={10.3389/fspas.2024.1384234},
ISSN={2296-987X}
}

@ARTICLE{li_multiple_flares_2018,
       author = {{Li}, Yutong and {Hu}, Shaoming and {Wiita}, Paul J. and {Gupta}, Alok C.},
        title = "{Statistical analysis of variability properties of the Kepler blazar W2R 1926+42}",
      journal = {\mnras},
         year = 2018,
        month = jul,
       volume = {478},
       number = {1},
        pages = {172-182},
          doi = {10.1093/mnras/sty1082},
archivePrefix = {arXiv},
       eprint = {1804.10245},
 primaryClass = {astro-ph.GA},
       adsurl = {https://ui.adsabs.harvard.edu/abs/2018MNRAS.478..172L}
}

@article{sarkar_double_lorentz_2021,
    author = {Sarkar, Arkadipta and Gupta, Alok C and Chitnis, Varsha R and Wiita, Paul J},
    title = {Multiwaveband quasi-periodic oscillation in the blazar 3C 454.3},
    journal = {\mnras},
    volume = {501},
    number = {1},
    pages = {50-61},
    year = {2021},
    month = {02},
    issn = {0035-8711},
    doi = {10.1093/mnras/staa3211},
    url = {https://doi.org/10.1093/mnras/staa3211},
    eprint = {https://academic.oup.com/mnras/article-pdf/501/1/50/34934361/staa3211.pdf},
}

@article{nalewajko_flares_shapes_2013,
    author = {Nalewajko, Krzysztof},
    title = {The brightest gamma-ray flares of blazars},
    journal = {\mnras},
    volume = {430},
    number = {2},
    pages = {1324-1333},
    year = {2013},
    month = {04},
    issn = {0035-8711},
    doi = {10.1093/mnras/sts711},
    url = {https://doi.org/10.1093/mnras/sts711},
    eprint = {https://academic.oup.com/mnras/article-pdf/430/2/1324/9383405/sts711.pdf},
}

@article{bhatta_flares_2023,
    author = {Bhatta, Gopal and Zola, Staszek and Drozdz, M and Reichart, Daniel and Haislip, Joshua and Kouprianov, Vladimir and Matsumoto, Katsura and Sonbas, Eda and Caton, D and Pajdosz-Śmierciak, Urszula and Simon, A and Provencal, J and Góra, Dariusz and Stachowski, Grzegorz},
    title = {Catching profound optical flares in blazars},
    journal = {\mnras},
    volume = {520},
    number = {2},
    pages = {2633-2643},
    year = {2023},
    month = {04},
    issn = {0035-8711},
    doi = {10.1093/mnras/stad280},
    url = {https://doi.org/10.1093/mnras/stad280},
    eprint = {https://academic.oup.com/mnras/article-pdf/520/2/2633/51031991/stad280.pdf},
}

@article{blinov_exponential_flares_2018,
    author = {Blinov, D. and Pavlidou, V. and Papadakis, I. and Kiehlmann, S. and Liodakis, I. and Panopoulou, G. V. and Angelakis, E. and Baloković, M. and Hovatta, T. and King, O. G. and Kus, A. and Kylafis, N. and Mahabal, A. and Maharana, S. and Myserlis, I. and Paleologou, E. and Papamastorakis, I. and Pazderski, E. and Pearson, T. J. and Ramaprakash, A. and Readhead, A. C. S. and Reig, P. and Tassis, K. and Zensus, J. A.},
    title = {RoboPol: connection between optical polarization plane rotations and gamma-ray flares in blazars},
    journal = {Monthly Notices of the Royal Astronomical Society},
    volume = {474},
    number = {1},
    pages = {1296-1306},
    year = {2018},
    month = {02},
    issn = {0035-8711},
    doi = {10.1093/mnras/stx2786},
    url = {https://doi.org/10.1093/mnras/stx2786},
    eprint = {https://academic.oup.com/mnras/article-pdf/474/1/1296/22367606/stx2786.pdf},
}

@ARTICLE{fenimore_fred_1996,
       author = {{Fenimore}, Edward E. and {Madras}, Claudine D. and {Nayakshin}, Sergei},
        title = "{Expanding Relativistic Shells and Gamma-Ray Burst Temporal Structure}",
      journal = {\apj},
         year = 1996,
        month = dec,
       volume = {473},
        pages = {998},
          doi = {10.1086/178210},
archivePrefix = {arXiv},
       eprint = {astro-ph/9607163},
 primaryClass = {astro-ph},
       adsurl = {https://ui.adsabs.harvard.edu/abs/1996ApJ...473..998F}
}

@article{yan_fred_2018,
    author = {Yan, Zhen and Xie, Fu-Guo},
    title = {A decades-long fast-rise-exponential-decay flare in low-luminosity AGN NGC 7213},
    journal = {Monthly Notices of the Royal Astronomical Society},
    volume = {475},
    number = {1},
    pages = {1190-1197},
    year = {2018},
    month = {03},
    issn = {0035-8711},
    doi = {10.1093/mnras/stx3259},
    url = {https://doi.org/10.1093/mnras/stx3259},
    eprint = {https://academic.oup.com/mnras/article-pdf/475/1/1190/23564997/stx3259.pdf},
}

@ARTICLE{valtaoja_multi_exponential_1999,
       author = {{Valtaoja}, E. and {L{\"a}hteenm{\"a}ki}, A. and {Ter{\"a}sranta}, H. and {Lainela}, M.},
        title = "{Total Flux Density Variations in Extragalactic Radio Sources. I. Decomposition of Variations into Exponential Flares}",
      journal = {\apjs},
         year = 1999,
        month = jan,
       volume = {120},
       number = {1},
        pages = {95-99},
          doi = {10.1086/313170},
       adsurl = {https://ui.adsabs.harvard.edu/abs/1999ApJS..120...95V}
}

@ARTICLE{vlasyuk_multi_exponential_2024,
       author = {{Vlasyuk}, V.~V. and {Sotnikova}, Y.~V. and {Volvach}, A.~E. and {Mufakharov}, T.~V. and {Kovalev}, Y.~A. and {Spiridonova}, O.~I. and {Khabibullina}, M.~L. and {Kovalev}, Y.~Y. and {Mikhailov}, A.~G. and {Stolyarov}, V.~A. and {Kudryavtsev}, D.~O. and {Mingaliev}, M.~G. and {Razzaque}, S. and {Semenova}, T.~A. and {Kudryashova}, A.~K. and {Bursov}, N.~N. and {Trushkin}, S.~A. and {Popkov}, A.~V. and {Erkenov}, A.~K. and {Rakhimov}, I.~A. and {Kharinov}, M.~A. and {Gurwell}, M.~A. and {Tsybulev}, P.~G. and {Moskvitin}, A.~S. and {Fatkhullin}, T.~A. and {Emelianov}, E.~V. and {Arshinova}, A. and {Iuzhanina}, K.~V. and {Andreeva}, T.~S. and {Volvach}, L.~N. and {Ghosh}, A.},
        title = "{Multiwavelength variability of the blazar AO 0235+164}",
      journal = {\mnras},
         year = 2024,
        month = dec,
       volume = {535},
       number = {3},
        pages = {2775-2799},
          doi = {10.1093/mnras/stae2491},
archivePrefix = {arXiv},
       eprint = {2411.01497},
 primaryClass = {astro-ph.HE},
       adsurl = {https://ui.adsabs.harvard.edu/abs/2024MNRAS.535.2775V}
}

@ARTICLE{hovatta_multi_exponential_2009,
       author = {{Hovatta}, T. and {Valtaoja}, E. and {Tornikoski}, M. and {L{\"a}hteenm{\"a}ki}, A.},
        title = "{Doppler factors, Lorentz factors and viewing angles for quasars, BL Lacertae objects and radio galaxies}",
      journal = {\aap},
         year = 2009,
        month = feb,
       volume = {494},
       number = {2},
        pages = {527-537},
          doi = {10.1051/0004-6361:200811150},
archivePrefix = {arXiv},
       eprint = {0811.4278},
 primaryClass = {astro-ph},
       adsurl = {https://ui.adsabs.harvard.edu/abs/2009A&A...494..527H}
}

@ARTICLE{veronesi_gaussian_expo_2025,
       author = {{Veronesi}, Niccol{\`o} and {van Velzen}, Sjoert and {Rossi}, Elena Maria},
        title = "{AGN flares as counterparts to LIGO/Virgo mergers: no confident causal connection in spatial correlation analysis}",
      journal = {\mnras},
         year = 2025,
        month = jan,
       volume = {536},
       number = {3},
        pages = {3112-3122},
          doi = {10.1093/mnras/stae2787},
archivePrefix = {arXiv},
       eprint = {2405.05318},
 primaryClass = {astro-ph.HE},
       adsurl = {https://ui.adsabs.harvard.edu/abs/2025MNRAS.536.3112V}
}

@ARTICLE{graham_gaussian_expo_2020,
       author = {{Graham}, M.~J. and {Ford}, K.~E.~S. and {McKernan}, B. and {Ross}, N.~P. and {Stern}, D. and {Burdge}, K. and {Coughlin}, M. and {Djorgovski}, S.~G. and {Drake}, A.~J. and {Duev}, D. and {Kasliwal}, M. and {Mahabal}, A.~A. and {van Velzen}, S. and {Belecki}, J. and {Bellm}, E.~C. and {Burruss}, R. and {Cenko}, S.~B. and {Cunningham}, V. and {Helou}, G. and {Kulkarni}, S.~R. and {Masci}, F.~J. and {Prince}, T. and {Reiley}, D. and {Rodriguez}, H. and {Rusholme}, B. and {Smith}, R.~M. and {Soumagnac}, M.~T.},
        title = "{Candidate Electromagnetic Counterpart to the Binary Black Hole Merger Gravitational-Wave Event S190521g$^{*}$}",
      journal = {\prl},
         year = 2020,
        month = jun,
       volume = {124},
       number = {25},
          eid = {251102},
        pages = {251102},
          doi = {10.1103/PhysRevLett.124.251102},
archivePrefix = {arXiv},
       eprint = {2006.14122},
 primaryClass = {astro-ph.HE},
       adsurl = {https://ui.adsabs.harvard.edu/abs/2020PhRvL.124y1102G}
}

@ARTICLE{bic_schwarz,
       author = {{Schwarz}, Gideon},
        title = "{Estimating the Dimension of a Model}",
      journal = {Annals of Statistics},
         year = 1978,
        month = jul,
       volume = {6},
       number = {2},
        pages = {461-464},
       adsurl = {https://ui.adsabs.harvard.edu/abs/1978AnSta...6..461S}
}

@ARTICLE{moraitis_lorentzian_2011,
       author = {{Moraitis}, K. and {Mastichiadis}, A.},
        title = "{X-ray variability patterns in blazars}",
      journal = {\aap},
         year = 2011,
        month = jan,
       volume = {525},
          eid = {A40},
        pages = {A40},
          doi = {10.1051/0004-6361/201015871},
archivePrefix = {arXiv},
       eprint = {1010.4511},
 primaryClass = {astro-ph.HE},
       adsurl = {https://ui.adsabs.harvard.edu/abs/2011A&A...525A..40M}
}

@ARTICLE{saito_flares_2013,
       author = {{Saito}, S. and {Stawarz}, {\L}. and {Tanaka}, Y.~T. and {Takahashi}, T. and {Madejski}, G. and {D'Ammando}, F.},
        title = "{Very Rapid High-amplitude Gamma-Ray Variability in Luminous Blazar PKS 1510-089 Studied with Fermi-LAT}",
      journal = {\apjl},
         year = 2013,
        month = mar,
       volume = {766},
       number = {1},
          eid = {L11},
        pages = {L11},
          doi = {10.1088/2041-8205/766/1/L11},
archivePrefix = {arXiv},
       eprint = {1302.0335},
 primaryClass = {astro-ph.HE},
       adsurl = {https://ui.adsabs.harvard.edu/abs/2013ApJ...766L..11S}
}

@ARTICLE{marcher_turbulent_2014,
       author = {{Marscher}, Alan P.},
        title = "{Turbulent, Extreme Multi-zone Model for Simulating Flux and Polarization Variability in Blazars}",
      journal = {\apj},
         year = 2014,
        month = jan,
       volume = {780},
       number = {1},
          eid = {87},
        pages = {87},
          doi = {10.1088/0004-637X/780/1/87},
archivePrefix = {arXiv},
       eprint = {1311.7665},
 primaryClass = {astro-ph.HE},
       adsurl = {https://ui.adsabs.harvard.edu/abs/2014ApJ...780...87M}
}

@ARTICLE{marscher_gear_1985,
       author = {{Marscher}, A.~P. and {Gear}, W.~K.},
        title = "{Models for high-frequency radio outbursts in extragalactic sources, with application to the early 1983 millimeter-to-infrared flare of 3C 273.}",
      journal = {\apj},
         year = 1985,
        month = nov,
       volume = {298},
        pages = {114-127},
          doi = {10.1086/163592},
       adsurl = {https://ui.adsabs.harvard.edu/abs/1985ApJ...298..114M}
}

@ARTICLE{saple_2026,
       author = {{Marcotulli}, Lea and {Torres-Alb{\`a}}, N{\'u}ria},
        title = "{SAPLE: Swift Analysis Pipeline for Lightcurve Extraction}",
      journal = {arXiv e-prints},
         year = 2026,
        month = may,
          eid = {arXiv:2605.10423},
        pages = {arXiv:2605.10423},
          doi = {10.48550/arXiv.2605.10423},
archivePrefix = {arXiv},
       eprint = {2605.10423},
 primaryClass = {astro-ph.IM},
       adsurl = {https://ui.adsabs.harvard.edu/abs/2026arXiv260510423M}
}

@article{Kass_Raftery_1995,
  author  = {Kass, Robert E. and Raftery, Adrian E.},
  title   = {Bayes Factors},
  journal = {Journal of the American Statistical Association},
  year    = {1995},
  volume  = {90},
  number  = {430},
  pages   = {773--795},
  doi     = {10.1080/01621459.1995.10476572}
}

\clearpage
\appendix
\noindent
\setcounter{table}{0}
\setcounter{figure}{0}
\renewcommand{\thetable}{A\arabic{table}}
\renewcommand{\thefigure}{A\arabic{figure}}

\subsection{Figures}\label{sec:appendix}
This appendix presents additional figures supporting the profile-model analysis described in the main text. The figures are organized into three groups: examples of the empirical profile templates, and the best-fit profiles for each wavelength band.

\subsubsection{Profile-model templates}

\begin{figure*}
        \centering
        \includegraphics[scale=0.55]{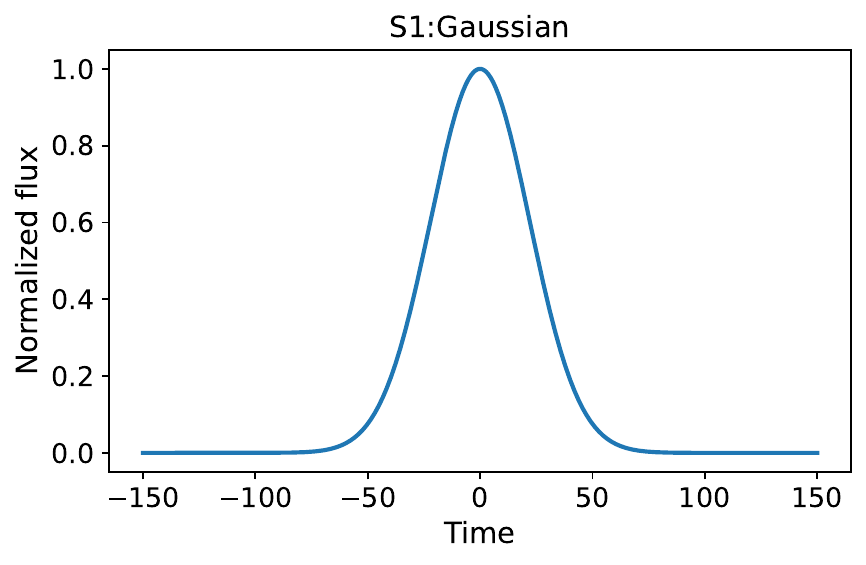}
        \includegraphics[scale=0.55]{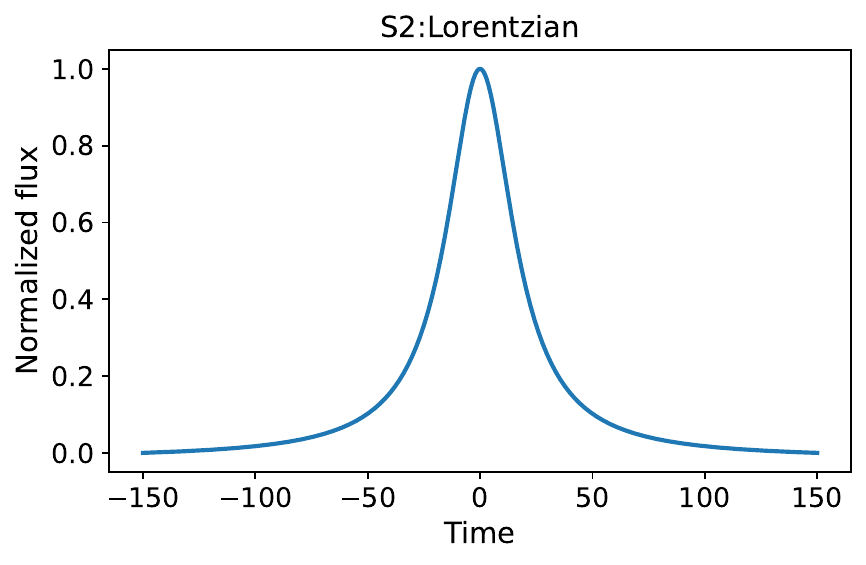}
        \includegraphics[scale=0.55]{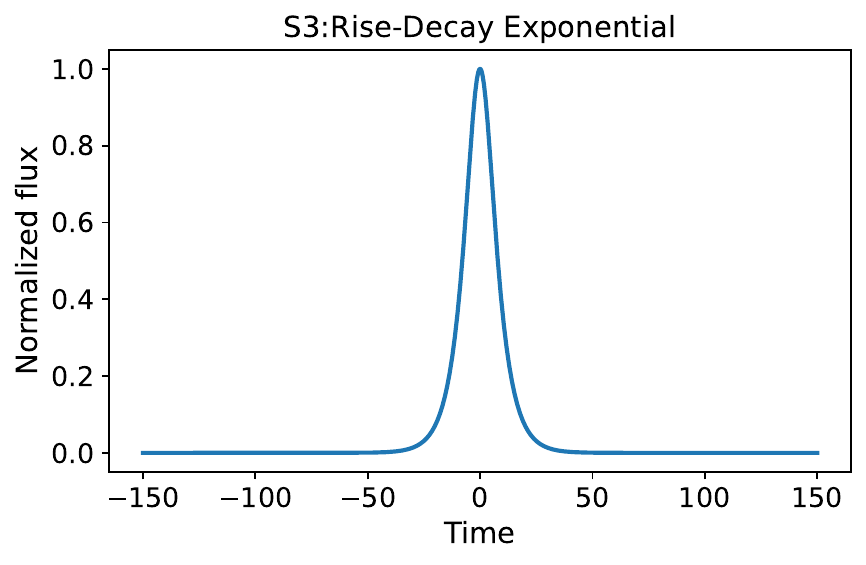}
        \includegraphics[scale=0.55]{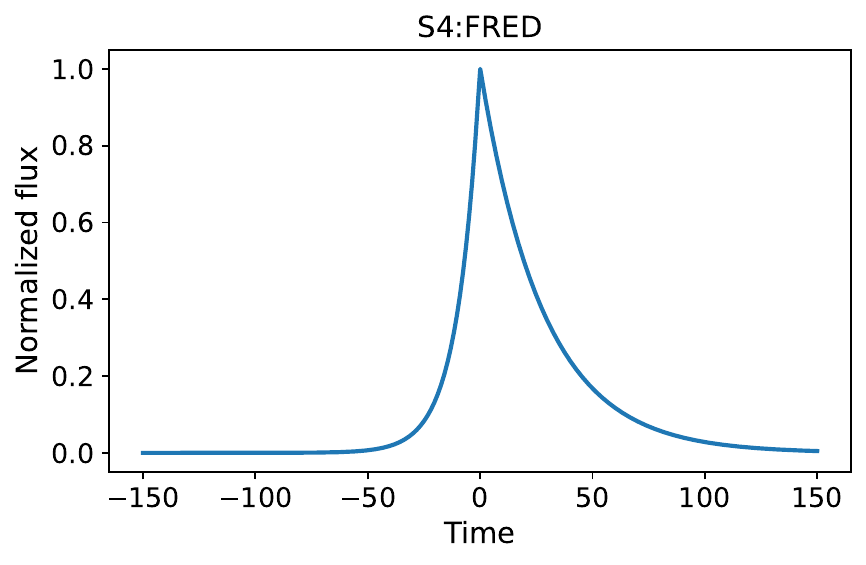}
        \includegraphics[scale=0.55]{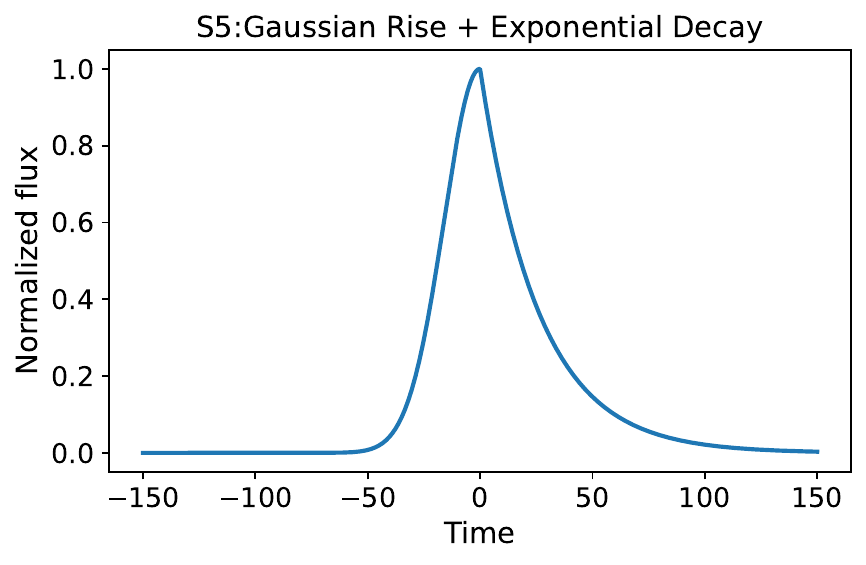}
        \caption{Examples of the empirical profiles used to model single-peaked oscillations. The panels show the single Gaussian profile (S1), single Lorentzian profile (S2), exponential rise--decay profile (S3), FRED profile (S4), and Gaussian-rise plus exponential-decay profile (S5).} \label{fig:example_profiles_single_peak}
\end{figure*}

\begin{figure*}
        \centering
        \includegraphics[scale=0.55]{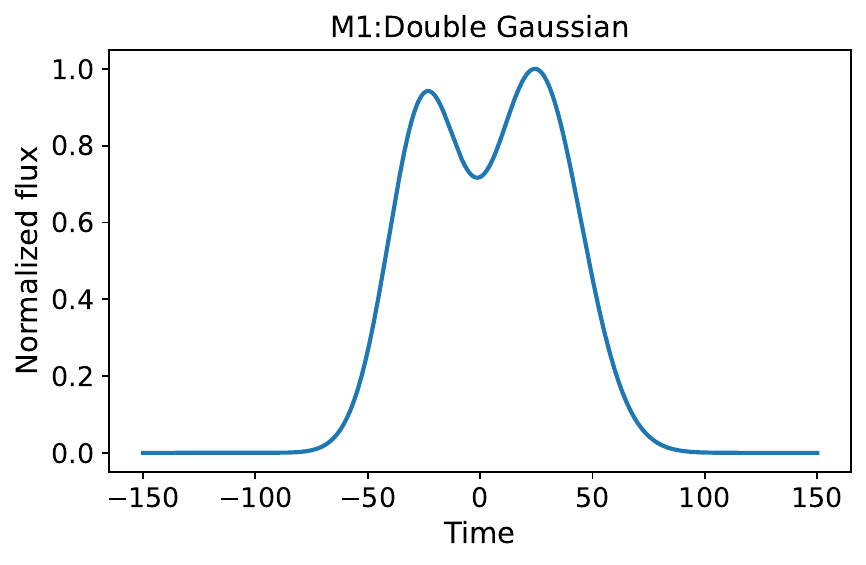}
        \includegraphics[scale=0.55]{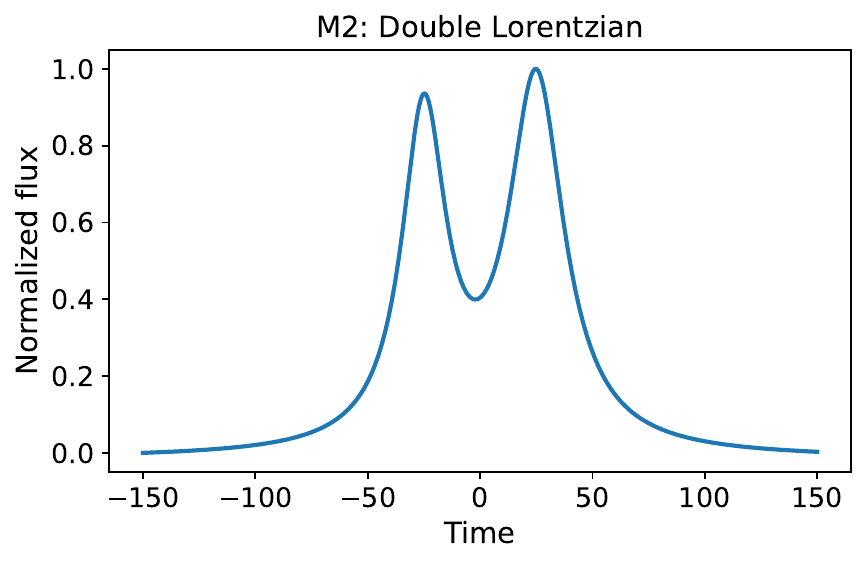}
        \includegraphics[scale=0.55]{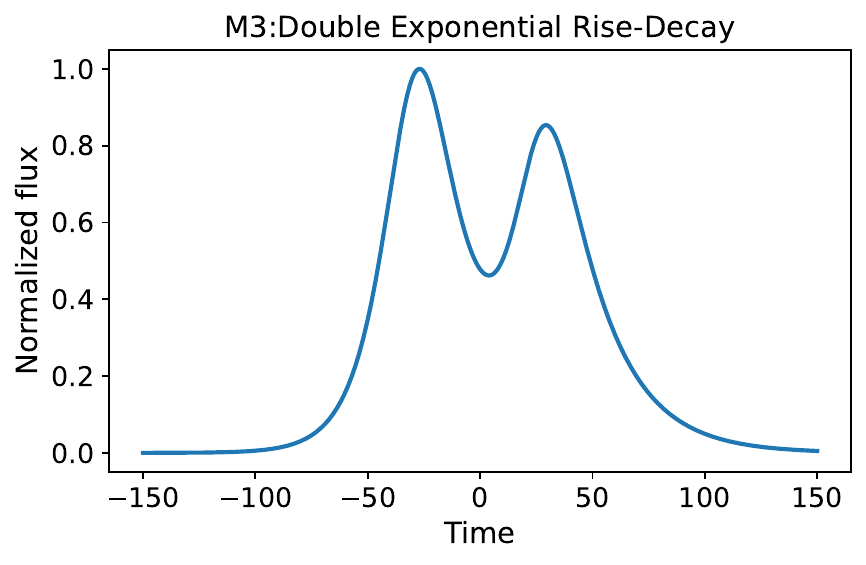}
        \includegraphics[scale=0.55]{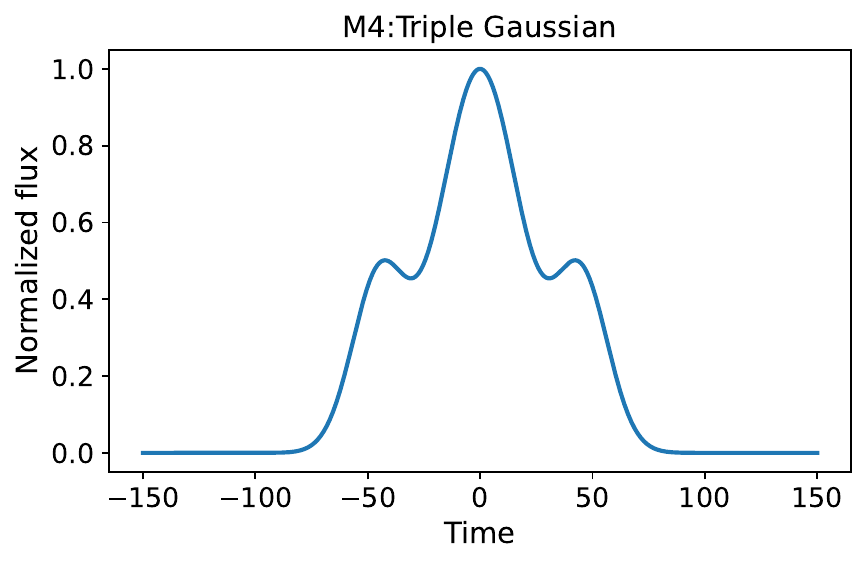}
        \includegraphics[scale=0.55]{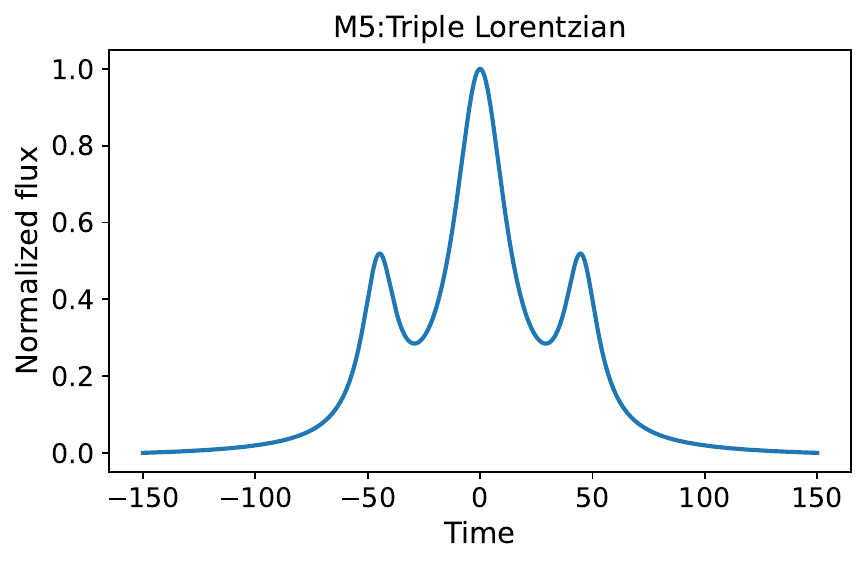}
        \includegraphics[scale=0.55]{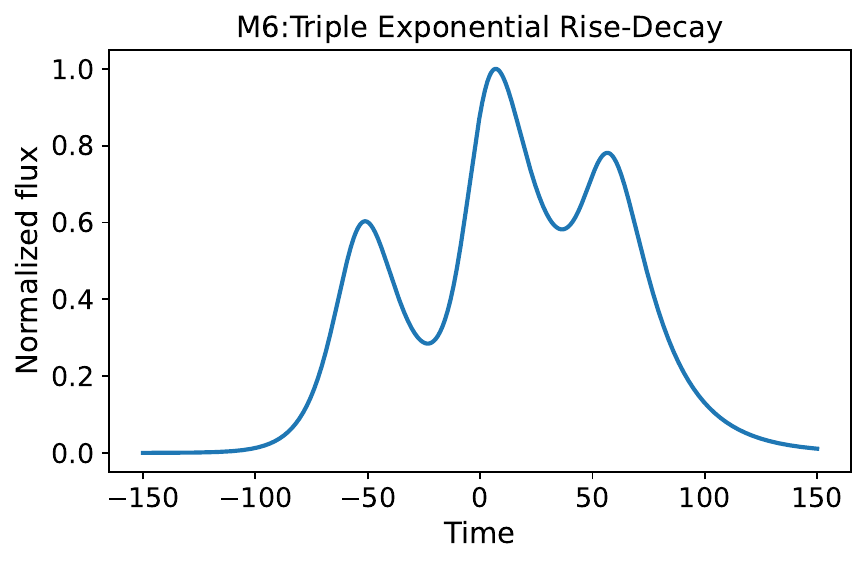}
        \includegraphics[scale=0.55]{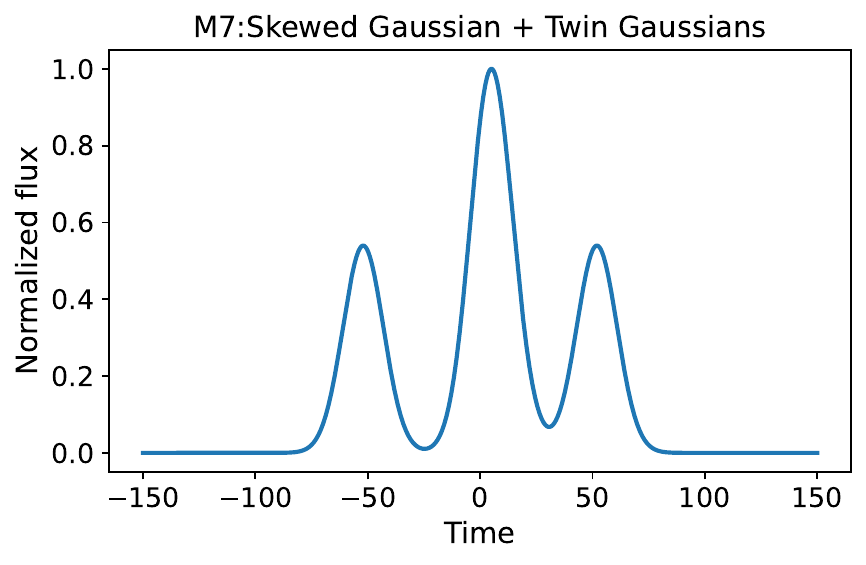}        
        \includegraphics[scale=0.55]{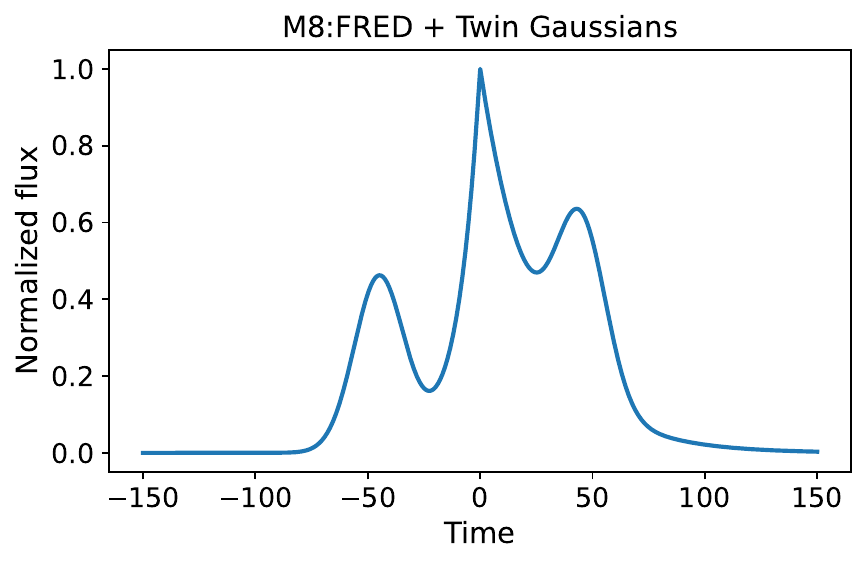}
        \caption{Examples of the empirical profiles used to model multiple-peaked oscillations. The panels show the double-Gaussian profile (M1), double-Lorentzian profile (M2), double exponential rise--decay profile (M3), triple-Gaussian profile (M4), triple-Lorentzian profile (M5), triple exponential rise--decay profile (M6), skewed-Gaussian plus twin-Gaussian profile (M7), and FRED plus twin-Gaussian profile (M8).} \label{fig:example_profiles_multi_peak}
\end{figure*}

This subsection shows representative examples of the empirical functions used in the profile analysis. The templates are divided into single-peaked (Figure \ref{fig:example_profiles_single_peak}) and multiple-peaked profiles (Figure \ref{fig:example_profiles_multi_peak}), following the model classification introduced in the main text.

\clearpage

\subsubsection{Best-fit profiles by wavelength band}

\begin{figure}[H]
        \centering
        \includegraphics[scale=0.28]{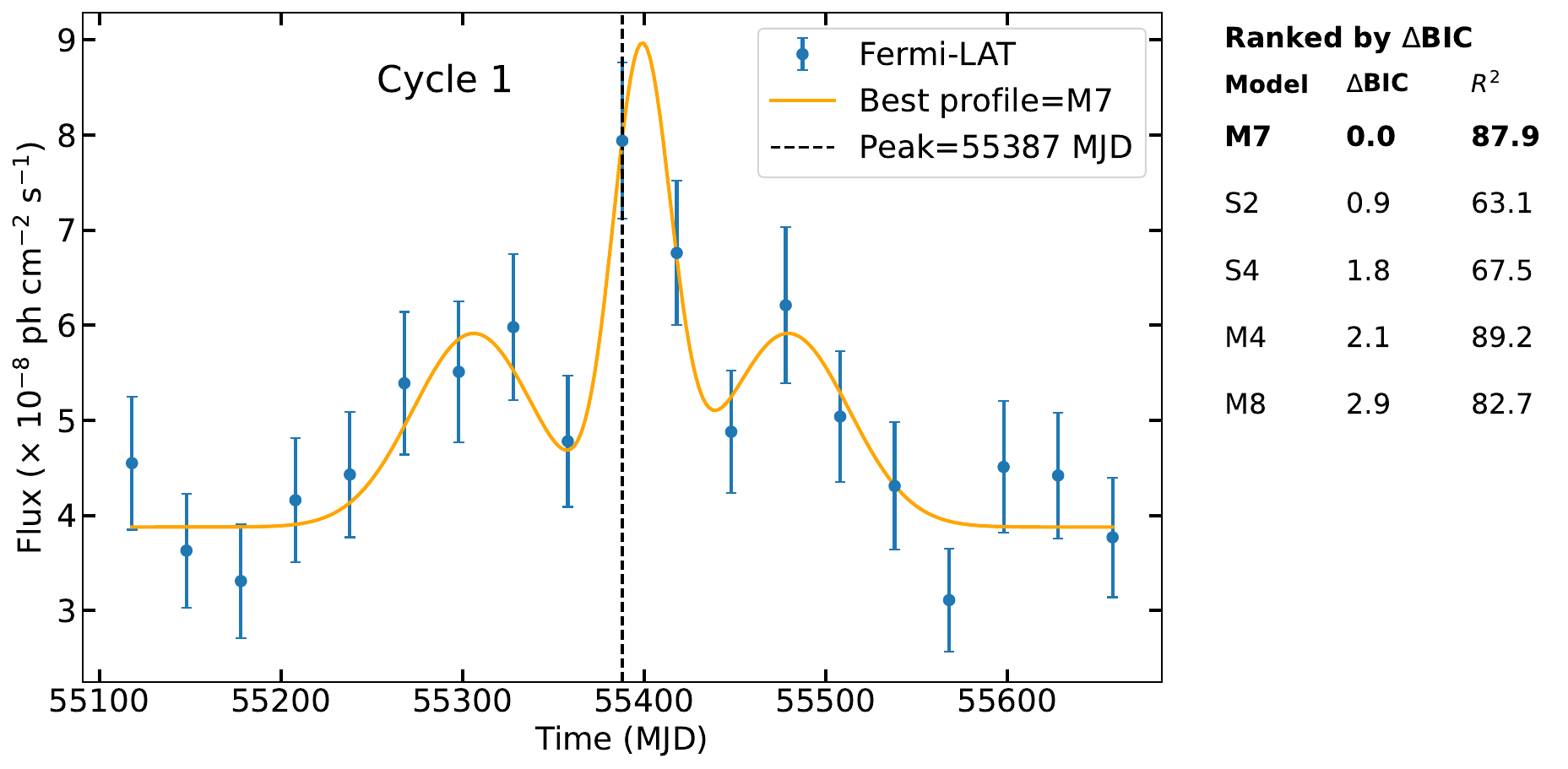}
        \includegraphics[scale=0.28]{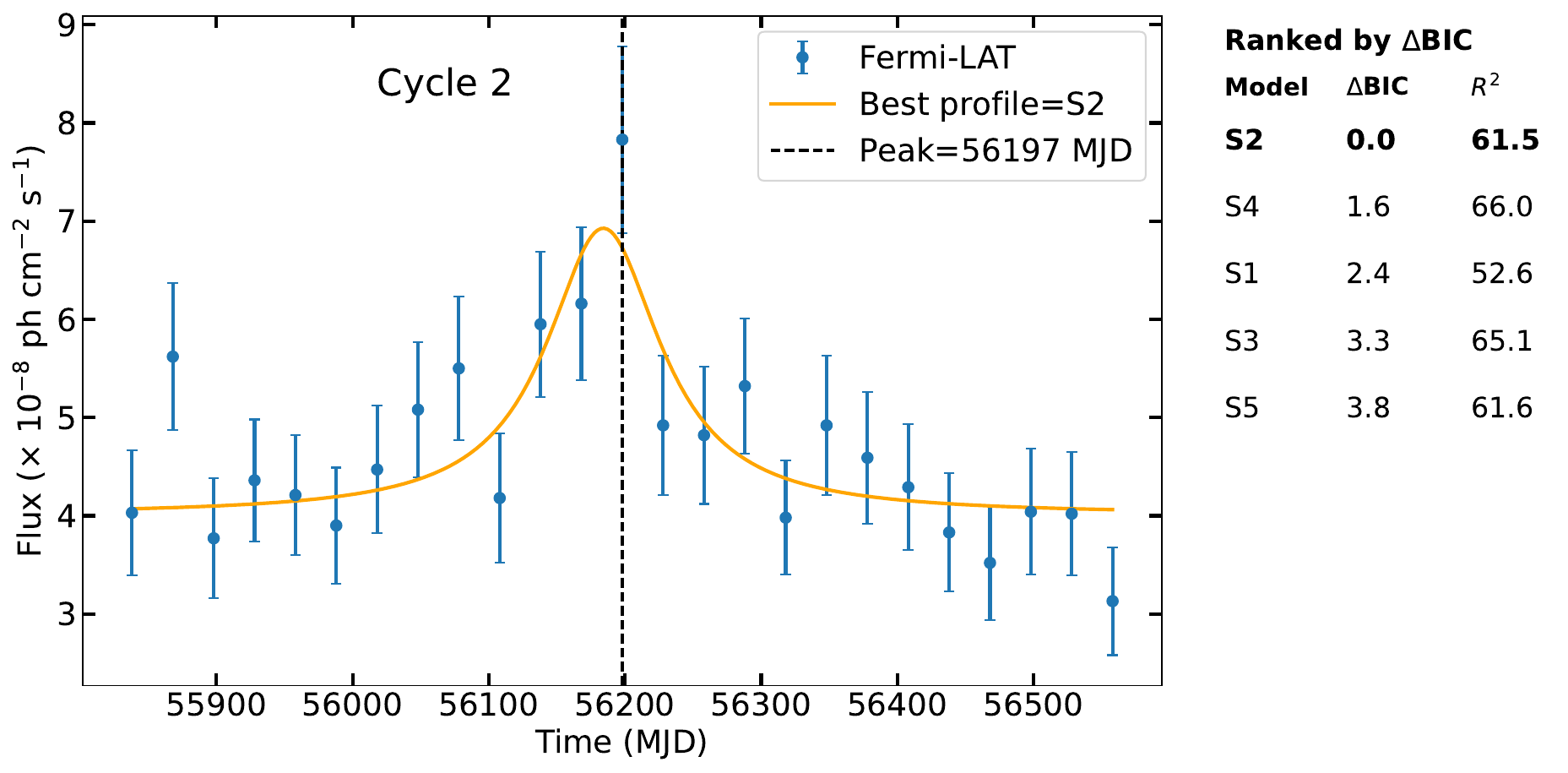}
        \includegraphics[scale=0.28]{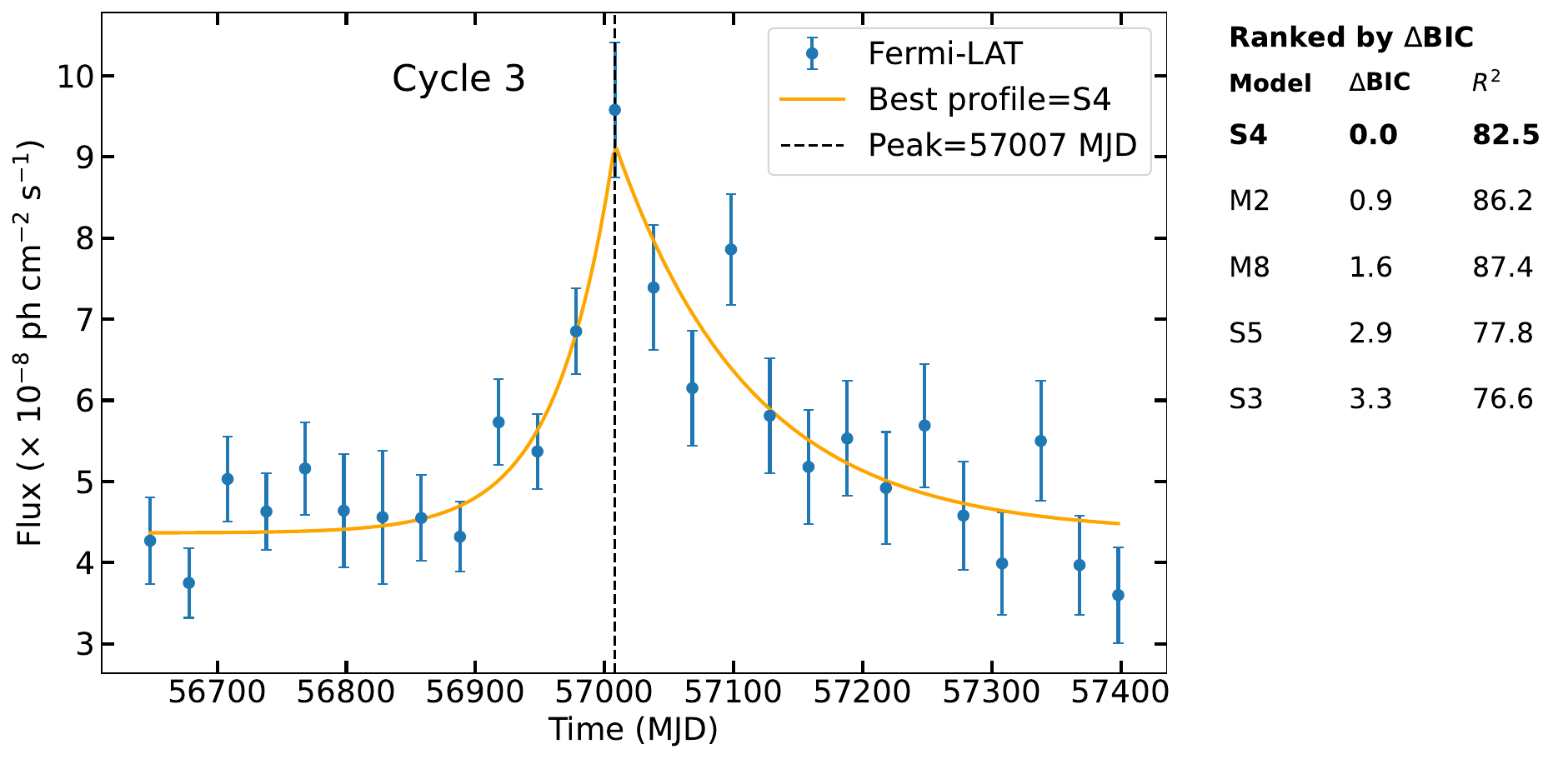}
        \includegraphics[scale=0.28]{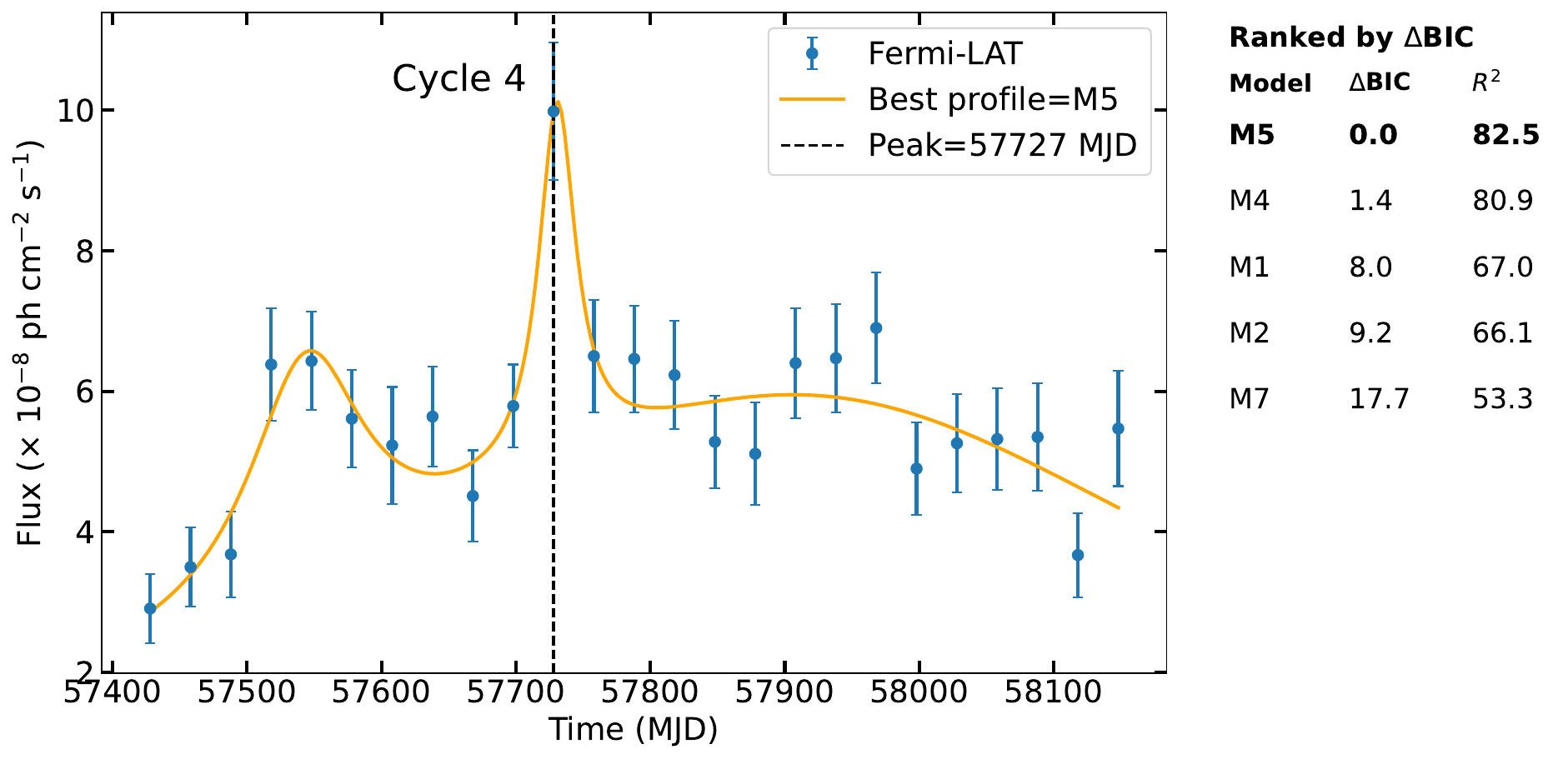}
        \includegraphics[scale=0.28]{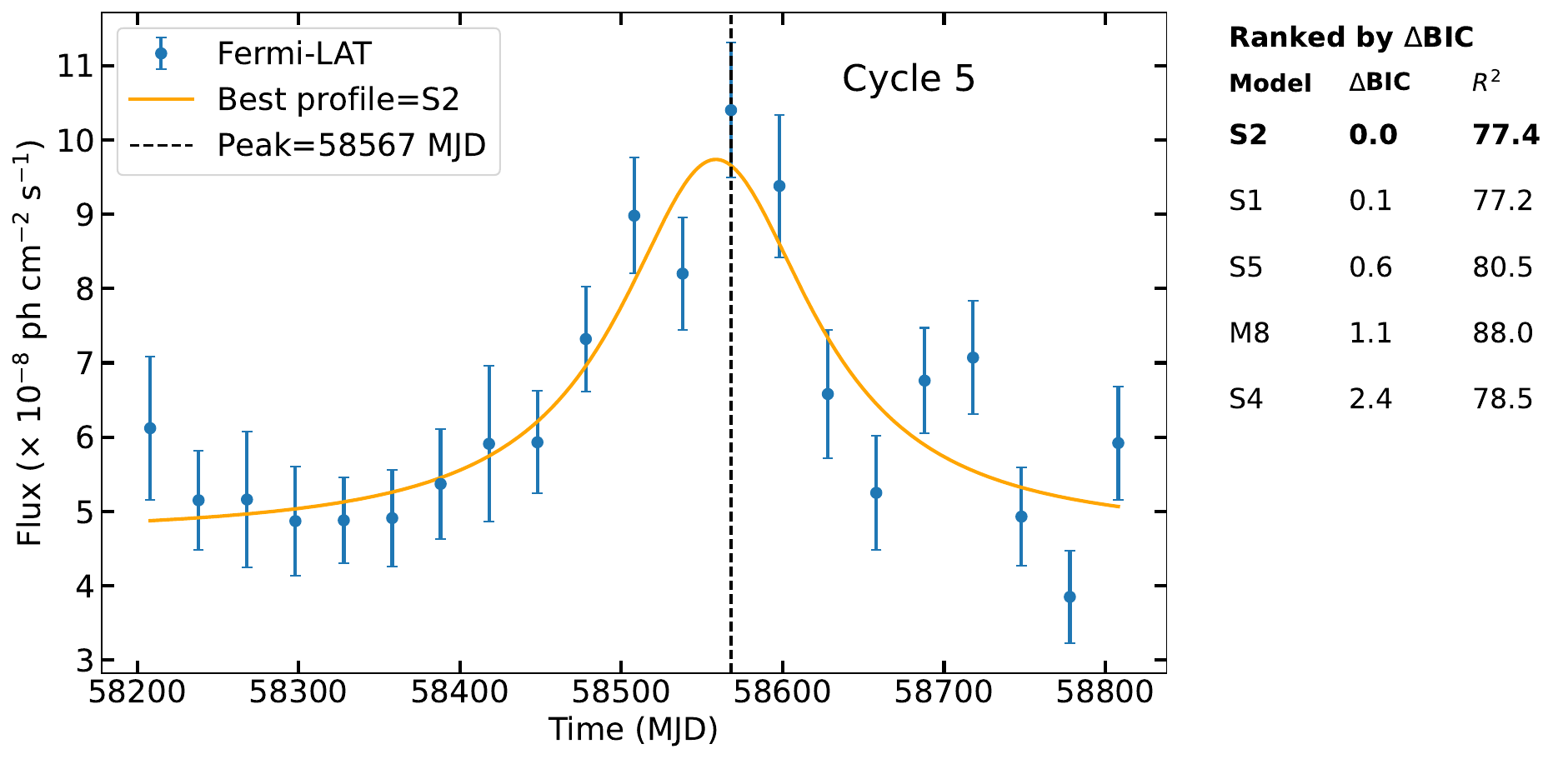}
        \includegraphics[scale=0.28]{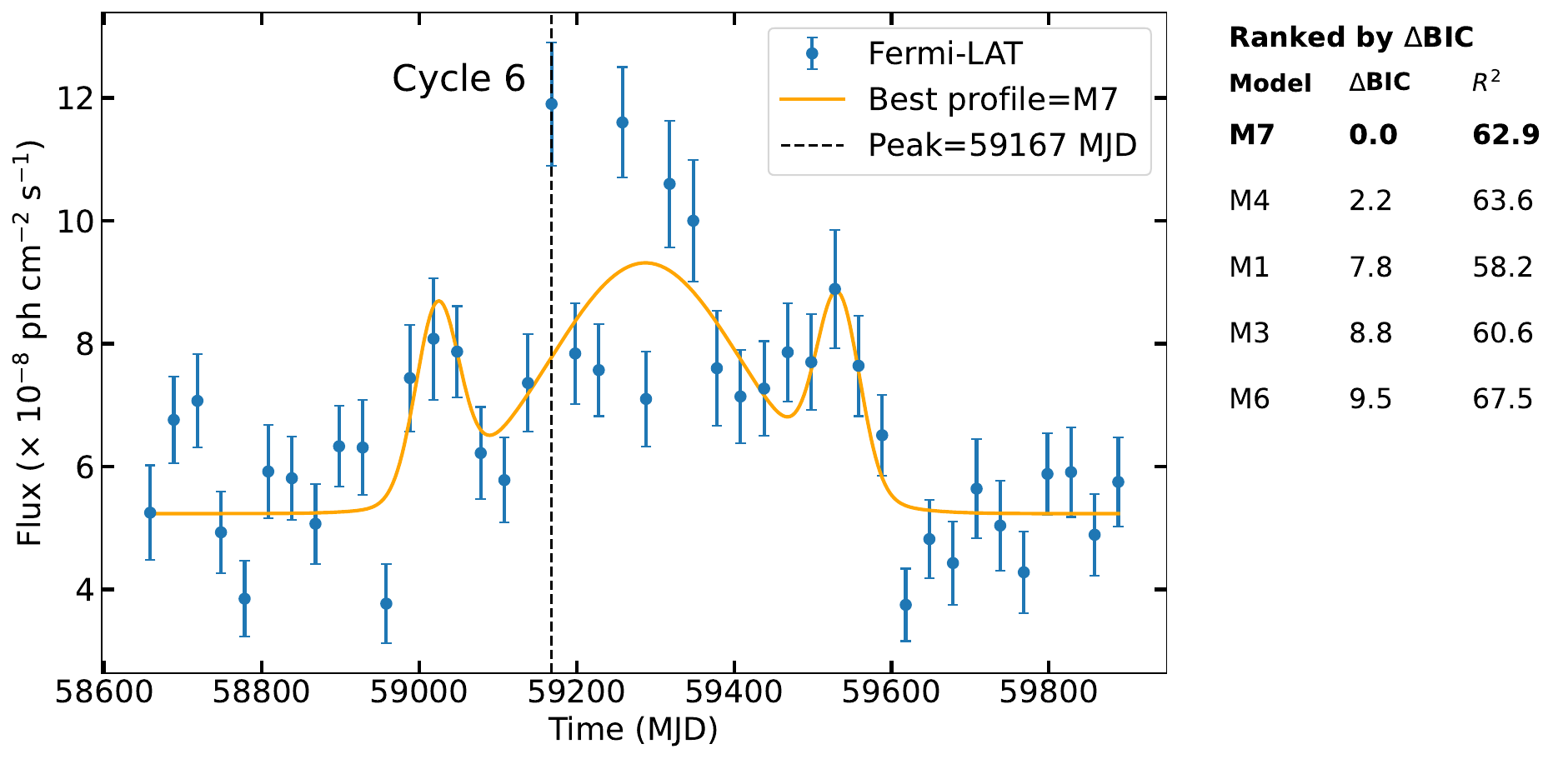}
        \includegraphics[scale=0.28]{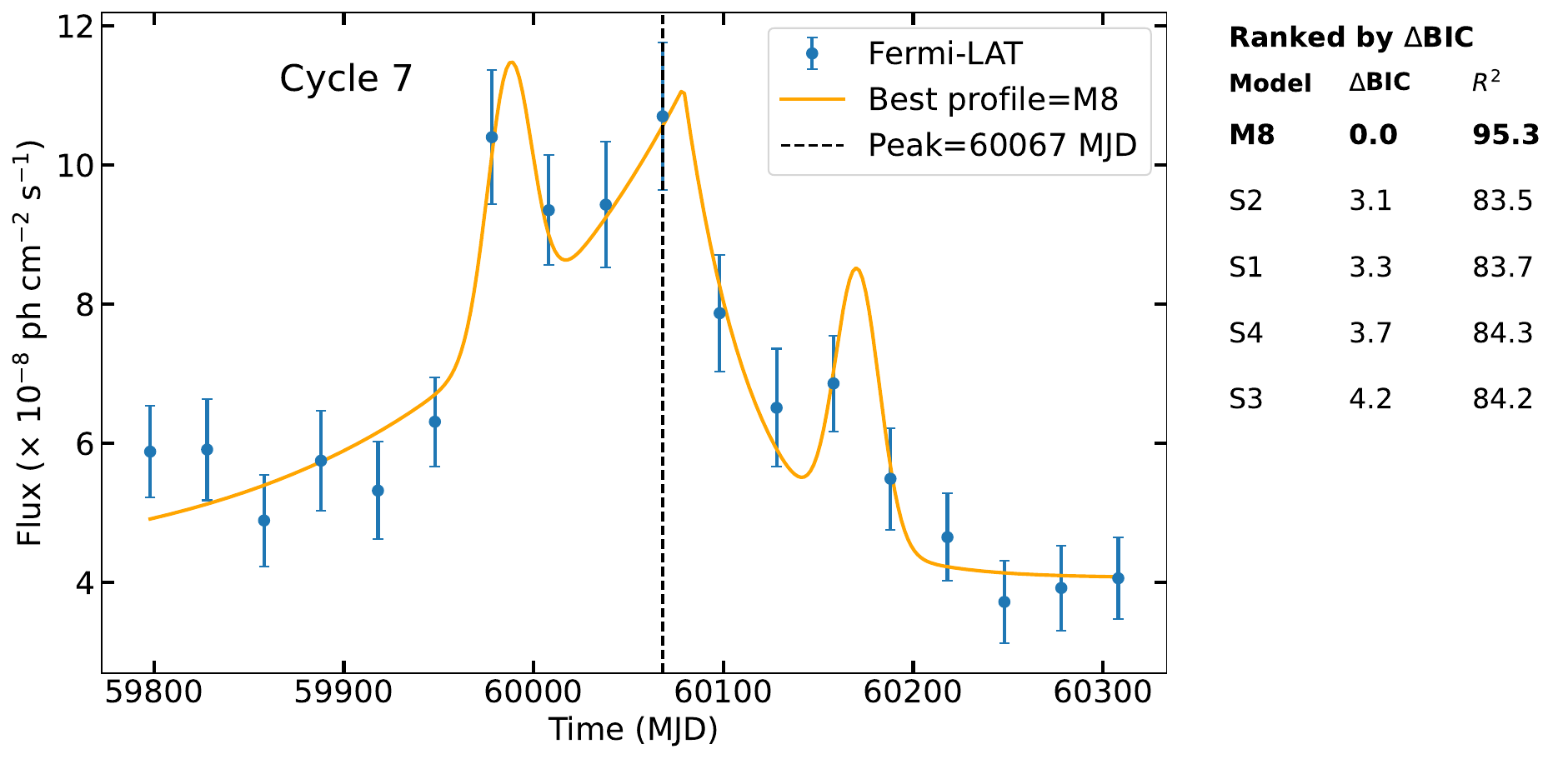}
        \includegraphics[scale=0.28]{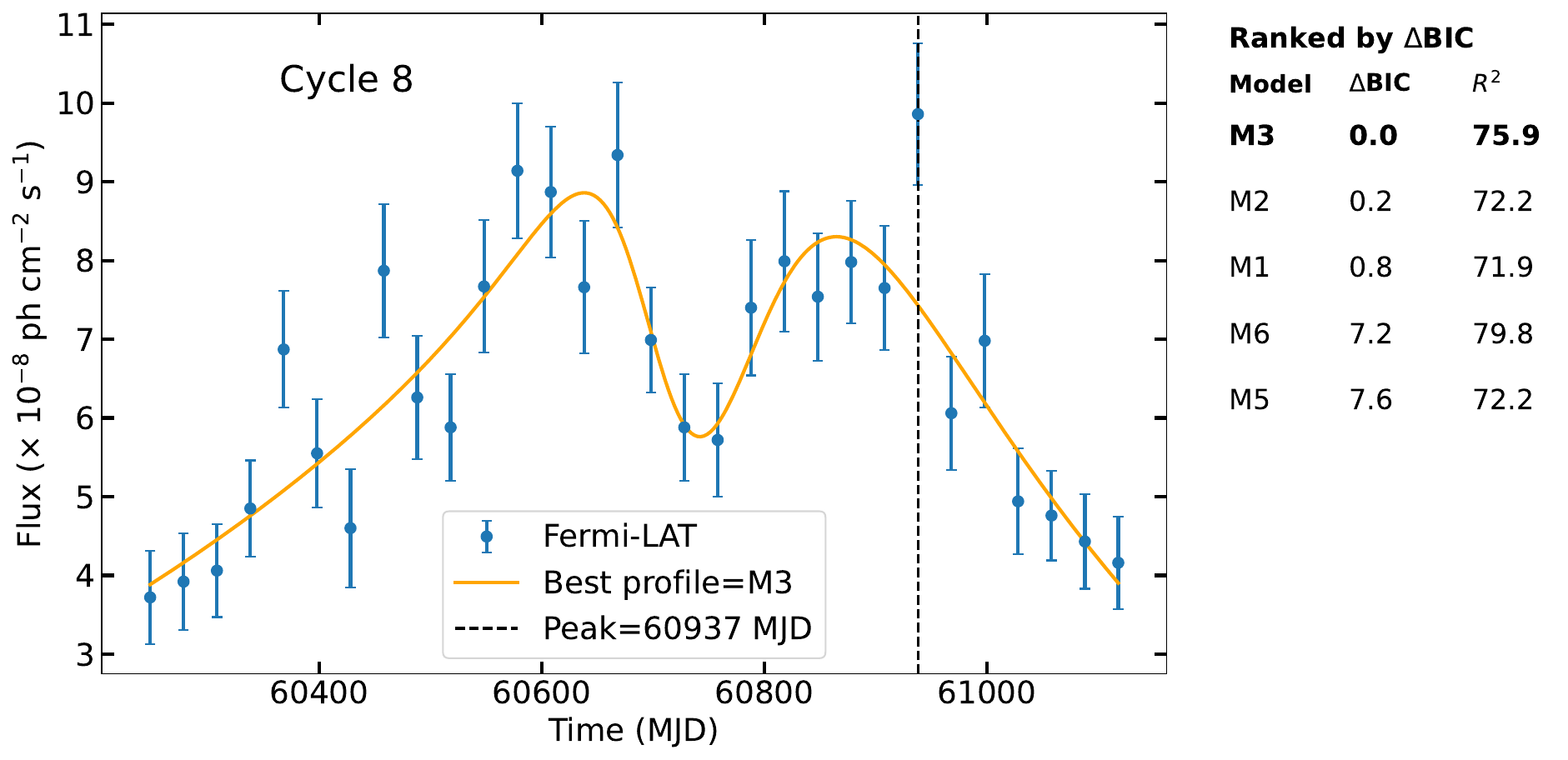}
        \caption{Best-fit profiles for the identified $\gamma$-ray oscillations listed in Table~\ref{tab:oscillation_formology}. Each panel shows the observed light curve, the best-fit profile, the MJD of the maximum observed flux within the cycle, and the corresponding BIC and $R^{2}$ values. The side ranking reports the relative model comparison using $\Delta{\rm BIC}={\rm BIC}-{\rm BIC}_{\rm min}$, where $\Delta{\rm BIC}=0$ identifies the best-ranked model for that oscillation. Models with $\Delta{\rm BIC}<2$ are considered statistically comparable to the best fit ($\S$\ref{sec:profile_comparision}). The oscillation profile could be $S2$ (Lorentzian), $S4$ (FRED), $M3$ (double exponential rise decay), $M5$ (triple Lorentzian), $M7$ (skewed Gaussian plus twin Gaussian), and $M8$ (FRED plus twin Gaussian)} \label{fig:best_fit_gamma}
\end{figure}

\begin{figure*}
        \centering
        \includegraphics[scale=0.274]{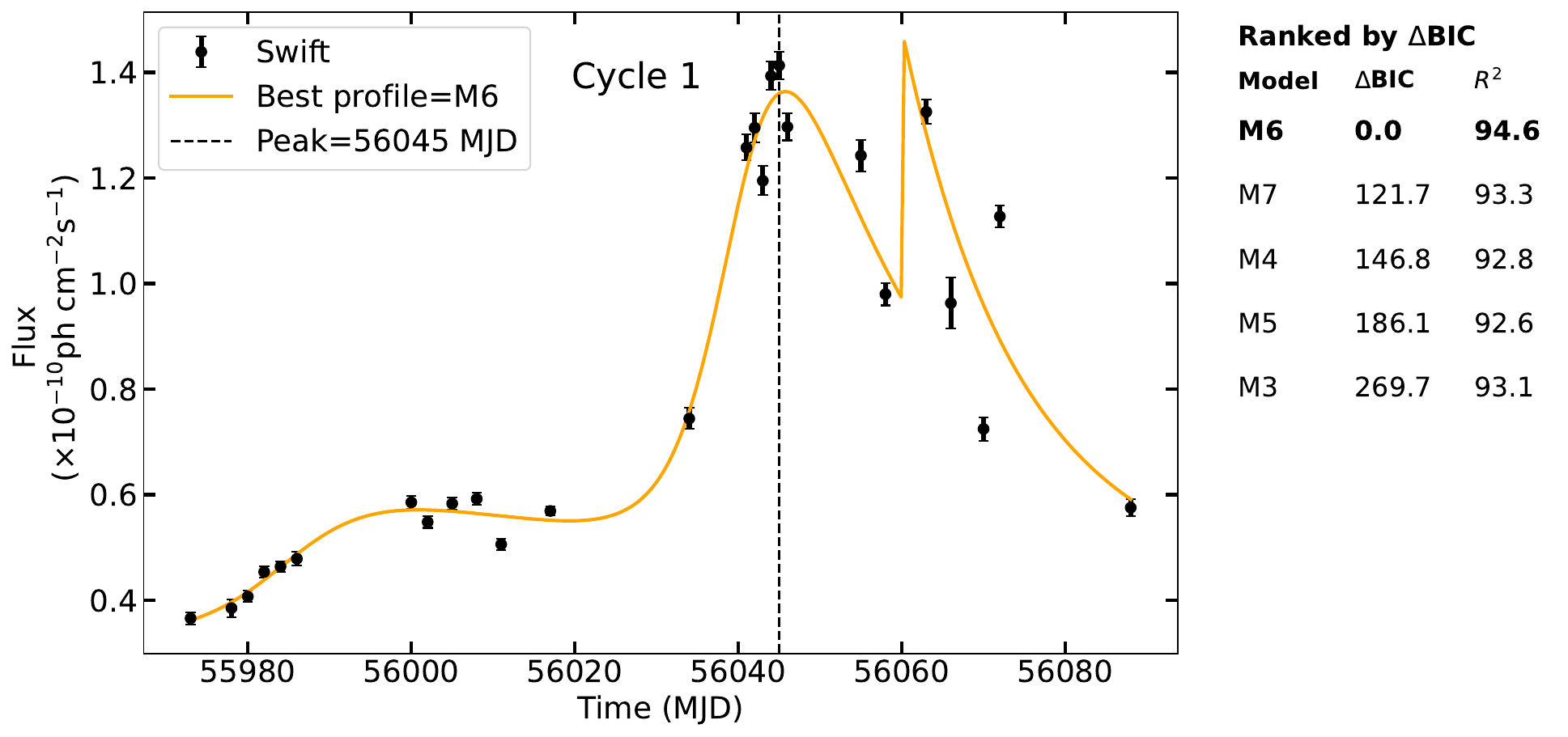}
        \includegraphics[scale=0.274]{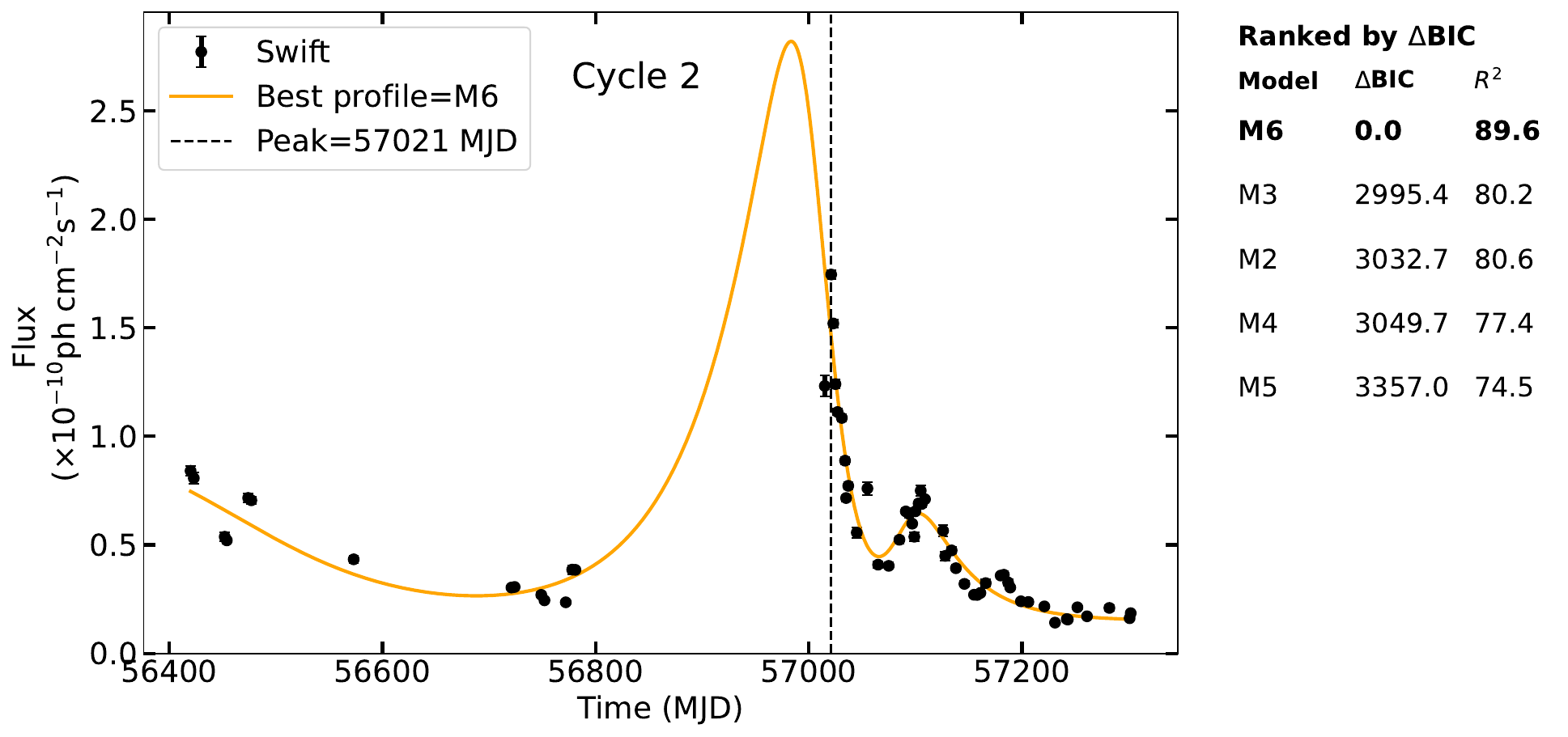}
        \includegraphics[scale=0.274]{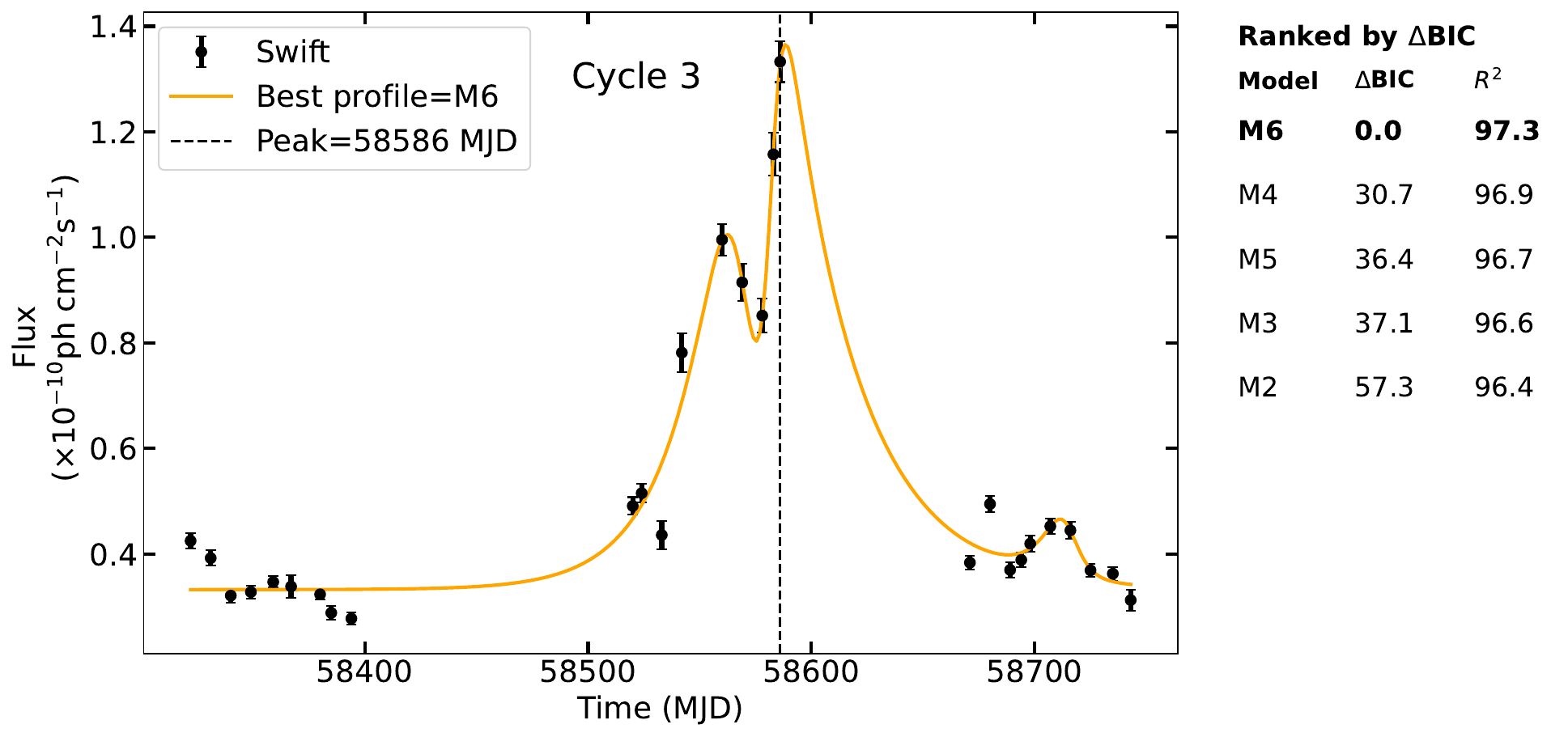}
        \includegraphics[scale=0.274]{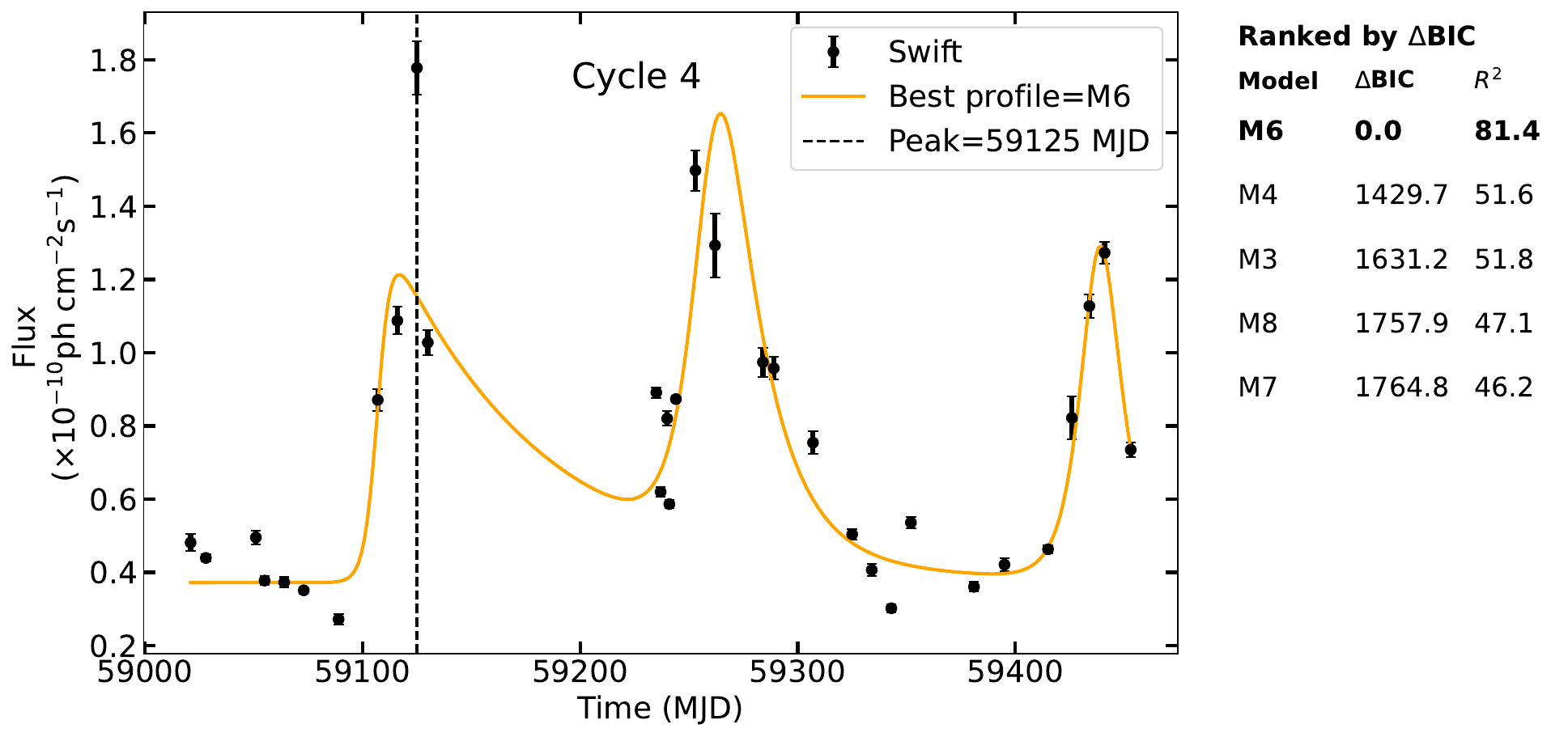}
        \includegraphics[scale=0.274]{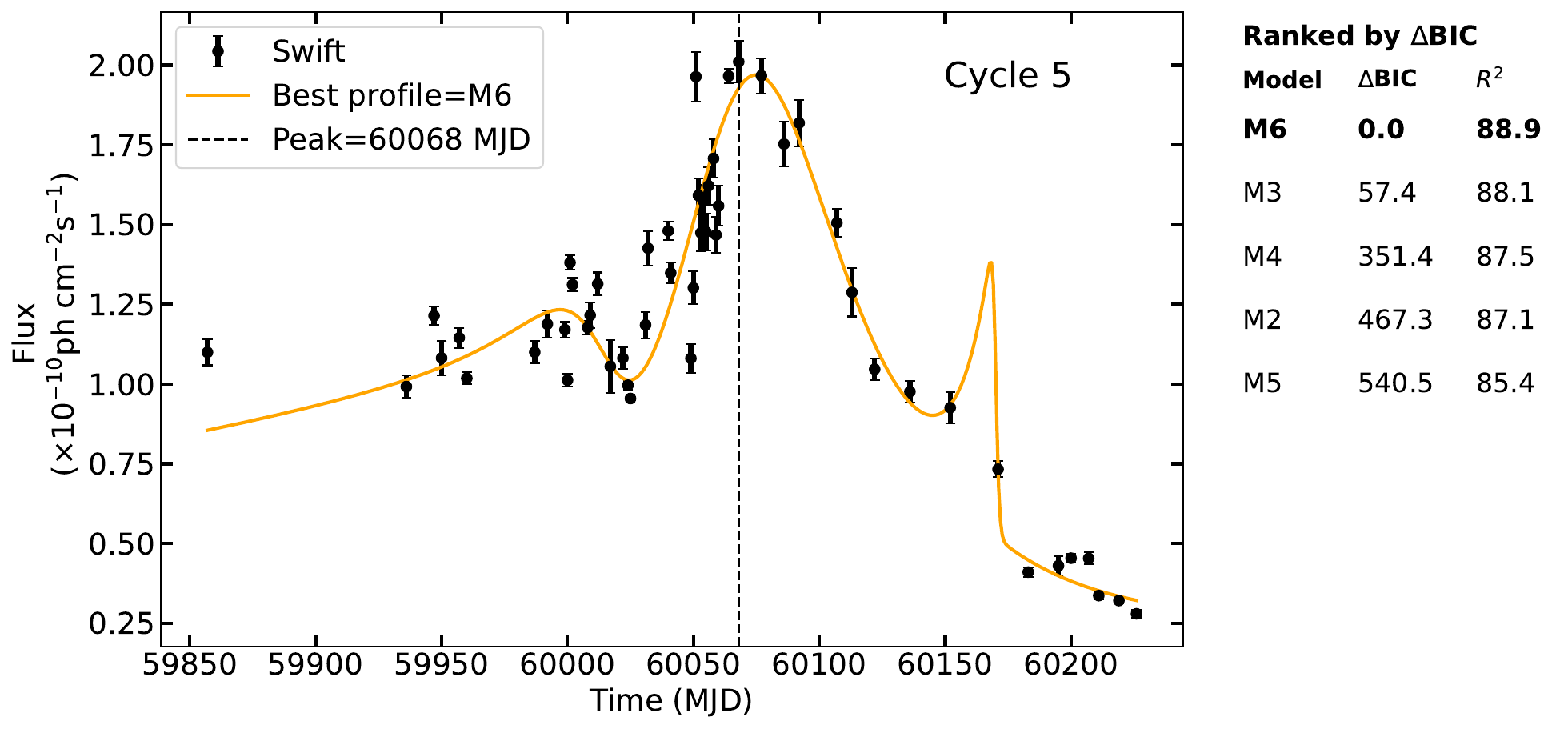}
        \caption{Best-fit profiles for the identified X-ray oscillations listed in Table~\ref{tab:oscillation_formology_xray}. Each panel shows the observed light curve, the best-fit profile, the MJD of the maximum observed flux within the cycle, and the corresponding BIC and $R^{2}$ values. The side ranking reports the relative model comparison using $\Delta{\rm BIC}={\rm BIC}-{\rm BIC}_{\rm min}$, where $\Delta{\rm BIC}=0$ identifies the best-ranked model for that oscillation. Models with $\Delta{\rm BIC}<2$ are considered statistically comparable to the best fit \citep[][]{Kass_Raftery_1995}. The oscillation profile is $M6$, triple exponential.} \label{fig:best_fit_xray}
\end{figure*}

\begin{figure*}
        \centering
        \includegraphics[scale=0.274]{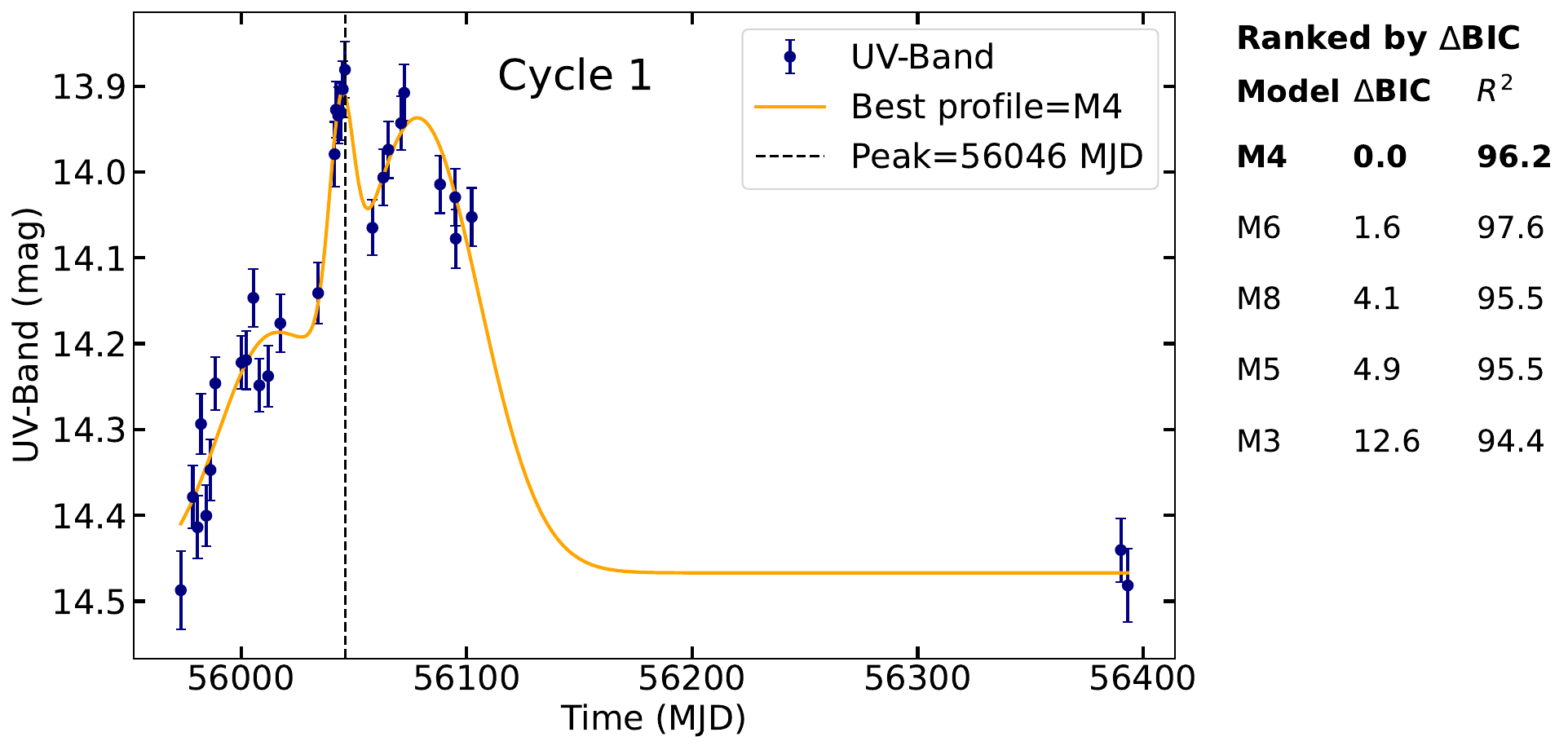}
        \includegraphics[scale=0.274]{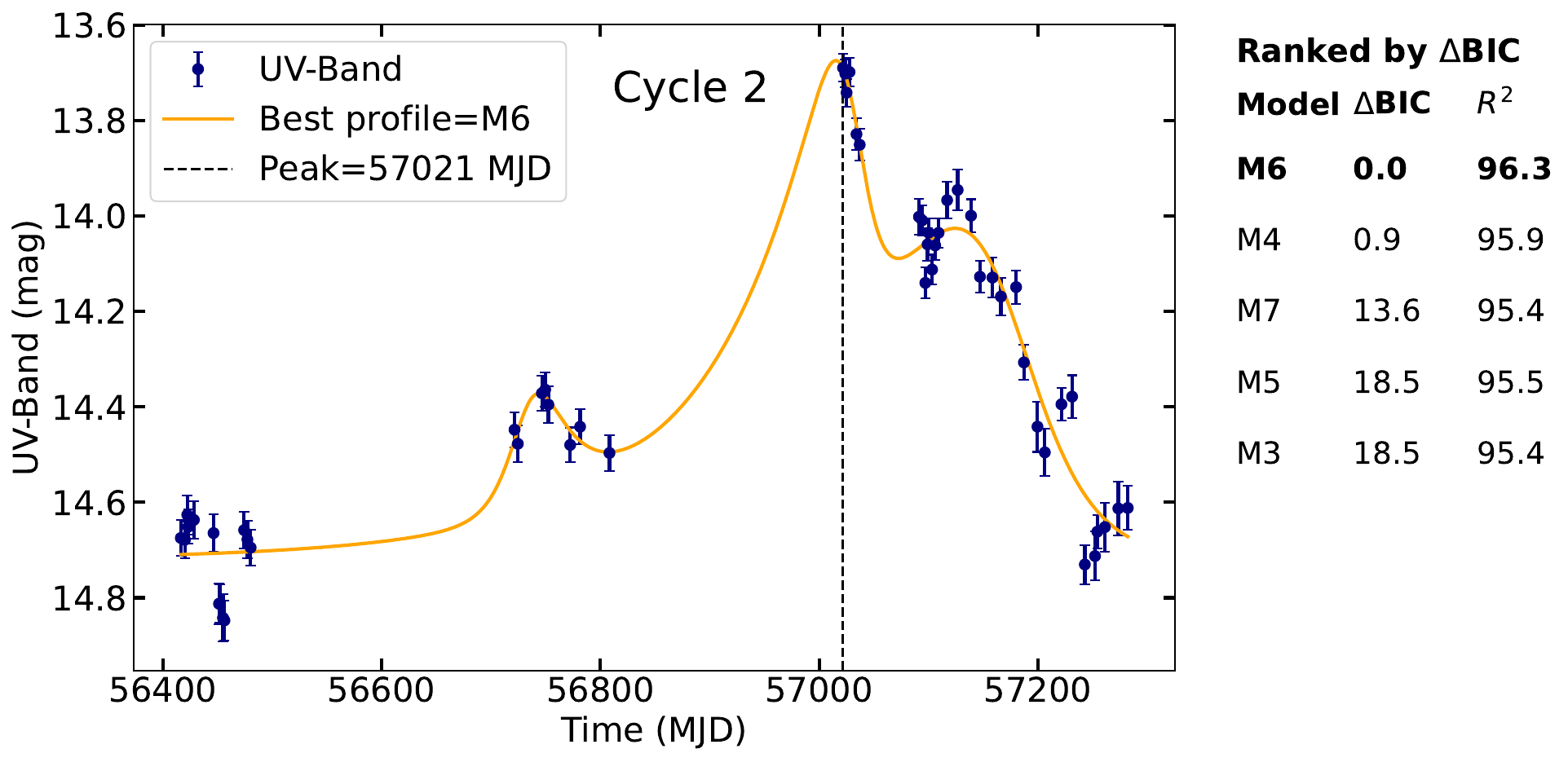}
        \includegraphics[scale=0.274]{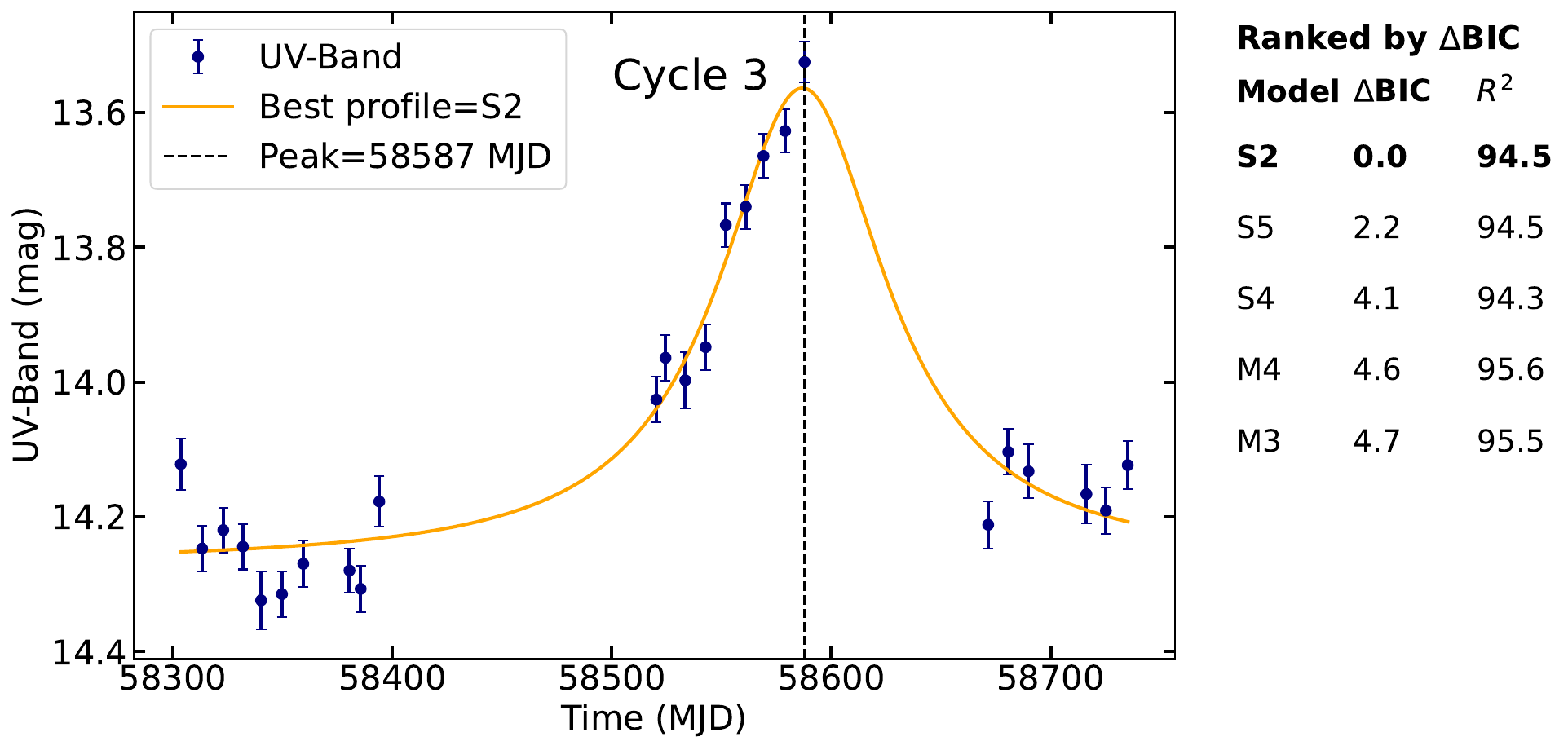}
        \includegraphics[scale=0.274]{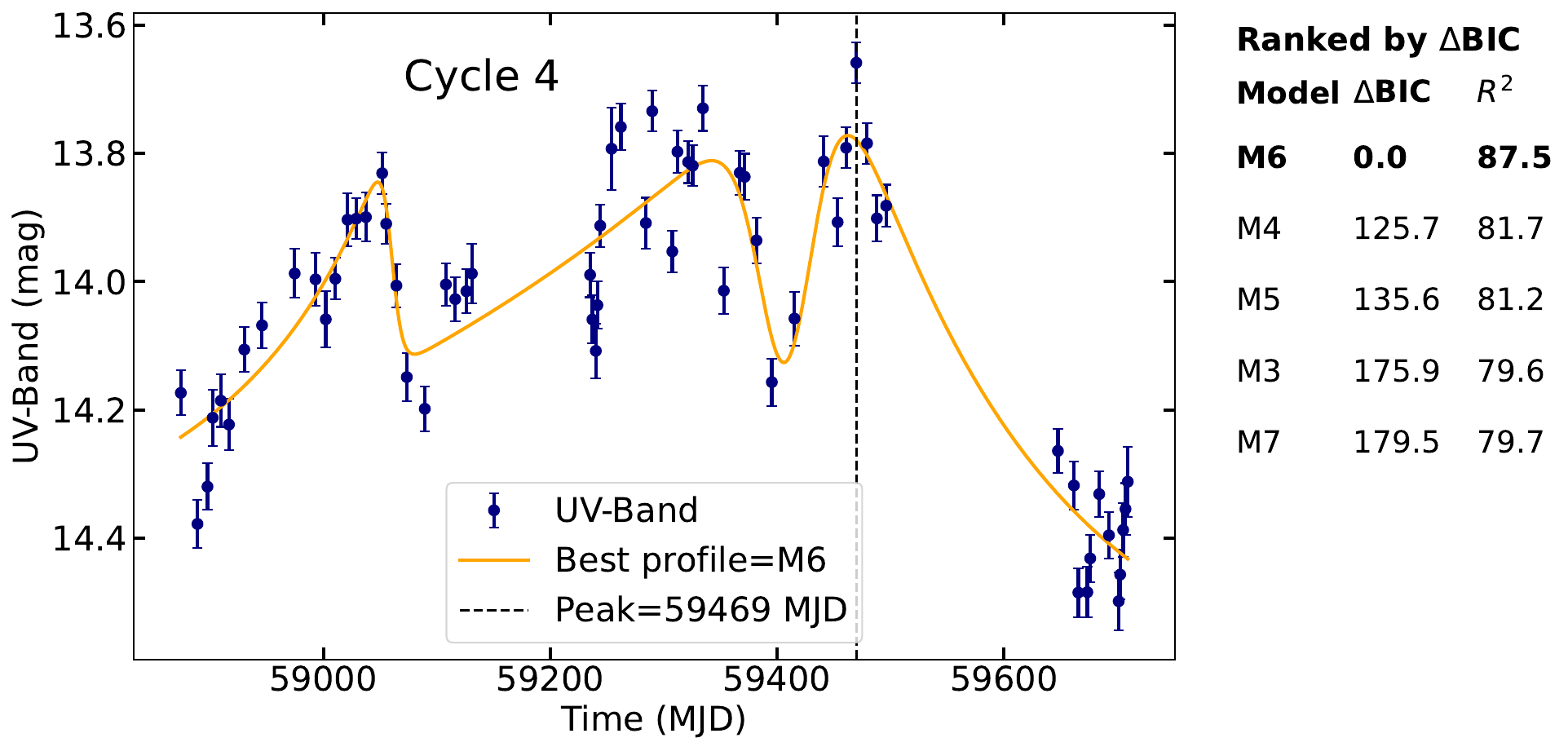}
        \includegraphics[scale=0.274]{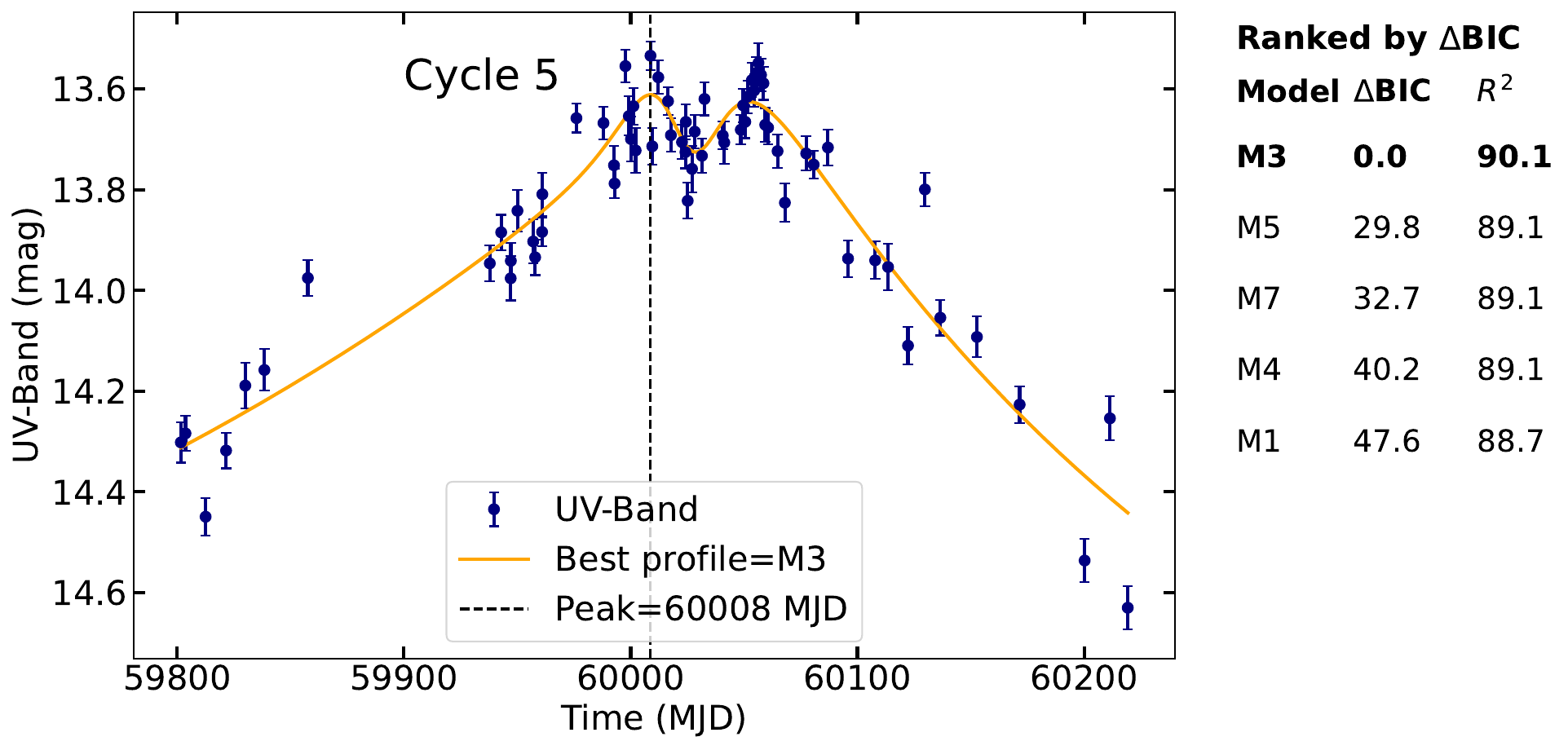}
        \caption{Best-fit profiles for the identified UV oscillations listed in Table~\ref{tab:oscillation_formology_uv}. Each panel shows the observed light curve, the best-fit profile, the MJD of the maximum observed brightness (minimum magnitude) within the cycle, and the corresponding BIC and $R^{2}$ values. The side ranking reports the relative model comparison using $\Delta{\rm BIC}={\rm BIC}-{\rm BIC}_{\rm min}$, where $\Delta{\rm BIC}=0$ identifies the best-ranked model for that oscillation. Models with $\Delta{\rm BIC}<2$ are considered statistically comparable to the best fit ($\S$\ref{sec:profile_comparision}). The oscillation profile could be $S2$ (Lorentzian), $M3$, double exponential, $M4$ (triple Gaussian), and $M6$ (triple exponential)} \label{fig:best_fit_uv}
\end{figure*}

\begin{figure*}
        \centering
        \includegraphics[scale=0.274]{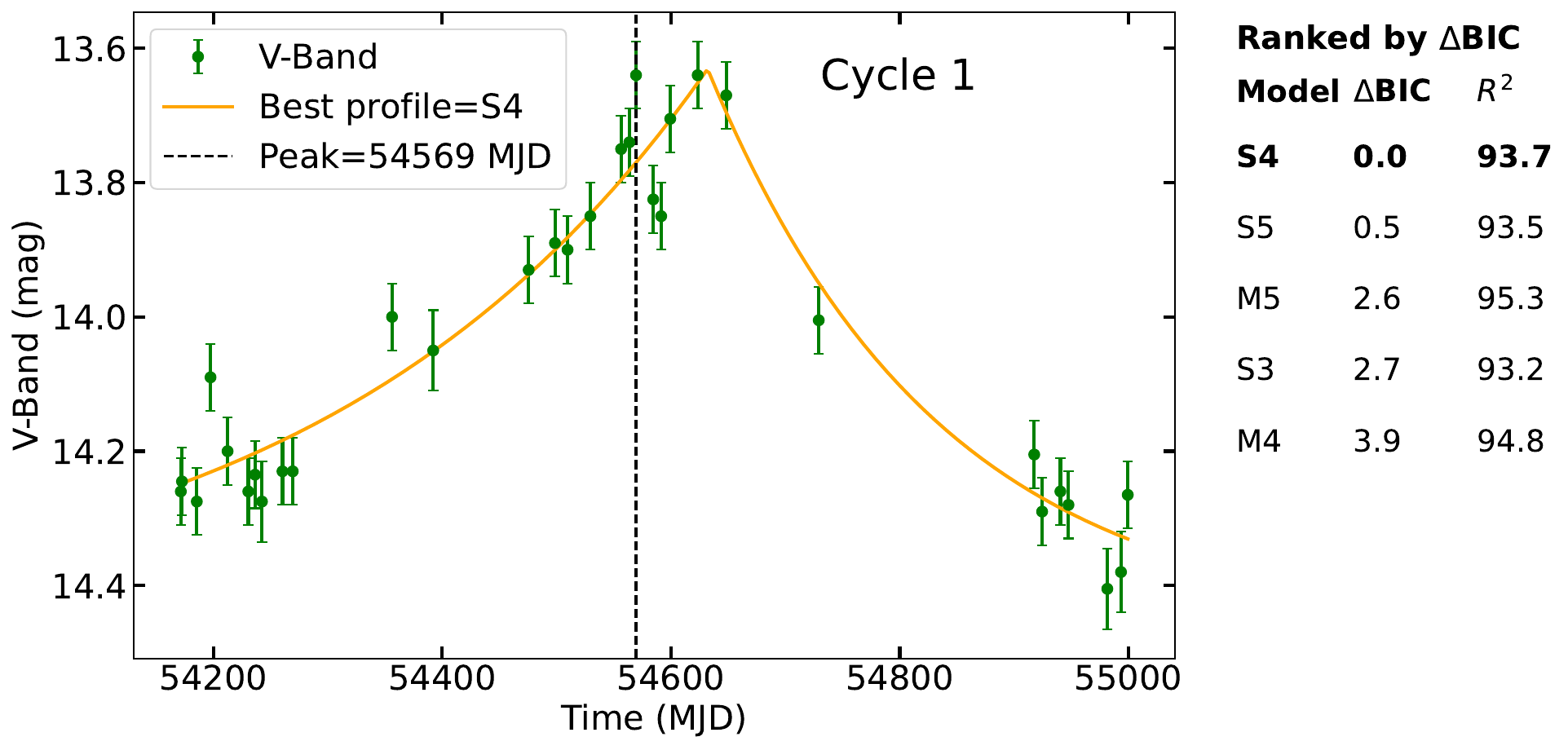}
        \includegraphics[scale=0.274]{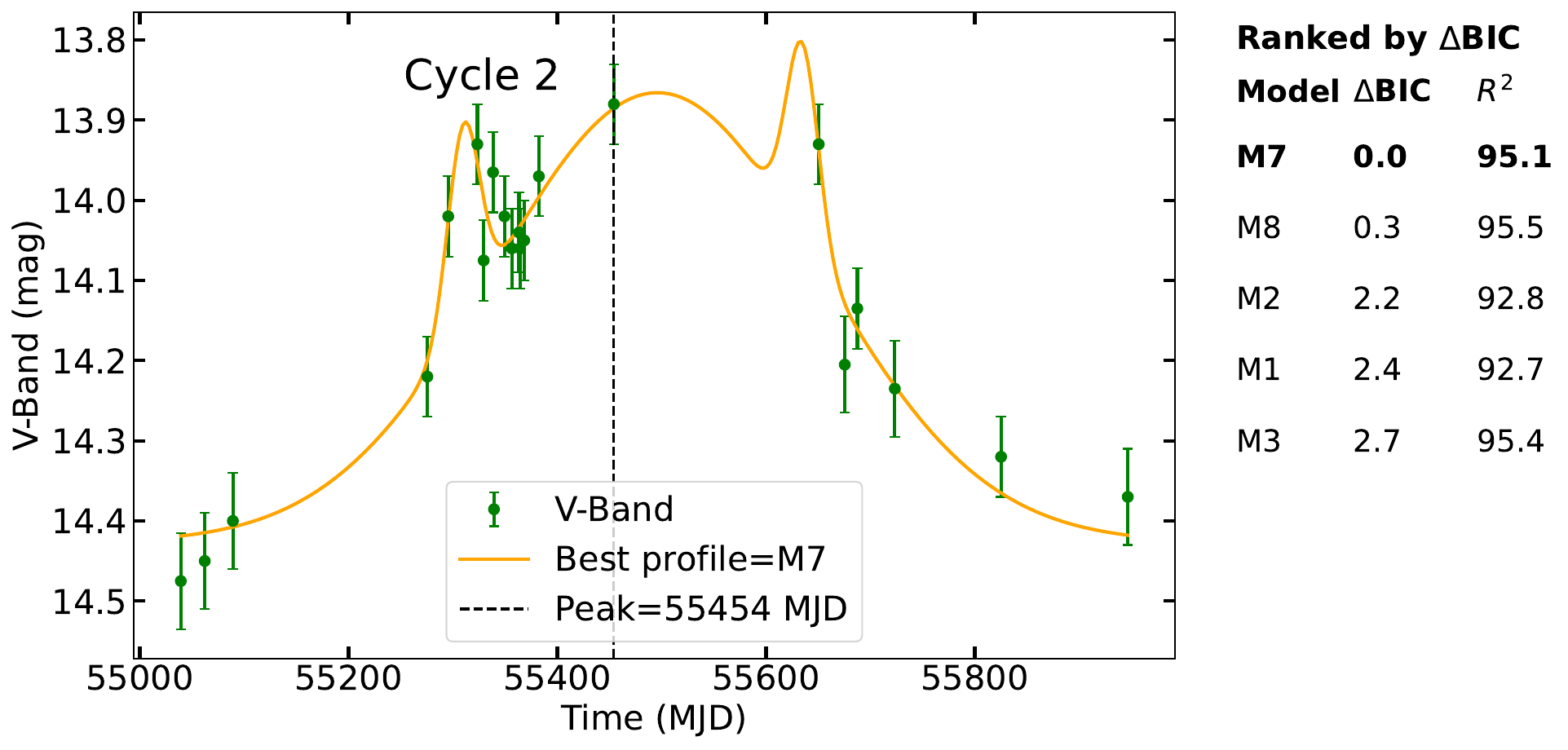}
        \includegraphics[scale=0.274]{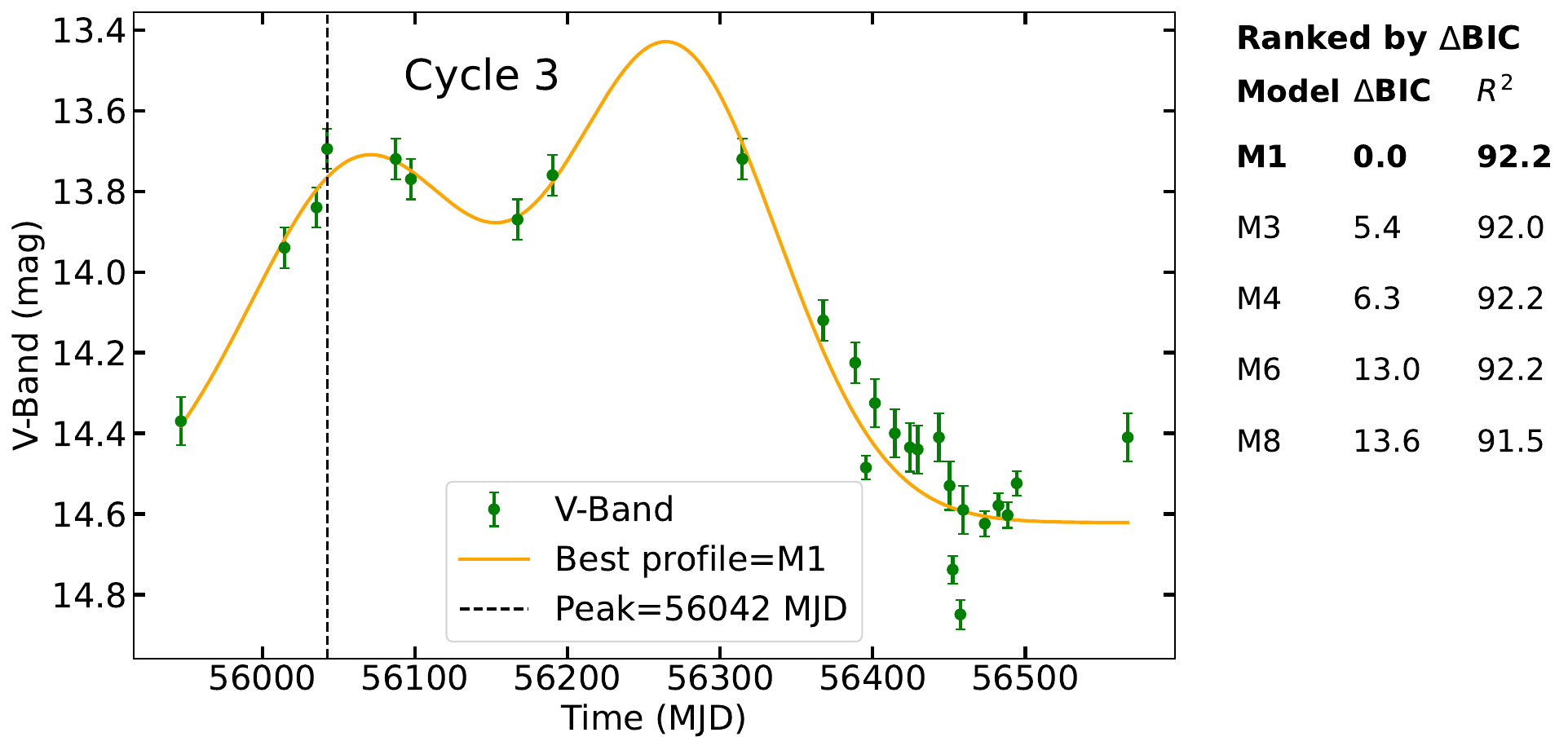}
        \includegraphics[scale=0.274]{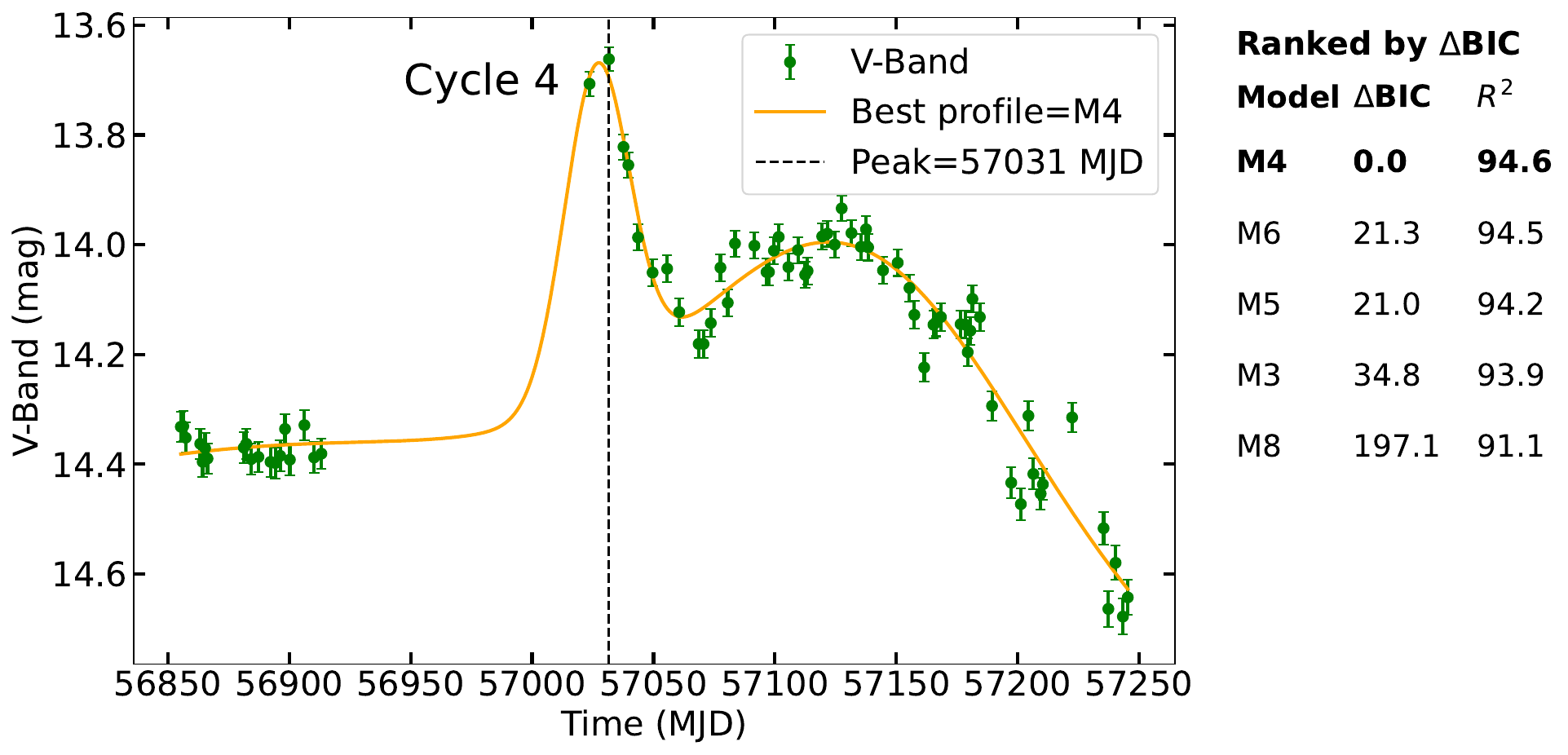}
        \includegraphics[scale=0.274]{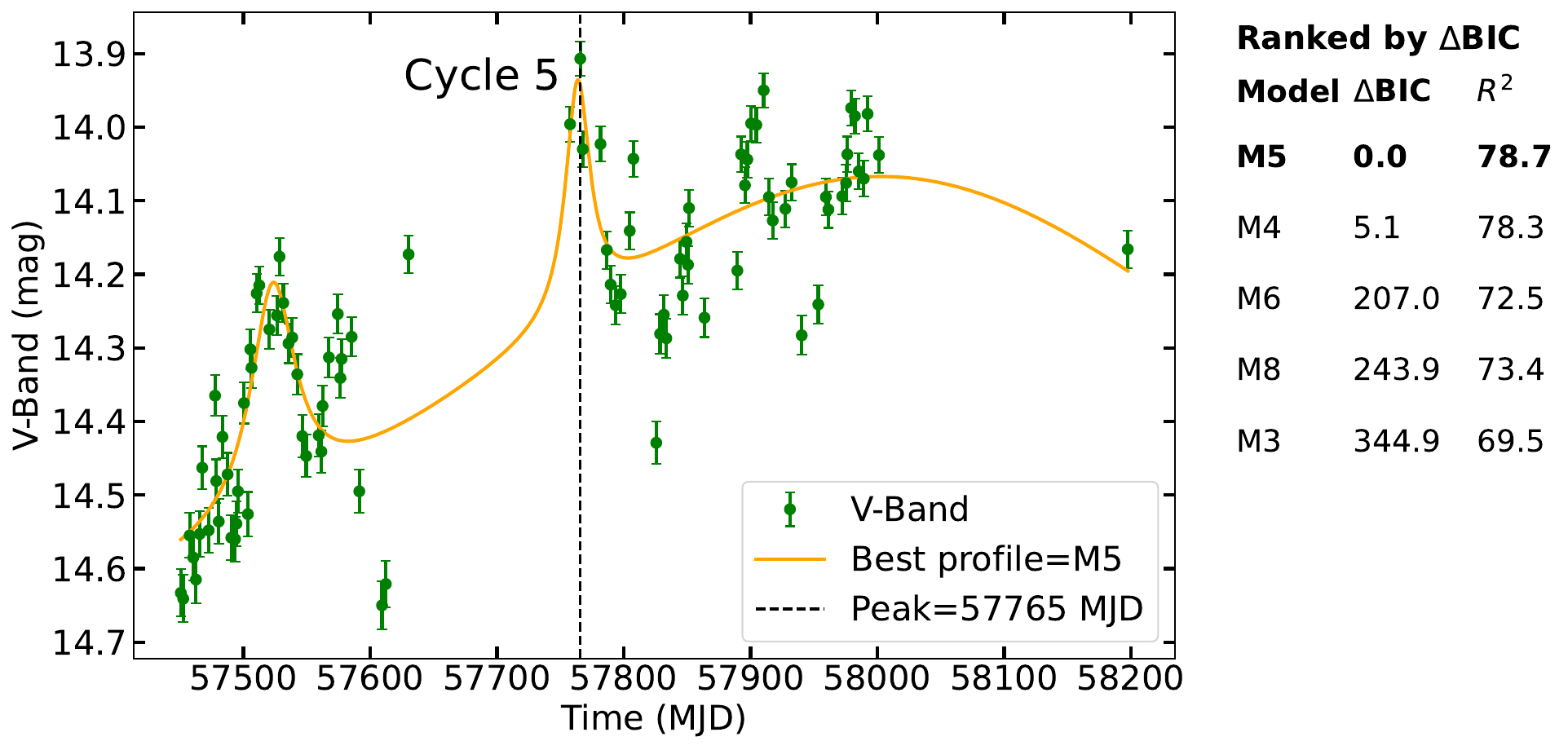}
        \includegraphics[scale=0.274]{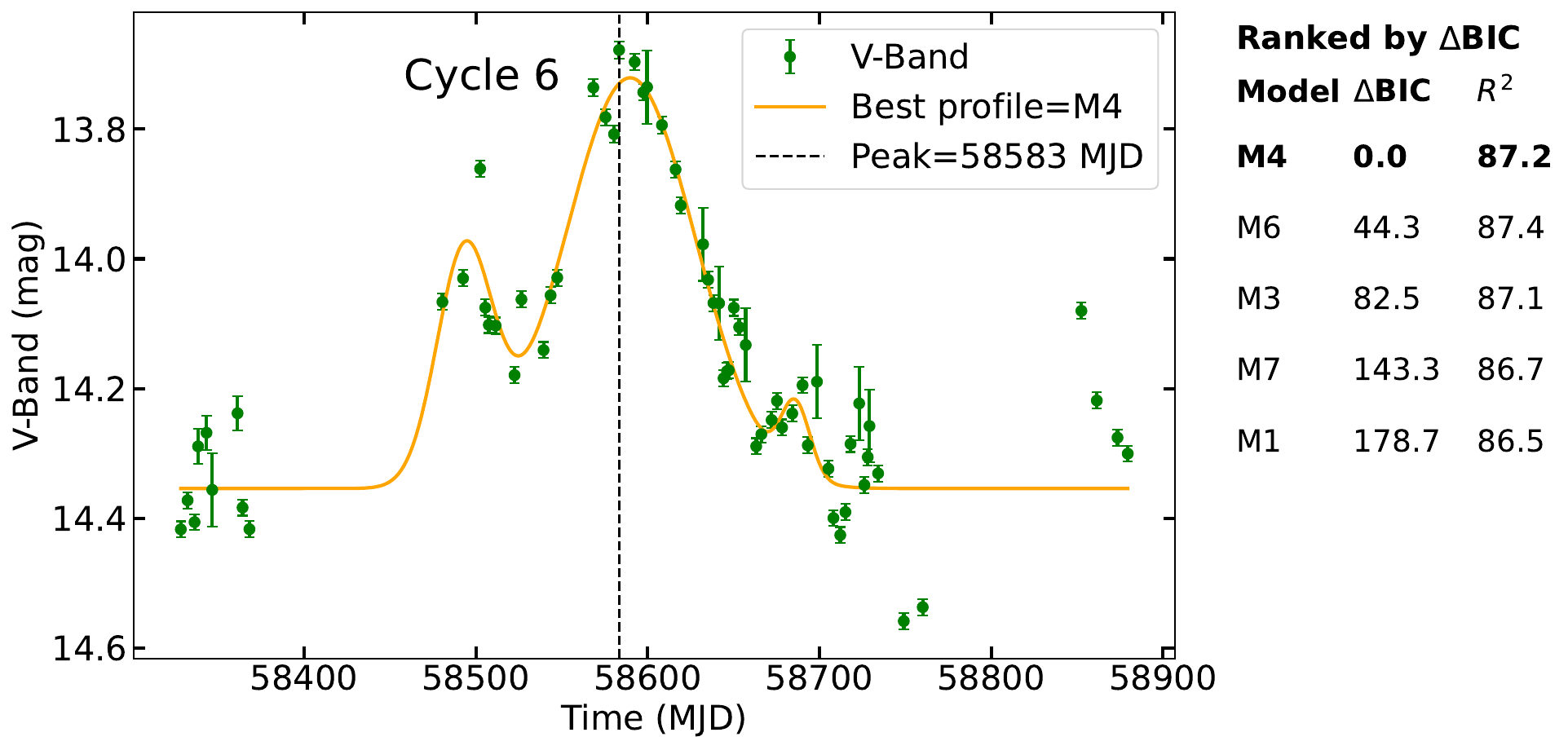}
        \includegraphics[scale=0.274]{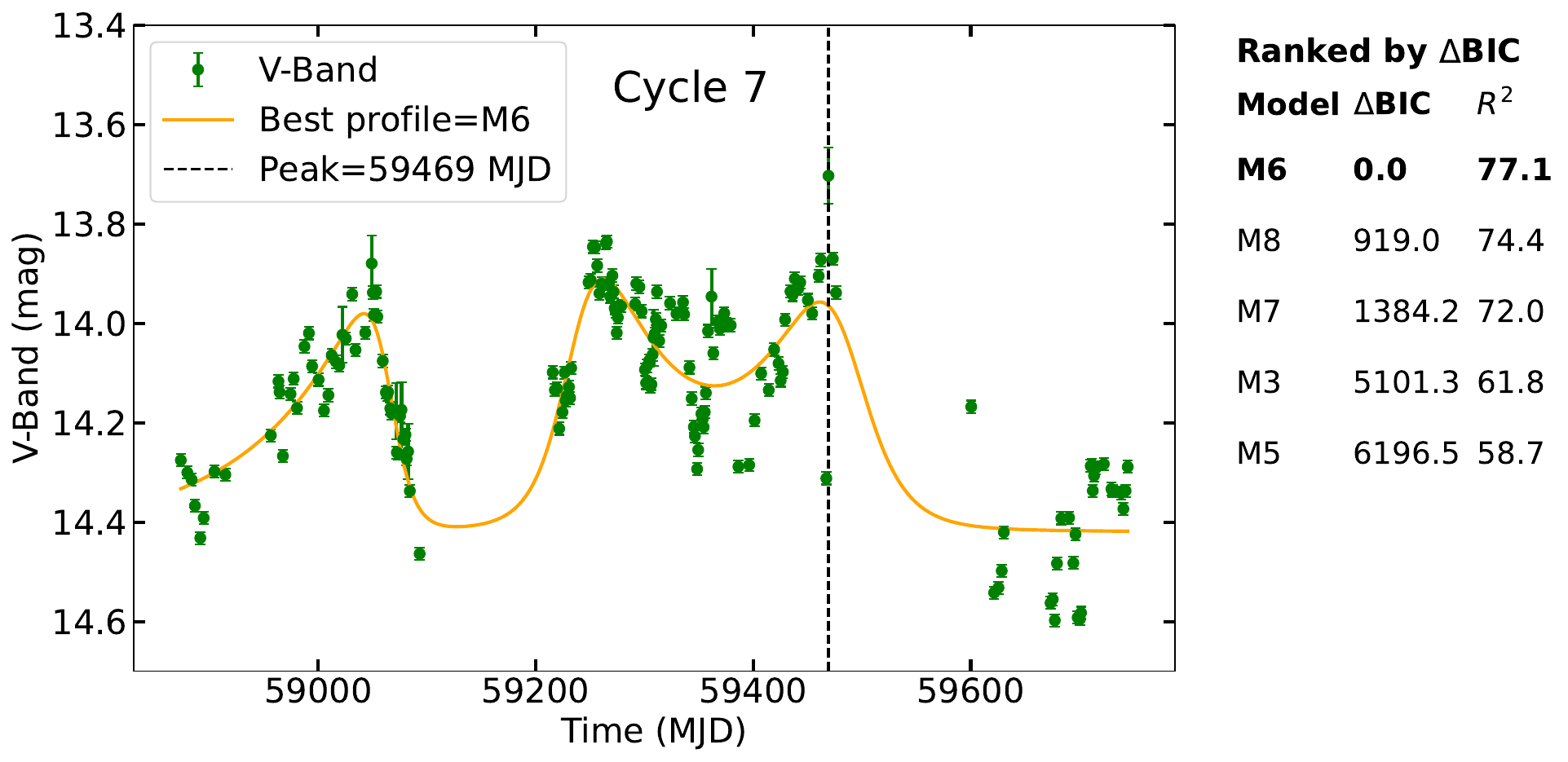}
        \includegraphics[scale=0.274]{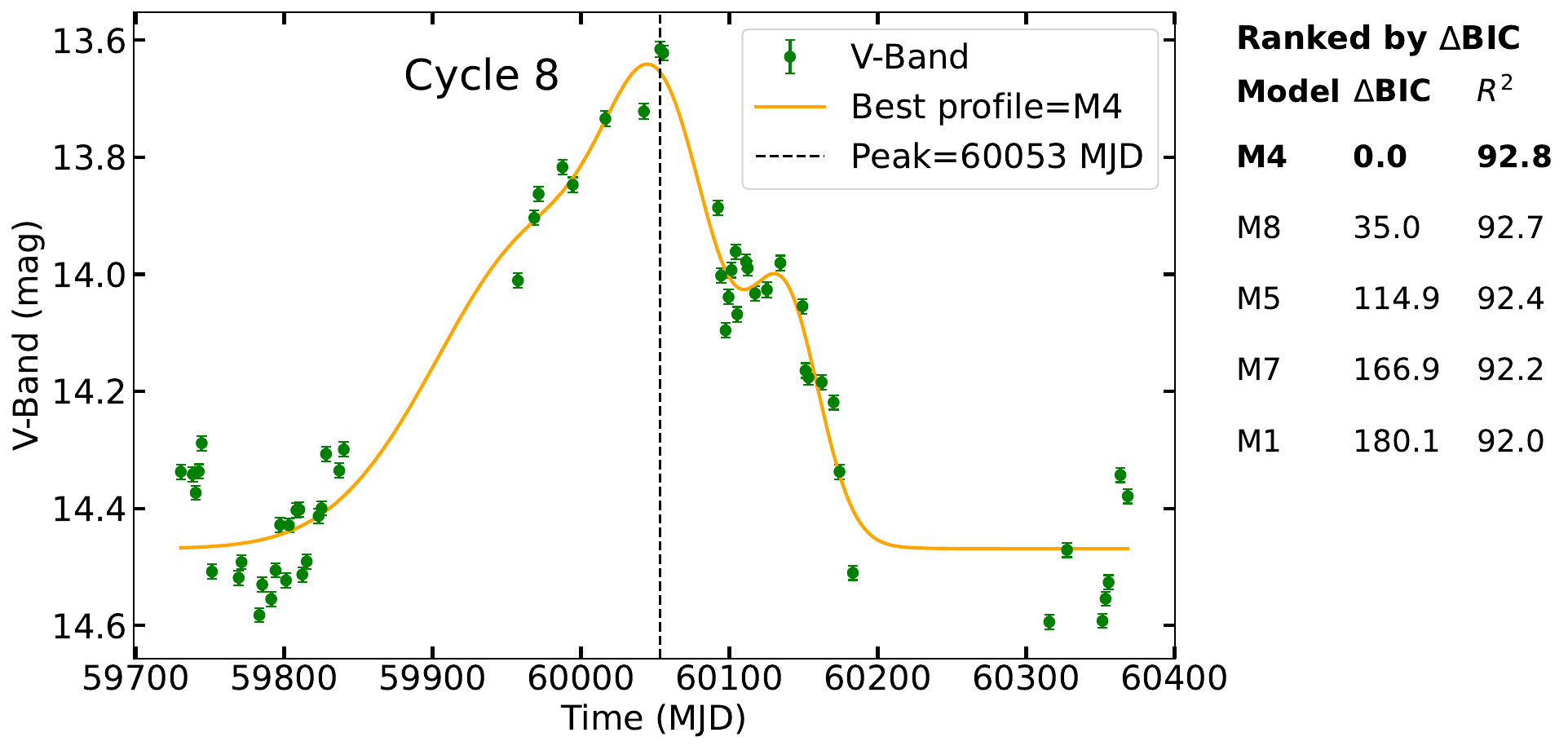}
        \caption{Best-fit profiles for the identified optical oscillations listed in Table~\ref{tab:oscillation_formology_optical}. Each panel shows the observed light curve, the best-fit profile, the MJD of the maximum observed brightness (minimum magnitude) within the cycle, and the corresponding BIC and $R^{2}$ values. The side ranking reports the relative model comparison using $\Delta{\rm BIC}={\rm BIC}-{\rm BIC}_{\rm min}$, where $\Delta{\rm BIC}=0$ identifies the best-ranked model for that oscillation. Models with $\Delta{\rm BIC}<2$ are considered statistically comparable to the best fit ($\S$\ref{sec:profile_comparision}). The oscillation profile could be $S4$ (FRED), $M1$ (double Gaussian), $M4$ (triple Gaussian), $M5$ (triple Lorentzian), $M6$ (triple exponential), and $M7$ (skewed Gaussian plus twin Gaussian)} \label{fig:best_fit_optical}
\end{figure*}

This subsection presents the individual best-fit profiles obtained for each identified oscillation in the $\gamma$-ray (Figure \ref{fig:best_fit_gamma}), X-ray (Figure \ref{fig:best_fit_xray}), UV (Figure \ref{fig:best_fit_uv}), and optical (Figure \ref{fig:best_fit_optical}) light curves. For each cycle, the figures show the observed data, the preferred empirical profile, the main peak location, the relative ranking of the closest competing models according to $\Delta{\rm BIC}$, and the corresponding $R^{2}$.

\clearpage

\subsection{Tables}

\FloatBarrier

This section presents the tables associated with the characterization of the oscillation profiles of PG~1553+113. These tables summarize the best-fit profile and morphological parameters, and properties of the contemporaneous MWL oscillation tuples discussed in the main text.

\begin{table*}
\centering
\caption{List of the parameters to characterize the best profile fit of each oscillation identified in the $\gamma$ rays. The table shows the day localization of the main peak time, oscillation profile, BIC, and R$^{2}$, the main peak amplitude (Flux $\times 10^{-8}$ ph cm$^{-2}$ s$^{-1}$), the rise and decay time of the oscillation, the symmetry of the peak (if it is $\approx$0 the peak is considered symmetric). The reported main-peak time corresponds to the epoch of the maximum observed flux within each cycle. In the case that the oscillation had multiple peaks, we show the structure fraction to measure how much of the oscillation is contributed by the secondary peaks. The oscillation profile could be $S2$ (Lorentzian), $S4$ (FRED), $M3$ (double exponential rise decay), $M5$ (triple Lorentzian), $M7$ (skewed Gaussian plus twin Gaussian), and $M8$ (FRED plus twin Gaussian).} \label{tab:oscillation_formology}
\resizebox{\textwidth}{!}{%
\begin{tabular}{ccccccccccc}
\hline\hline
Cycle Number &  Main  Peak &  Profile   &  BIC & R$^{2}$ &   Peak Amplitude  &   Width  &   Rise/Decay Time & Symmetry &  Structure Fraction\\
             &    [MJD]    &            &      &   [\%]  &   [Flux]    &    [MJD]  &     [MJD]         &      &   [\%]             \\
\hline
Cycle 1 & 55387 & M7 & 31.3 & 87.9 & 5.1 & 284.6 & 148.5/136.1 & -0.04 & 47.1 \\
Cycle 2 & 56197 & S2 & 31.1 & 61.5 & 3.4 & 180.2 & 90.2/90.1 & 0.0 & -- \\
Cycle 3 & 57007 & S4 & 39.7 & 82.5 & 4.7 & 322.1 & 96.5/225.6 & 0.41 & -- \\
Cycle 4 & 57727 & M5 & 44.9 & 82.5 & 6.2 & 631.6 & 235.9/395.6 & 0.25 & 59.8 \\
Cycle 5 & 58567 & S2 & 34.8 & 77.4 & 4.7 & 338.1 & 169.1/169.1 & 0.0 & -- \\
Cycle 6 & 59167 & M7 & 119.5 & 62.9  & 9.5 & 631.3 & 324.1/307.1 & -0.02 & 60.1 \\
Cycle 7 & 60067 & M8 & 35.3 & 95.3 & 7.4 & 378.5 & 175.1/203.4 & 0.07 & 54.3 \\
Cycle 8 & 60937 & M3 & 63.6 & 75.9 & 4.7 & 752.3 & 313.3/438.9 & 0.16 & 43.1 \\
\hline
\end{tabular}%
}
\end{table*}
\begin{table*}
\centering
\caption{List of the parameters to characterize the best profile fit of each oscillation identified in the X-rays. The table shows the day localization of the main peak, oscillation profile, BIC, and R$^{2}$, the main peak amplitude (Flux $\times 10^{-10}$ ph cm$^{-2}$ s$^{-1}$), the rise and decay time of the oscillation, the symmetry of the peak (if it is $\approx$0 the peak is considered symmetric). In the case that the oscillation had multiple peaks, we show the structure fraction to measure how much of the oscillation is contributed by the secondary peaks. The oscillation profile is $M6$, triple exponential.} \label{tab:oscillation_formology_xray}
\resizebox{\textwidth}{!}{%
\begin{tabular}{ccccccccccc}
\hline\hline
Cycle Number &  Main  Peak &  Profile   &  BIC & R$^{2}$ &   Peak Amplitude  &   Width  &   Rise/Decay Time & Symmetry &  Structure Fraction\\
             &    [MJD]    &            &      &   [\%]  &   [Flux]    &    [MJD]  &     [MJD]         &      &   [\%]             \\
\hline
Cycle 1 & 56045 & $M6$ & 395.6 & 94.6 & 0.9 & 59.3 & 31.7/27.6 & 0.07 & 36.3 \\
Cycle 2 & 57021 & $M6$ & 1653.7 & 89.6 & 2.4 & 196.8 & 135.8/61.1 & -0.37 & 31.4 \\
Cycle 3 & 58586 & $M6$ & 231.7 & 97.3 & 1.0 & 148.7 & 73.3/75.43 & 0.01 & 51.3 \\
Cycle 4 & 59125 & $M6$ & 1165.7 & 81.4 & 1.1 & 296.6 & 175.2/121.4 & -0.18 & 56.1 \\
Cycle 5 & 60068 & $M6$ & 822.6 & 88.9 & 1.4 & 211.6 & 136.2/75.4 & -0.29 & 37.7 \\
\hline
\end{tabular}%
}
\end{table*}
\begin{table*}
\centering
\caption{List of the parameters to characterize the best profile fit of each oscillation identified in the UV. The table shows the day localization of the main peak, oscillation profile, BIC, and R$^{2}$, the main peak amplitude relative to the estimated local baseline, expressed in magnitudes, the rise and decay time of the oscillation, the symmetry of the peak (if it is $\approx$0 the peak is considered symmetric). In the case that the oscillation had multiple peaks, we show the structure fraction to measure how much of the oscillation is contributed by the secondary peaks. The oscillation profile could be $S2$ (Lorentzian), $M3$ (double exponential), $M4$ (triple Gaussian), and $M6$ (triple exponential).} \label{tab:oscillation_formology_uv}
\resizebox{\textwidth}{!}{%
\begin{tabular}{ccccccccccc}
\hline\hline
Cycle Number &  Main  Peak &  Profile   &  BIC & R$^{2}$ &   Peak Amplitude  &   Width  &   Rise/Decay Time & Symmetry &  Structure Fraction\\
             &    [MJD]    &            &      &   [\%]  &   [mag]    &    [MJD]  &     [MJD]         &      &   [\%]             \\
\hline
Cycle 1 & 56046 & M4 & 67.8 & 96.2 & 0.53 & 320.8 & 67.1/253.7 & 0.58 & 55.9 \\
Cycle 2 & 57021 & M6 & 187.8 & 96.3 & 1.01 & 543.3 & 314.7/228.5 & -0.16 & 55.3 \\
Cycle 3 & 58587 & S2 & 69.1 & 94.5 & 0.66 & 217.7 & 108.9/108.8 & 0.0 & -- \\
Cycle 4 & 59469 & M6 & 325.9 & 87.5 & 0.63 & 736.8 & 588.1/148.7 & -0.59 & 52.3 \\
Cycle 5 & 60008 & M3 & 368.5 & 90.1 & 0.75 & 354.4 & 187.6/166.8 & -0.05 & 46.1 \\
\hline
\end{tabular}%
}
\end{table*}

\clearpage

\begin{table*}
\centering
\caption{List of the parameters to characterize the best profile fit of each oscillation identified in the optical band. The table shows the day localization of the main peak, oscillation profile, BIC, and R$^{2}$, the main peak amplitude relative to the estimated local baseline, expressed in magnitudes, the rise and decay time of the oscillation, the symmetry of the peak (if it is $\approx$0 the peak is considered symmetric). In the case that the oscillation had multiple peaks, we show the structure fraction to measure how much of the oscillation is contributed by the secondary peaks. The oscillation profile could be $S4$ (FRED), $M1$ (double Gaussian), $M4$ (triple Gaussian), $M5$ (triple Lorentzian), $M6$ (triple exponential), and $M7$ (skewed Gaussian plus twin Gaussian). The value $--$ denotes not converging value in the fit process.} \label{tab:oscillation_formology_optical}
\resizebox{\textwidth}{!}{%
\begin{tabular}{ccccccccccc}
\hline\hline
Cycle Number &  Main  Peak &  Profile   &  BIC & R$^{2}$ &   Peak Amplitude  &   Width  &   Rise/Decay Time & Symmetry &  Structure Fraction\\
             &    [MJD]    &            &      &   [\%]  &   [mag]    &    [MJD]  &     [MJD]         &        &   [\%]             \\
\hline
Cycle 1 & 54569 & S4 & 60.9 & 93.7 & 0.65 & 552.5 & 442.3/110.1 & -0.61 & --  \\
Cycle 2 & 55454 & M7 & 35.6 & 95.1  & 0.62 & 630.1 & 451.9/178.3 & -0.43 & 46.9 \\
Cycle 3 & 56042 & M1 & 155.9 & 92.2 & 1.07 & 448.1 & 317.8/130.2 & -0.42 & 43.2 \\
Cycle 4 & 57031 & M4 & 327.7 & 94.6 & 0.77 & 326.1 & 149.9/176.1 & 0.08 & 45.4 \\
Cycle 5 & 57765 & M5 & 1016.3 & 78.7 & 0.39 & 484.4 & 315.7/169.4 & -0.30 & 40.9 \\
Cycle 6 & 58583 & M4 & 1964.7 & 87.2 & 0.65 & 233.8 & 126.8/106.9 & -0.08 & 42.3 \\
Cycle 7 & 59469 & M7 & 7914.5 & 77.1 & 0.46 & 340.8 & -- & -- & 63.3 \\
Cycle 8 & 60053 & M4 & 2058.6 & 92.8 & 1.23 & 342.9 & 207.4/135.5 & -0.2 & 62.6 \\
\hline
\end{tabular}%
}
\end{table*}
\begin{table*}
\centering
\caption{Results of the analysis of contemporaneous MWL oscillation tuples. For each tuple and energy band, the table reports the oscillation cycle, the MJD locations of the identified subpeaks, the temporal separations between the identified subpeaks, and the relative amplitudes normalized to the strongest component in each band. The reported peak times correspond to the centers of the fitted model components. For multi-component profiles, the center of the highest-amplitude component may differ from the maximum observed flux and from the maximum of the total fitted profile because of component overlap. Missing components are indicated by ``--''.} \label{tab:tuple_5}
\begin{tabular}{ccccccccccc}
\hline\hline
Tuple &  Energy Band &  Cycle & Peak Time  &  Peak Separation &   Amplitude ratios  \\
      &              &        &    [MJD]   &    [MJD]         &       [\%]           \\
\hline
Tuple 1 & \makecell{$\gamma$ rays \\ X-rays \\ UV \\ Optical} & \makecell{3 \\ 2 \\ 2 \\ 4} & \makecell{(--, 57008, --) \\ (--, 56985, 57102) \\ (56744, 57015, 57124) \\ (--, 57028, 57115)} & \makecell{(--,--,--) \\ (--, 117) \\ (271, 109) \\ (--, 87)} & \makecell{(--, 100, --) \\ (-, 100, 10.9) \\ (30.5, 100, 64.9) \\ (--, 100, 48.7)} \\
\hline
Tuple 2 & \makecell{$\gamma$ rays \\ X-rays \\ UV \\ Optical} & \makecell{5 \\ 3 \\ 3 \\ 6} & \makecell{(58561, --, --) \\ (58562, 58588, 58712) \\ (--, 58587, --) \\ (58495, 58590, 58685)} & \makecell{(--, --) \\ (26, 124) \\ (--,--) \\ (95, 95)} & \makecell{(100, --, --) \\ (64.9, 100, 12.4) \\ (--, 100, --) \\ (60.3, 100, 21.8)} \\
\hline
Tuple 3 & \makecell{$\gamma$ rays \\ X-rays \\ UV \\ Optical} & \makecell{6 \\4 \\4 \\7} & \makecell{(59025, 59286, 59530) \\ (59117, 59264, 59439) \\ (59048, 59342, 59461) \\ (59043, 59261, 59461)} & \makecell{(261, 244) \\ (147, 175) \\ (294, 119) \\ (218, 200)} & \makecell{(84.2, 100, 88.2) \\ (64.9, 100, 71.2) \\ (81.4, 89.9, 100) \\ (83.9, 100, 89.3)} \\
\hline
Tuple 4 & \makecell{$\gamma$ rays \\ X-rays \\ UV \\ Optical} & \makecell{7 \\5 \\ 5 \\8} & \makecell{(59987, 60078, 60169)\\ (59996, 60075, 60168) \\ (60009, 60051, --) \\ (--, 60044, 60130)} & \makecell{(91, 91)\\ (79, 93) \\ (42, --) \\ (--, 86)} & \makecell{(100, 97.9, 45.1) \\ (29.8, 100, 44.1) \\ (100, 98.4, --) \\ (--, 100, 39.2) } \\
\hline
\end{tabular}%
\end{table*}

\end{document}